\documentclass[twocolumn,showpacs, showkeys,nofootinbib,preprintnumbers,amsmath,amssymb,superscriptaddress,prd]{revtex4-1}

\usepackage{lipsum}
\usepackage[utf8]{inputenc}
\usepackage{amsmath,ulem}
\usepackage{cases,color}
\usepackage{indentfirst}
\usepackage{natbib}
\usepackage{appendix} 
\usepackage[T1]{fontenc}
\usepackage{bbold}
\usepackage[table]{xcolor}
\usepackage{mathrsfs}
\usepackage{amssymb}
\usepackage{multirow}
\usepackage{slashed}
\usepackage{braket}
\usepackage{graphicx}
\usepackage{float}
\usepackage{cases}
\usepackage{hyperref}

\newcommand{\secref}[1]{Sec.~\ref{#1}}
\newcommand{\be}{\begin{eqnarray}}
\newcommand{\ee}{\end{eqnarray}}
\newcommand{\expm}{e^{-\beta\left(E_p-\mu\right)}}

\newcommand{\expmmm}{e^{-3\beta\left(E_p-\mu\right)}}
\newcommand{\expp}{e^{-\beta\left(E_p+\mu\right)}}
\newcommand{\expppp}{e^{-3\beta\left(E_p+\mu\right)}}
\newcommand{\exppp}{e^{-2\beta\left(E_p+\mu\right)}}
\newcommand{\expmT}{e^{-\left(E_p-\mu\right)/T}}
\newcommand{\expmmT}{e^{-2\left(E_p-\mu\right)/T}}
\newcommand{\expmmmT}{e^{-3\left(E_p-\mu\right)/T}}
\newcommand{\exppT}{e^{-\left(E_p+\mu\right)/T}}
\newcommand{\exppppT}{e^{-3\left(E_p+\mu\right)/T}}
\newcommand{\expppT}{e^{-2\left(E_p+\mu\right)/T}}
\newcommand{\Zed}{\mathbb{Z}}
\newcommand{\quarkdensity}{|\langle{\bar q}q\rangle|^{1/3}}

\def\blr#1{\left(#1\right)}
\def\slr#1{\left[#1\right]}
\newcommand{\bea}{\begin{eqnarray}}
\newcommand{\eea}{\end{eqnarray}}
\newcommand{\bse}{\begin{subequations}}
\newcommand{\ese}{\end{subequations}}
\def\eq#1{(\ref{#1})}
\def\Eq#1{Eq.~(\ref{#1})}
\def\Eqs#1{Eqs.~(\ref{#1})}
\def\Fig#1{Fig.~\ref{#1}}
\newcommand{\mrm}[1]{\mathrm{#1}}
\def\mc#1{\mathcal{#1}}
\def\Phib{\bar{\Phi}}
\def\Phir{\Phi_\mrm{r}}
\def\Phii{\Phi_\mrm{i}}
\def\qbar{\bar{q}}
\def\qqb{{q\bar{q}}}
\def\Ref#1{Ref.~\cite{#1}}
\def\Refs#1{Refs.~\cite{#1}}
\def\Sec#1{Sec.~\ref{#1}}

\hypersetup{
colorlinks=true,
citecolor=blue,
citebordercolor=red,
linktoc=all,
linkcolor=blue
}

\begin{document} 

\title{Quark and Polyakov-loop correlations in effective models\\at zero and non-vanishing density}

\author{Hubert Hansen}\email{hansen@ipnl.in2p3.fr}
\affiliation{Univ.~Lyon, Universit\'e Claude Bernard Lyon 1, CNRS/IN2P3, IP2I Lyon, F-69622, Villeurbanne, France}

\author{Rainer Stiele}\email{rainer.stiele@to.infn.it}
\affiliation{INFN - Sezione di Torino, Via Pietro Giuria 1, I-10125 Torino, Italy}
\address{Univ.~Lyon, ENS de Lyon, D\'epartement de Physique, F-69342, Lyon, France}
\affiliation{Univ.~Lyon, Universit\'e Claude Bernard Lyon 1, CNRS/IN2P3, IP2I Lyon, F-69622, Villeurbanne, France}

\author{Pedro Costa}\email{pcosta@uc.pt}
\affiliation{CFisUC - Center for Physics of the University of Coimbra, Department of Physics, University of Coimbra, P-3004 - 516  Coimbra, Portugal}

\begin{abstract}
	The aim of this work is to shed light on some lesser known aspects of Polyakov-loop--extended chiral models  (namely the Polyakov-loop extended Nambu--Jona-Lasinio and Quark-Meson models), especially on the correlation of the quark sector with the Polyakov loop.
	We show that the ordering of chiral and Polyakov-loop transitions and their difference in temperature as seen in lattice QCD calculations could be realized with a critical scale of the Polyakov-loop potential that is larger than the one in pure gauge theory.
	The comparison  of the results for the Polyakov-loop susceptibility obtained using the self-consistent medium-dependent quark mass with those obtained while keeping these masses at a fixed value allows to disentangle chiral-symmetry restoration and center-symmetry breaking effects. Furthermore, a confined chirally restored phase is identified by a plateau in the quark contribution to thermodynamics and by sigma and pion spectral functions that coincide but have a small width.
	We also discuss that, for some large chemical potential values, the explicit center-symmetry breaking is so strong that statistical deconfinement is realized at infinitely small temperatures.
	Both the missing sensitivity of the Polyakov loop to the quark mass, except at close to the chiral transition, and the Polyakov loop being zero at zero temperature at all chemical potentials, can be interpreted as indications of a missing mechanism which accounts for the quark back-reaction on the Polyakov loop.
\end{abstract}

\thanks{}

\maketitle

%%% %%% %%% %%% %%% %%% %%% %%% %%% %%% %%% %%% %%% %%% %%% %%% %%% %%% %%% %%%
\section{Introduction} \label{sec:Introduction}
%%% %%% %%% %%% %%% %%% %%% %%% %%% %%% %%% %%% %%% %%% %%% %%% %%% %%% %%% %%%

Recently, and in the near future, there is and will be a big experimental effort to explore the phase diagram of strongly-interacting matter be it by the energy scan program at RHIC \cite{Ye:2018oqi}  
and the NA61/SHINE experiment at CERN \cite{Andronov:2018ccl} or at the future facilities NICA at JINR \cite{Kekelidze:2016wkp} and FAIR at the GSI site \cite{Senger:2017oqn}. These laboratory experiments at low and intermediate baryon densities are complemented by the detection of gravitational waves of inspiraling neutron stars that allow us to learn about the nature of QCD matter at very large densities \cite{TheLIGOScientific:2017qsa}.
All these experimental measurements require a theoretical counterpart to interpret and analyze their results. First principles calculations are not yet able to be this match in the medium and large baryon density region. Calculations on a discretized space-time lattice face the infamous sign problem \cite{Soltz:2015ula} and methods which circumvent it as complex Langevin dynamics are still limited to fundamental theoretical investigations but are not yet connected to phenomenology \cite{Aarts:2016qrv,Bloch:2017sex,Kogut:2019qmi}. First principles continuum calculations using the Functional Renormalization Group, perturbation theory or variational approaches are still in progress towards the true number of quark flavors and quark masses \cite{Mitter:2014wpa,Maelger:2018vow,Maelger:2019cbk,Quandt:2018bbu}.

Therefore, frameworks which are based on chiral symmetry, center symmetry, and eventually scale symmetry are widely used as alternatives. These symmetries of the QCD Lagrangian are related to fundamental properties of strongly interacting matter, namely, the appearance of constituent quark masses and confinement in the hadronic phase (and conversely the liberation of light quarks in the transition to the quark-gluon plasma).
The interaction between constituent quarks that gives them their mass due to the spontaneous breaking of chiral symmetry can be described, for example, as a point-like interaction or by the exchange of a meson. The former leads to what is called the Nambu--Jona-Lasinio (NJL) model \cite{Nambu:1961tp} and the latter to the Quark-Meson (QM) model \cite{GellMann:1960np,Metzger:1993cu,Jungnickel:1995fp}. Extended by the order parameter of center-symmetry breaking, the Polyakov loop, these frameworks allow for the phenomenological exploration of the phase diagram for strongly-interacting matter \cite{Mocsy:2003qw,Fukushima:2003fw}.

Even though the Lagrangian of such models itself is invariant under chiral transformations for massless quarks, the appearance of a non-vanishing quark condensate $\langle\bar{q} q\rangle$ breaks chiral-symmetry spontaneously. Therefore, the quark condensate $\langle\bar{q} q\rangle$ is an order parameter for chiral-symmetry breaking.
The relation between quark masses, chiral symmetry, and quark condensate can be exploited to explain the generation of constituent quark masses by spontaneous chiral-symmetry breaking. The constituent quark mass of up and down quarks that are confined in protons or neutrons is of the order of one third of the mass of these nucleons, $\mc{O}\sim300\,\mrm{MeV}$, which is significantly larger than their current quark masses, $\mc{O}\sim5\,\mrm{MeV}$.
While the non-zero current quark masses are responsible for explicit chiral-symmetry breaking, the constituent quark masses are dynamically generated by spontaneous chiral-symmetry breaking, $m\sim|\langle\bar{q} q\rangle^{1/3}|$. With increasing temperature and/or density of the quarks, the restoration of chiral symmetry takes place. In the chiral limit (vanishing of current quark masses) this is a phase transition of second order which turns at the tricritical point into a first-order one. With explicit chiral-symmetry breaking, when the non-zero current quark masses are taken into account, the second-order phase transition becomes washed out into a crossover which can turn into a first-order phase transition at a critical endpoint (CEP). The chiral phase structure of these models is, for example, discussed in Refs.~\cite{Scavenius:2000qd,Buballa:2003qv,Costa:2008yh}.

For what concerns center symmetry and deconfinement,
the Polyakov-loop field $\Phi(\vec x)$ is the appropriate order  parameter to
study the SU$(N_c)$ phase structure and it is associated with $\Zed_{N_c}$, the
center of SU$(N_c)$ \cite{Polyakov:1978vu}.
In the pure gauge sector, the corresponding phase transition that occurs at high temperatures is related to
color deconfinement such that the Polyakov loop is an order
parameter for the deconfinement transition.
The spontaneous breaking of center symmetry can be described by a Polyakov-loop potential that
represents the effective glue potential at finite temperature.
Coupling dynamical quarks to the Polyakov-loop field breaks explicitly the center symmetry. Furthermore, one obtains in this way also information on
the deconfinement of quarks.
In a strict sense, however, quarks are not confined in these models: gluons
are not dynamical and the gluon field is treated as a static background gauge field that does not change the fact that NJL and QM models do not confine quarks.
Nonetheless the thermal  contributions coming from one and two (anti-)quarks are suppressed below the
transition temperature (but are not vanishing) \cite{Hansen:2006ee}, thanks to the Polyakov loop.
This is the so-called ``statistical confinement''.

In this work, we will discuss several aspects of the correlation between quarks 
and the Polyakov loop. We will also study how to disentangle the effects of 
the restoration of chiral symmetry from the effects of the breaking of center 
symmetry in order to have a deeper knowledge on how both are correlated.
This allows for a better understanding of the current status of the comparison 
between the results of these kinds of models and those obtained in lattice QCD 
(LQCD) calculations, while at the same time shedding further light on certain 
lesser known results of the models.

The paper is organized as follows.
In the next section we will give a summarized introduction to the Polyakov-loop--extended Nambu--Jona-Lasinio (PNJL) model and to the Polyakov-loop--extended Quark-Meson (PQM) model, putting special emphasis on the correlation between quarks and the Polyakov loop. In \Sec{sec:transition_scales} we discuss the correlation of the transition scales of restoration of chiral symmetry and of the breaking of center symmetry and how it could be possible to reproduce the splitting that is seen in LQCD calculations at vanishing densities. Section \ref{sec:thermaleffect} contains the analysis of the effect of the kinetic contribution of quarks to the correlation between quarks and the Polyakov loop by comparing results for a medium-dependent and for a constant quark mass. The analysis of the effects that can be found in such a study is extended in Secs.~\ref{sec:MFPhaseDiag} and \ref{sec:CondensPloopcorr}. In \Sec{sec:FermiMomEffect} we analyze and discuss the correlation between the quark contribution and the Polyakov loop due to the chemical potential. Section \ref{sec:CombindedEffects} discusses the combination of both effects which is complemented by an analysis of a confined chirally restored phase in \Sec{sec:CCS}.

When we do quantitative comparisons with LQCD results, we include the
strange quark (\Fig{fig:PQMvsLQCD}) but otherwise we will present
results for two flavors. Focusing on two light flavors allows us to
disentangle non-trivial correlations in the strange sector from
effects due to the statistical confinement giving a more comprehensive
view of the physical phenomenon.

%%% %%% %%% %%% %%% %%% %%% %%% %%% %%% %%% %%% %%% %%% %%% %%% %%% %%% %%% %%%
\section{The PNJL and PQM models in SU(2)} \label{sec:PNJLmodel}
%%% %%% %%% %%% %%% %%% %%% %%% %%% %%% %%% %%% %%% %%% %%% %%% %%% %%% %%% %%%

In this section we introduce PNJL and PQM models to study the thermodynamics 
of QCD at finite temperature and density, in the grand canonical ensemble.

The lack of confinement in the NJL and QM models does not allow to study the very important deconfinement transition at non-zero temperature.  As explained in Sec.~\ref{sec:Introduction}, the Polyakov loop can be used as an order parameter for this transition. We will introduce it in the NJL and QM models and describe its features.

Throughout  this  section, we  will discuss, in  particular, that if  the Polyakov loop brings  some   sort  of   confinement  (``statistical confinement''), it  is not  a real  confinement. The Fock space structure of this model still  contains  quark  degrees  of  freedom and only the quark occupation numbers will be modified.   
As a result, when we discuss confinement we mean the statistical one, not the true one.

%%% %%% %%% %%% %%% %%% %%% %%% %%% %%% %%% %%% %%% %%% %%% %%% %%% %%% %%% %%%
\subsection{Pure gauge sector at finite temperature}

In the pure gauge sector, the phase transition that occurs is related to the deconfinement of color at high temperatures.
Following the arguments given in \cite{Meisinger:1995ih,Pisarski:2000eq,Meisinger:2001cq,Mocsy:2003qw} to study the SU$(N_c)$ phase structure, the appropriate order  parameter (associated with $\Zed_{N_c}$,  center of SU$(N_c)$) is the Polyakov-loop field $\Phi(\vec x)$:
\be
L(\vec      x)      &\equiv&     {\mathrm{Tr}}_c  
\mathcal{P}\exp\left[i\int_{0}^{\beta}d\tau\,
A_4\left(\vec{x},\tau\right)\right], \\
\Phi(\vec  x) &\equiv&  \frac 1  {N_c} \,  \langle\!\langle L(\vec  x)
\rangle\!\rangle_\beta .
\ee
In the above, $A_4 = i A^0$ is the temporal component of the Euclidean gauge field  $(\vec{A}, A_4)$, in  which the strong  coupling constant $g_S$ has been  absorbed, $\mathcal{P}$ denotes path  ordering, and the usual notation for the thermal  expectation value has  been introduced with $\beta = 1/T$ and
the Boltzmann constant set to one ($k_B \equiv 1$).
In general, the Polyakov-loop field is a complex scalar field $\Phi = \Phir + i \Phii$ that simplifies to $\Phi=\Phib=\Phir$ in pure gauge theory. 

An effective potential that respects the  $\Zed_3$ symmetry of the original  Lagrangian may be built and it is conveniently chosen  to reproduce results obtained  in pure gauge lattice  calculations. In  this approximation, the Polyakov-loop field  $\Phi(\vec{x})$ is simply set to be equal to its expectation value $\Phi=$const., which minimizes the potential  \cite{Pisarski:2000eq,Ratti:2005jh},

\begin{eqnarray}
\frac{\mathcal{U}_{Poly}\left(\Phi,\bar\Phi;T\right)}{T^4} &=&-\frac{b_2\left(T\right)}{2}\bar\Phi \Phi
-\frac{b_3}{6}\left(\Phi^3+
{\bar\Phi}^3\right)\nonumber\\
&+&\frac{b_4}{4}\left(\bar\Phi \Phi\right)^2,
\label{Ueff}
\end{eqnarray}
where
\be
b_2\left(T\right)=a_0+a_1\left(\frac{T_0}{T}\right)+a_2\left(\frac{T_0}{T}
\right)^2+a_3\left(\frac{T_0}{T}\right)^3.
\label{eq:b2}
\ee
$T_0 = 270$ MeV is the critical temperature for the deconfinement phase transition   according  to   pure   gauge   lattice  results \cite{Kaczmarek:2002mc}. 
The transition is from $\Phi=0$ (confined phase, $T< T_0$) to the deconfined phase ($\Phi\to1$).

A fit of the coefficients $a_i,~b_i$ has been performed in Ref.~\cite{Ratti:2005jh} in order to reproduce thermodynamics lattice data \cite{Boyd:1996bx} in the pure gauge sector: $a_0 = 6.75$, $a_1 = -1.95$, $a_2 = 2.625$, $a_3 = -7.44$, $b_3 = 0.75$, $b_4 = 7.5$.  
By minimizing this potential, it was possible to compute the Polyakov-loop expectation value in good agreement with lattice gauge findings \cite{Kaczmarek:2002mc}. 

A popular alternative to this potential is the logarithmic one \cite{Roessner:2006xn} 
that reads
%%%%%%
\begin{multline}
    \frac{\mathcal{U}_{Log}\left(\Phi,\bar\Phi;T\right)}{T^4}
    =-\frac{a\left(T\right)}{2}\bar\Phi \Phi\\
		+b(T)\mbox{ln}[1-6\bar\Phi \Phi  +4(\bar\Phi^3+ \Phi^3)-3(\bar\Phi
    \Phi)^2],
    \label{UeffLog}
\end{multline}
%%%%%%
where
%%%%%%
\be
  a\left(T\right)&=&a_0+a_1\left(\frac{T_0}{T}\right)+a_2\left(\frac{T_0}{T}
  \right)^2,
\ee
%%%%%
and
%%%%%
\be
    b(T)&=&b_3\left(\frac{T_0}{T}\right)^3.
\ee
%%%%%

The parameters of the effective potential $\mathcal{U}_{Log}$ are given by $a_0=3.51$, $a_1=
-2.47$, $a_2=15.2$, and $b_3=-1.75$. With these parameters, this  effective potential exhibits  the  feature  of  a transition from color confinement to color deconfinement through a stronger first-order phase transition than $\mathcal{U}_{Poly}$ with the parameters of \Ref{Ratti:2005jh} (see, e.g., Fig.~8 of \Ref{Haas:2013qwp}).

%%% %%% %%% %%% %%% %%% %%% %%% %%% %%% %%% %%% %%% %%% %%% %%% %%% %%% %%% %%%
\subsection{Coupling between quarks and the gauge sector}
The PNJL and PQM models  aim at describing, in a  simple way, two of the characteristic phenomena of QCD, namely deconfinement and chiral-symmetry breaking \cite{Mocsy:2003qw,Fukushima:2003fw,Ratti:2005jh}.
We start from the two-flavor  NJL and QM  description of  quarks (global   SU$_c(3)$  symmetric   and  chirally   invariant  point-like interaction or meson exchange, respectively), coupled  in a minimal way  to the Polyakov loop  via the following Lagrangian 
(the range of applicability of these models is limited to temperatures $T\lesssim 2.5~T_c$):
\bse
\label{lagr:PNJLPQM}
\be
\mc{L}_\mrm{PNJL/PQM} &=& \mc{L}_\mrm{chiral}^\mrm{NJL/QM} +\bar{q}\slr{i\gamma_{\mu}\left(D^{\mu} + \mu\,\delta^{\mu0}\right)}q\nonumber\\ 
&-&\mathcal{U}\left(\Phi[A],\bar\Phi[A];T\right),
\ee
\begin{equation}\label{eq:lagrNJL}
	{\mc{L}_\mrm{chiral}^\mrm{NJL}	= G_S \slr{ \blr{\qbar q}^2 + \blr{\qbar i \gamma_5 \vec{\tau} q }^2 } - \hat{m}_0 \blr{\qbar q}}
\end{equation}
\be
	\mc{L}_\mrm{chiral}^\mrm{QM} &=& \qbar \slr{g_S \blr{ \sigma + i \gamma_5 \vec{\tau} \vec{\pi} }} q \nonumber \\
	&+&h\sigma - \frac{\lambda^2}{4} \blr{ \sigma^2 + \vec{\pi}^2 - v^2 }^2.
	\label{eq:lagrQM}
\ee
\ese
where $\mu$ is the chemical potential of the quarks and
$q=\left(q_u,q_d\right)$ are the quark fields and where the covariant
derivative reads $ D^{\mu} = \partial^\mu - i A ^ \mu$ with
$A^\mu (x)= g_{S} {\cal A}^\mu_a(x) \frac
{\lambda_a}2$ and in the Polyakov gauge
$A^\mu=\delta^{\mu}_{0}A^0$, with $A^0 = -iA_4$.
The two-flavor  current quark  mass matrix  is $\hat{m}_0=\mbox{diag}(m_u,m_d)$ which breaks chiral symmetry explicitly in the NJL model (we work in the isospin symmetric limit and consequently $m_u=m_d=m_0$). In the QM model, the explicit breaking of chiral symmetry is realized by the linear tilt $h\sigma$ of the Mexican hat potential of the meson fields $\sigma$ and $\pi$ which is the last term of \Eq{eq:lagrQM}.
$G_S$ in Eq.~(\ref{eq:lagrNJL}) is the coupling strength of the chirally symmetric four-fermion interaction and $g_S$  {in Eq.~(\ref{eq:lagrQM})} is the quark-meson Yukawa coupling strength.
Finally, a tridimensional ultraviolet cut-off $\Lambda$ will also be introduced in \Eq{omega} in order to regularize divergent integrals that appear in the NJL model.
This cut-off will only be applied to the vacuum integrals. The thermal integrals do not need this cut-off to be finite and besides, as discussed in \cite{Costa:2009ae}, this prescription allows to get the correct Stefan-Boltzmann limit for thermodynamic quantities,  e.g., the pressure, as quark degrees of freedom are still present in the thermal bath at high temperatures.

The minimal coupling between quarks and the Polyakov loop is a first and simple way to take into account back-reactions of the quark sector to the gluonic sector.

In PNJL $m_0$, $G_S$  and $\Lambda$ (in PQM $\lambda$, $v^2$, $h$ and $g_S$) are taken as free parameters, despite the fact that  $m_0$ can be estimated experimentally.
The parameters of the pure NJL sector in Eq.~(\ref{lagr:PNJLPQM}) are fixed at zero temperature and density and have the following values:
$\Lambda =  651 $ MeV,  $G_S = 5.04$ GeV$^{-2}$,  and $m_0 =  5.5$ MeV (taken from \Ref{Ratti:2005jh}).
They are chosen to  reproduce the mass of the pion, $m_\pi = 139.3$ MeV, and its  decay constant $f_\pi  = 92.3$ MeV  (obtained in a  Hartree + Ring calculation)  as well  as  the chiral  condensate (order  parameter  of the  chiral symmetry) $\quarkdensity = 251$ MeV.
The constituent (or dynamical) quark mass in the Hartree (mean-field) approximation  is $m = 325$ MeV.
With the nucleon mass $M_N$ this means $m \simeq  M_N/3$.
The vacuum meson masses, pion decay constant, and constituent quark mass of the NJL model can be used to fix the parameters $\lambda$, $v^2$, $h$ and $g_S$, of the corresponding QM model.
The PNJL/PQM model  gives a simple  explanation of  the nucleon  as being  a state  composed of  three quarks but their  mass is the corresponding dynamical mass. Even with $m_0  = 0$ {($h=0$)}, $m$ has the same  magnitude: essentially the baryonic mass is due to the ``glue'',  the interaction energy carried out by  the gauge fields.
The mechanism of  mass generation via the spontaneous symmetry breaking  was one of the first interests of NJL and QM models.

Concerning the NJL parameters, several sets of parameters which fit physical observables in the vacuum can be chosen. In particular, the chosen set of parameters foresees the existence of a low-density phase of homogeneously distributed constituent quarks \cite{Costa:2009ae} which is unrealistic (see also \cite{Buballa:2003qv}). This reflects the missing of confinement in the NJL model. However, for the study that we are carrying out, this fact is not eminently relevant and the discussions we are going to make remain valid.
We discuss only transition lines that are not affected by this stability issue. Quantities depending on the correct choice of parameters are, for example, isentropic lines or the study of quark matter inside compact stars.

%%% %%% %%% %%% %%% %%% %%% %%% %%% %%% %%% %%% %%% %%% %%% %%% %%% %%% %%% %%%
\subsection{Statistical confinement at  finite temperature and density
in the Hartree approximation}

\subsubsection{Grand canonical potential, mean-field  equations, and the modified Fermi-Dirac distribution functions} 
\label{sec:GrandPotential}

In the the mean-field approximation, the usual techniques can be used
to obtain the PNJL/PQM grand canonical potential from the Hartree
propagator (see, for instance,
Refs.~\cite{Hansen:2006ee,Klevansky:1992qe,Ratti:2005jh}),
\bse
\label{omega} 
\begin{eqnarray}
\Omega(&\Phi,&\bar\Phi,m;T,\mu)= U_\mrm{chiral}^\mrm{NJL/QM}\blr{m} + {\cal U}\left(
\Phi,\bar{\Phi},T\right)\nonumber\\
&-& 2 N_c N_f\int_{0}^{\Lambda}\frac{\mathrm{d}^3p}{\left(2\pi\right)^3}\,{E_p}  \nonumber\\
&-& 2N_f\,T\int_{0}^{+\infty}\frac{\mathrm{d}^3p}{\left(2\pi\right)^3}
\left\{\mathrm{Tr}_c\ln\left[1+ L^\dagger \mathrm{e}^{-\left(E_p-\mu
\right)/T}\right] \right.\nonumber\\
&+&\left.\mathrm{Tr}_c\ln\left[1+ L\mathrm{e}^{-\left(E_p+\mu\right)/T}\right]
\right\}\;,
\label{eq:Omega} 
\end{eqnarray}
\bea
	{ U_\mrm{chiral}^\mrm{NJL}} &=& {\frac{\blr{m - m_0}}{4G_S}}\;, \\
	{ U_\mrm{chiral}^\mrm{QM}} &=& {\frac{\lambda^2}{4} \blr{ \sigma^2 - v^2 }^2  -h\sigma\,; \quad m = g_S \sigma}\;.
\eea
\ese
In the above  equation, $E_p=\sqrt{\vec{p}\,^2+m^2}$ is the Hartree single  quasi-particle  energy  (which  includes  the  constituent or dynamical quark mass $m$ and not  the current mass $m_0$). Performing the trace for $N_c = 3$ we define for simplicity,
\begin{eqnarray}
	z^+_\Phi(E_p)&=&\mathrm{Tr}_c\ln\left[1+ L^\dagger
         \mathrm{e}^{-\left(E_p-\mu\right)/T}\right]\nonumber\\
				&=&\ln\left\{ 1 + 3\bar\Phi\expmT + 3 \Phi \expmmT
         \right.\nonumber\\
				&+&\left. \expmmmT \right\},
\label{zplus}
\end{eqnarray}
\begin{eqnarray}
	z^-_\Phi(E_p)&=&\mathrm{Tr}_c\ln\left[1+ L
         \mathrm{e}^{-\left(E_p+\mu\right)/T}\right]\nonumber\\
				&=& \ln\left\{ 1 + 3\Phi\exppT  + 3\bar\Phi\expppT \right.\nonumber\\
				&+&\left. \exppppT \right\}.
\label{zmoins}
\end{eqnarray}

The solutions of  the mean-field equations are  obtained by 
solving the equations of motion for $m$, $\Phi$, and $\bar\Phi$, namely,
$\frac{\partial\Omega}{\partial\Phi}          =          0          $,
$\frac{\partial\Omega}{\partial\bar\Phi}      =      0      $,      and
$\frac{\partial\Omega}{\partial m}  = 0  $.\\
In the NJL model, the latter equation can be simplified to 
the  gap equation,
\begin{eqnarray}
  m-m_0&=&2G_S N_f
  N_c\int_{0}^{+\infty}\frac{\mathrm{d}^3p}{\left(2\pi\right)^3}\frac{2m}{E_p}\left[\theta(\Lambda^2 - p^2)\right. \nonumber \\
	&-& \left.f^+_\Phi(E_p)-f^-_\Phi(E_p)\right],
\end{eqnarray}
where the  modified Fermi-Dirac distribution  functions $f^+_\Phi$ and $f^-_\Phi$ have  been introduced:
\begin{multline}
  f^+_\Phi(E_p) 
          =\\\frac{ \left( \Phi + 2\bar\Phi \expm \right) \expm + \expmmm }
  {1 + 3\left( \Phi + \bar\Phi \expm \right) \expm + \expmmm} \label{fpPhi},
\end{multline}
\begin{multline}
  f^-_\Phi(E_p) 
         =\\\frac{ \left( \bar\Phi + 2\Phi \expp \right) \expp + \expppp }
  {1 + 3\left( \Phi + \bar\Phi \expp \right) \expp + \expppp}. \label{fmPhi}
\end{multline}

The equations presented above, which were introduced for the first time in \cite{Hansen:2006ee}, allow to straightforwardly generalize the results for the thermodynamics of the NJL/QM model to those of the PNJL/PQM model by replacing the usual Fermi-Dirac occupation numbers by the modified ones given by Eqs.~(\ref{fpPhi}) and (\ref{fmPhi}).

The explicit form of the mean-field equations for  $\Phi$ and $\bar\Phi$, which can be found for  example in \cite{Hansen:2006ee},  will be given in \secref{sec:backreaction} where  we will discuss  in detail the correlation between quarks and the Polyakov loop.

Here, we want to point out that in PNJL-model calculations, we choose not to use the Fock terms (exchange diagrams) because, if added, we would obtain the same equations with the replacement $G_S \rightarrow G_S(1 + 4/N_c)$. Indeed, for the local four-point interaction, exchange diagrams can always be rewritten in the form of direct diagrams via a Fierz transformation. 
The Hartree-Fock approximation is then equivalent to the Hartree approximation with the appropriate redefinition of the coupling constants \cite{Klevansky:1992qe,Buballa:2003qv}.
Since $G_S$ is a parameter to be fixed, it is not very important to include the Fock term.
We can also notice that the Hartree term is of order $O(N_c^1)$, the Fock term is  $O(N_c^0)$, and the ring approximation can  be shown to be also $O(N_c^0)$. As discussed in Ref.~\cite{Blin:1987hw}, the Hartree term may be seen as the first term in a $1/N_c$ expansion. Then, when going beyond mean-field approximation, the Fock term should  be added for coherence. 

Furthermore, it is important to mention that at non-zero chemical potential the kinetic quark-antiquark contribution in \Eq{eq:Omega} adds an imaginary part to the effective potential, $\Omega_\qqb^\mrm{th} = \Omega_\qqb^\mrm{R} + i\,\Omega_\qqb^\mrm{I}$. This is the manifestation of the fermion sign problem in the Polyakov-loop extension of the NJL and QM models \cite{Fukushima:2006uv,Rossner:2007ik,Mintz:2012mz}. One way to avoid the sign problem is to neglect the imaginary part of the effective potential as a lowest-order perturbative approximation \cite{Rossner:2007ik,Mintz:2012mz}. Doing this implies that the imaginary part of the Polyakov loop $\Phii$ is zero, i.e., $\Phi=\Phib$ also at $\mu\neq0$.
Here, we will follow the more common approach to circumvent the sign problem (see, e.g., \Ref{Ratti:2005jh}), that is, to redefine the Polyakov loop $\Phi$ and its complex conjugate $\Phib$ as two independent, real variables. But our results do not depend on this choice and also hold true for the aforementioned approach.

%%% %%% %%% %%% %%% %%% %%% %%% %%% %%% %%% %%% %%% %%% %%% %%% %%% %%% %%% %%%
\subsubsection{The influence of quarks on the gauge fields} \label{sec:backreaction}
%%% %%% %%% %%% %%% %%% %%% %%% %%% %%% %%% %%% %%% %%% %%% %%% %%% %%% %%% %%%

To study the influence of quarks on the gauge fields (contained in the
Polyakov loop definition) we will push the model in later sections to
a very high density scale, probably well beyond the range of validity
of the model. However, we find these calculations interesting as they
lead us to understand what ingredients are probably missing in the
model to get a more faithful representation of QCD in the high density
region.

Obviously, in QCD quark fields act upon gauge fields via quark loops
in the gluon self-energy. Such an effect is lacking in the NJL and QM
models since there are no dynamical gluon fields (although, in some sense the
strong interaction carried by the gluons is present in the model via
the NJL contact interaction).

In the PNJL and PQM models there is still no  dynamical gluons but there is an interaction between  the  static gauge  field  and  the  quarks via  the  covariant derivative $ D^{\mu} =  \partial^\mu - i A ^ \mu$. As  a result, the mean-field equation for $\Phi$  does have a dependence on quarks (see the previous section). Explicitly for two flavors, we have for $\Phi$
(the discussion for $\bar\Phi$ goes along the same lines) the following:
\begin{multline}
\frac{\partial\Omega}{\partial\Phi} = 0 \quad\Leftrightarrow\quad
0  =   {T^4}  \frac{\partial{\cal  U}}{\partial  \Phi}
\\-            6           T            \sum_{\left\{i=u,d\right\}}
\int\frac{\mathrm{d}^3p}{\left(2\pi\right)^3}  \left( \frac  {\exppp}{
e^{z^+_\Phi(E_p)} } +  \frac  {\expm}{e^{z^-_\Phi(E_p)}  }
\right),
\label{eq:domegadfi} 
\end{multline}
with $z^+_\Phi(E_p)$ and $z^-_\Phi(E_p)$ given by Eqs. (\ref{zplus}) and (\ref{zmoins}), respectively. The mean-field equations for $\Phi$ and $\Phib$ are the same in the PQM and PNJL models since they do not depend explicitly on the chiral part of the grand potential $U_\mrm{chiral}\blr{m}$.

The second term in Eq.~(\ref{eq:domegadfi}) adds some quark
corrections to the first term (that describes the pure gauge sector)
and interesting information can be extracted from this equation as we
will discuss later. Without, quarks the solution would be the pure
gauge one, i.e., ${\partial {\cal U}}/{\partial\Phi} = 0$. The model
would then have a first-order phase transition.

%%% %%% %%% %%% %%% %%% %%% %%% %%% %%% %%% %%% %%% %%% %%% %%% %%% %%% %%% %%%
\subsubsection{Statistical confinement} \label{sec:StatisticalConf}

In the modified Fermi-Dirac functions, Eqs.\ (\ref{fpPhi}) and (\ref{fmPhi}), entering in the grand potential, Eq.\ (\ref{omega}), the confinement mechanism that exists in Polyakov-loop--extended models can be seen. It is not a true confinement (the quarks are still asymptotically free and we will see that they can be produced, e.g.\ by the sigma meson decay in vacuum; see also Ref.~\cite{Hansen:2006ee}). We call this effect ``statistical confinement'' as it is due to the suppression of the one and two \mbox{(anti-)}quarks Boltzmann factors.

Indeed, it can be seen that in  the  grand  potential, Eq.~(\ref{omega}), the contributions coming from one and two \mbox{(anti-)}quarks are suppressed below $T_c$ (when $\Phi,\bar\Phi\to 0$, the confined phase of  the model) due to their coupling with $\Phi$ and $\bar\Phi$ but the three-\mbox{(anti-)}quarks Boltzmann factor is not.
The interpretation is that  there are still unconfined \mbox{(anti-)}quarks in the  vacuum part (the Dirac sea) of the grand potential but in the thermal bath only three-\mbox{(anti-)}quarks contributions are present (a reminder of the  fact that in  QCD only colorless  combinations can exist  in the confined  phase).  This  reduces significantly  the number  of \mbox{(anti-)}quarks  in the  thermal bath  since it requires three times more energy for the \mbox{(anti-)}quarks to be thermodynamically active.

As a result the Polyakov-loop extension corrects a problem of NJL and QM models.
It is known that,  at a given value  of $T$ and $\mu$, pure NJL and QM models always overestimate the density (see Ref.~\cite{Ratti:2005jh}), even if they merge for large temperatures with the PNJL/PQM model (when $\Phi\to1$).
At fixed  values of $T$ and $\mu$, the PNJL/PQM-model value for the density is much lower than in the NJL/QM-model  case.
In fact, all the possible contributions to the density turn out to be somehow suppressed: the one- and two-quark contributions because of $\Phi,\bar\Phi\to 0$,  while the thermal excitation  of three quarks has a negligible  Boltzmann factor. We would be  tempted to identify these  clusters  of three  dressed  \mbox{(anti-)}quarks  with precursors  of
\mbox{(anti-)}baryons but no binding for  these structures is provided by the model. In any case, it is encouraging that coupling the chiral Lagrangian (whose parameters are chosen to reproduce zero temperature properties) with  the Polyakov-loop field  (described by a pure  gauge effective potential) leads to results that point in the right direction at finite density.

Another important effect of statistical confinement is that the chiral transition will occur in a smaller temperature range than in the NJL/QM model. It is  seen from  Eqs.\ (\ref{zplus}) and (\ref{zmoins}) that before the transition, the pressure (for example) is kept low by the effect that we have just discussed. Then, quark  degrees of freedom are liberated in  a narrow temperature range  when $\Phi  \rightarrow 1$.  As  seen in Ref.~\cite{Costa:2010zw}, the transition in the PNJL/PQM model is indeed much ``faster'' that the one in the pure NJL/QM model.

Another connection between the gauge sector and quarks is discussed in \Ref{Song:2019qoh}, where a moderately strong vector repulsion between quarks parametrized by a four-fermion interaction in terms of non-perturbative gluon exchange in QCD in the Landau gauge is considered and the effects of quark masses estimated.

The  mesonic contribution to the pressure is also important (about 10\%) and  this effect has to be included in the mean-field description, e.g., with the  Beth-Uhlenbeck formalism \cite{Hufner:1994ma} to the PNJL model \cite{Blaschke:2014zsa} which is another way of doing RPA or by applying the Functional Renormalization Group framework to the QM model \cite{Jungnickel:1995fp} and to the PQM model \cite{Stiele:2014gks}. 

Nowadays this allows to fit well the most recent LQCD results with physical quark masses \cite{Herbst:2013ufa,Torres-Rincon:2017zbr}.
A  detailed discussion of the results in mean-field approximation can be found in \cite{Costa:2010zw}. For a review on modeling hadronic and quark matter see \Refs{Fukushima:2013rx,Fukushima:2017csk,Baym:2017whm}.

%%% %%% %%% %%% %%% %%% %%% %%% %%% %%% %%% %%% %%% %%% %%% %%% %%% %%% %%% %%%
\subsection{Coincidence between chiral and deconfinement transition}
%%% %%% %%% %%% %%% %%% %%% %%% %%% %%% %%% %%% %%% %%% %%% %%% %%% %%% %%% %%%

One of  the first successes of both, PNJL and PQM models, was the observation 
that without any  additional  tuning, chiral  and deconfinement transitions almost 
coincide in a small range of temperatures at zero density (see Fig.~\ref{fig:PNJLPhases}).   
It  was  an interesting  test as LQCD calculations predicted this coincidence
(even if recent LQCD results show that there is no perfect coincidence between them \cite{Borsanyi:2010bp}).

What is not  obvious is the fact that by mixing the deconfinement scale $T_0 = 270$ MeV 
and the  chiral restoration scale (about $220$ MeV in the NJL model and in
the QM model if the same mass of the $\sigma$ meson and of the constituent quarks as 
in the NJL model are used), the  two  transitions automatically coincide  at  
a  lower temperature.  The minimal coupling in PNJL/PQM models is enough to have some sort 
of back-reaction effect that produces this matching.

This is also a quite stable feature \cite{Ratti:2005jh}: we can take
$T_0 = 190$ MeV and still get a quite good coincidence. The motivation to
change the value of $T_0$ was to get a better agreement with the value of the
LQCD calculations for the chiral transition temperature. 
It can be justified since in an effective model we are mixing
physical sectors with different scales in a symmetry based  (Landau) framework.  It is then 
understandable that the absolute scale of the two sectors can be slightly adjusted
to  get a  better agreement  with phenomenology.  
In other words, we authorize ourselves to consider $T_0$ as a free
parameter of the model with a loose constraint on it coming from LQCD
calculations, exactly as it is done in the chiral sector with the
constituent quark mass $m_0$ despite the fact that there are some estimates
of it.

%%% %%% %%% %%% %%% %%% %%% %%% %%% %%% %%% %%% %%% %%% %%% %%% %%% %%% %%% %%%
\section{Adjusting chiral and deconfinement phase scales \label{sec:transition_scales}}
%%% %%% %%% %%% %%% %%% %%% %%% %%% %%% %%% %%% %%% %%% %%% %%% %%% %%% %%% %%%

\begin{figure*}
  \centering
  \includegraphics[width=0.47\textwidth]{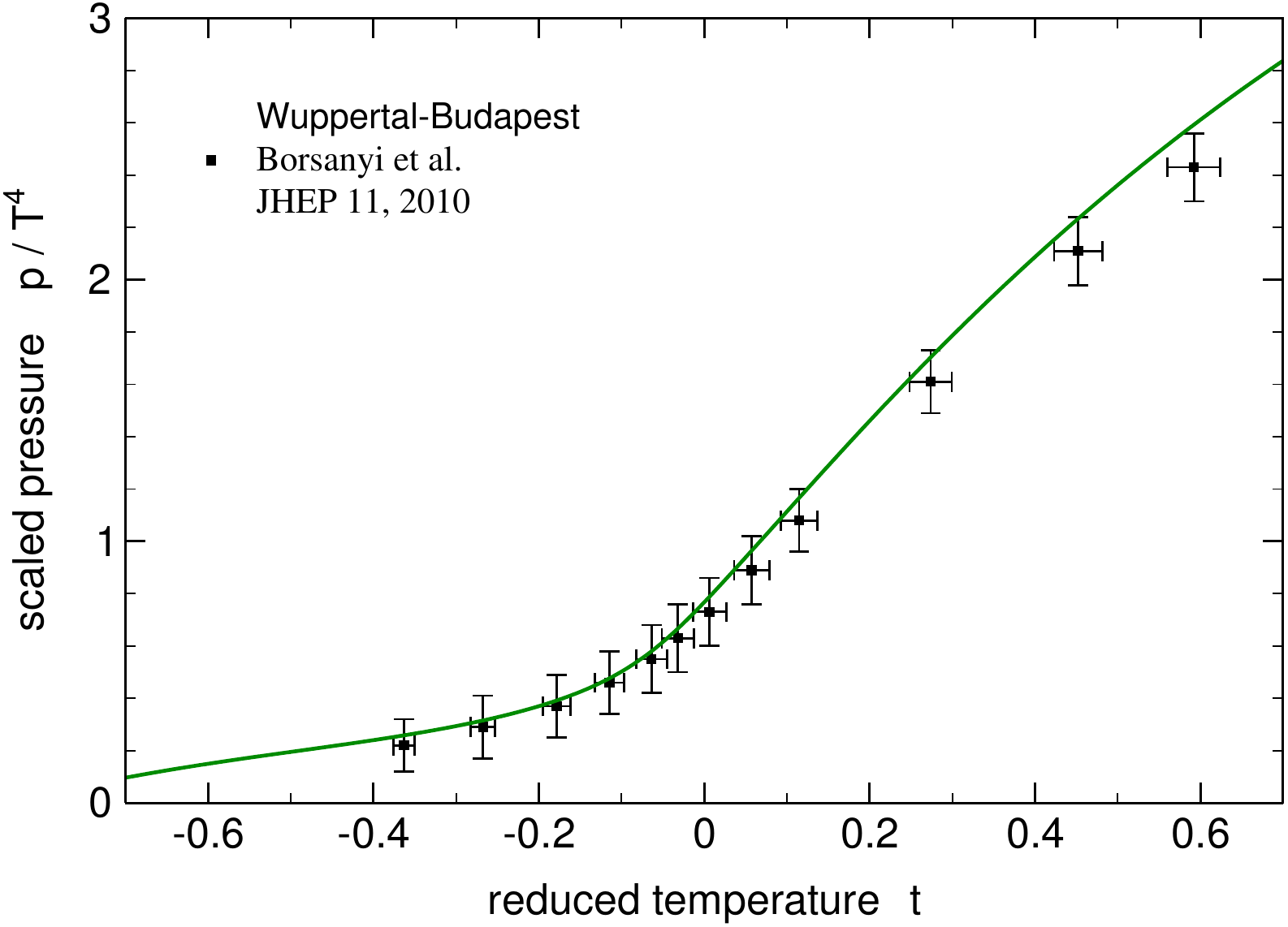}
  \hfill
  \includegraphics[width=0.47\textwidth]{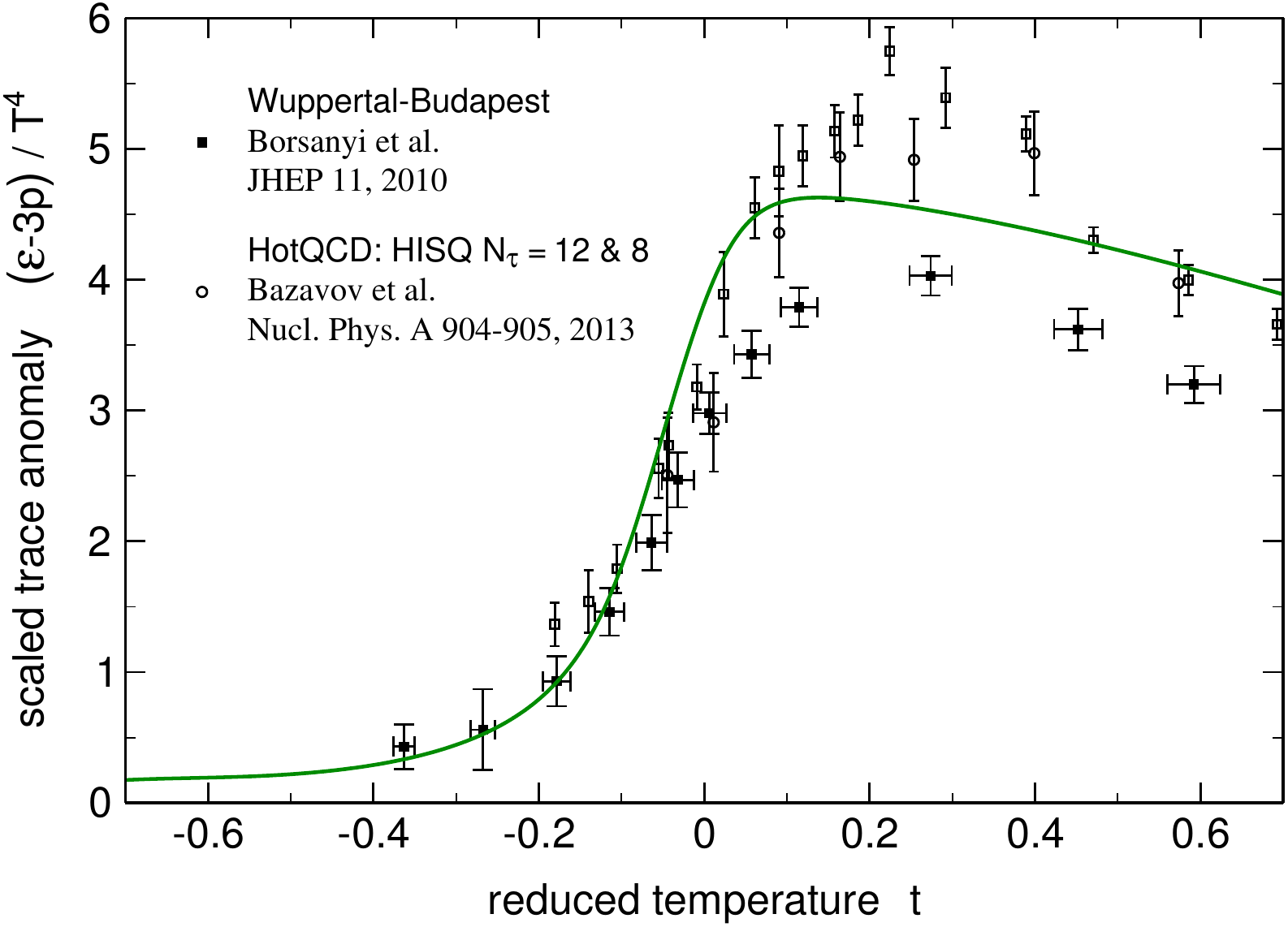}\\
  \includegraphics[width=0.47\textwidth]{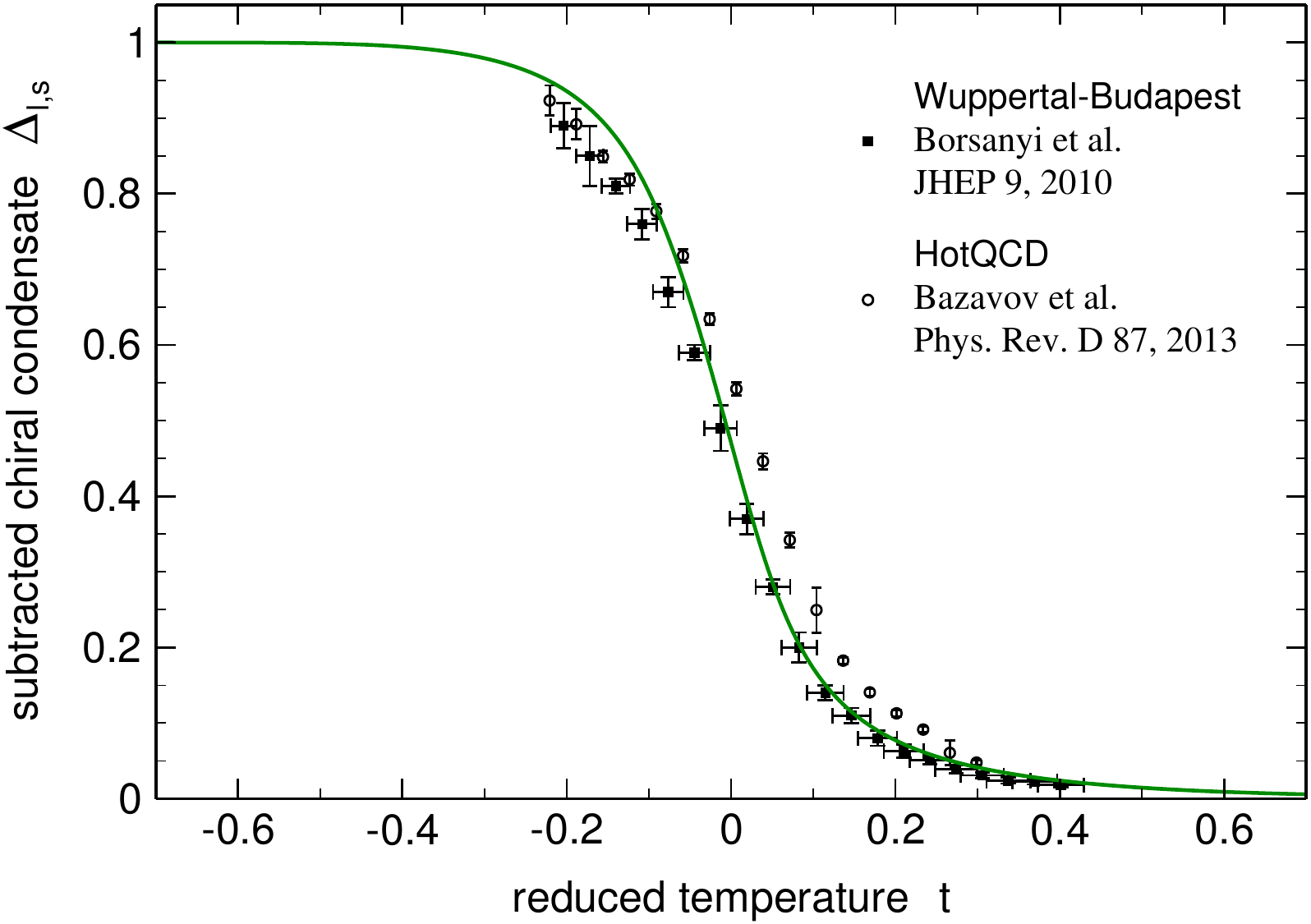}
  \hfill
  \includegraphics[width=0.47\textwidth]{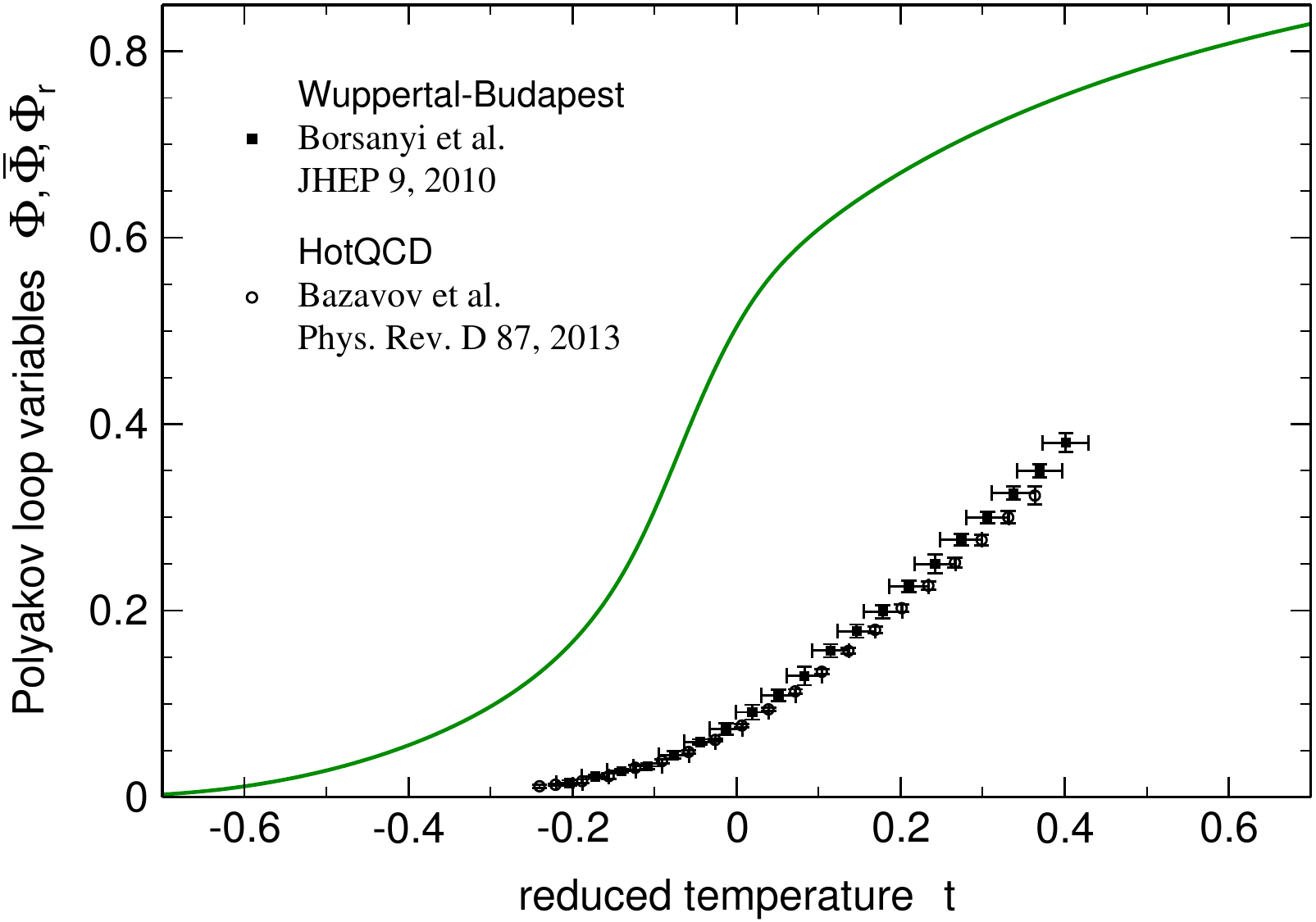}
  \caption{Pressure, trace anomaly, chiral condensate, and Polyakov loop as a 
	function of $T$ in a PQM model \cite{Stiele:2014gks} (for this quantitative 
	comparison, strange quarks have been taken into account) compared with the LQCD 
	data of Refs.~\cite{Borsanyi:2010bp,Borsanyi:2010cj,Bazavov:2012bp,Bazavov:2013yv}.
  }
  \label{fig:PQMvsLQCD}
\end{figure*}

The most problematic aspect for PNJL/PQM-type models is probably the fact that 
the difference between the chiral transition and the raise of $\Phi$ in the QGP 
in recent LQCD data is larger than previously seen. 
This is shown in Fig.~\ref{fig:PQMvsLQCD} for a PQM-model calculation but also 
holds true for the PNJL model as shown. e.g., in \Ref{Torres-Rincon:2017zbr}.
For this quantitative comparison with LQCD data we use the PQM model with 2+1 quark flavors \cite{Stiele:2014gks}.
%%%%%%%%%%%%%%%%%%%%%%
%%%%%%%%%%%%%%%%%%%%%%
In  this  figure, $t$  is  the  reduced  temperature adjusted  to  the
chiral crossover temperature  $T_c$: $t = T/T_c  - 1$.  The
pressure plot, or the quark number density one, shows that at $T_c$  ($t=0$) quark
degrees  of  freedom  are  liberated.  But  there  are  two  mechanisms
responsible  for  this  liberation:  the chiral restoration  and  the
deconfinement transition. This can be easily studied in effective models since
we have control on how these phenomena are coupled.

In Fig.~\ref{fig:PQMvsLQCD}, bottom left, it is possible to see that at $t=0$ 
chiral symmetry goes towards its restoration. There is a liberation of
thermodynamical degrees of freedom simply because  $m \rightarrow m_0$;
hence, the quark Boltzmann factor has a bigger contribution to the
pressure. This occurs at the chiral transition scale.
On the bottom right panel, it is seen that the results for $\Phi$ from LQCD and 
PQM-model calculations do not agree. The same occurs in the PNJL model 
(see Ref.~\cite{Torres-Rincon:2017zbr}).
The (statistical)  deconfinement occurs  at a higher temperature than in the PQM 
model (at $t=0$, the PQM model  gives $\Phi \simeq 0.5$ whereas it is only $0.1$ 
for LQCD results). This means that the (statistical) deconfinement scales differ 
in LQCD data and in the PQM/PNJL model.

One subtlety concerning the Polyakov loop is that there are different order parameters. The 
standard order parameter is the expectation value $\langle\Phi\slr{A_0}\rangle$. The 
functional dependence indicates that the Polyakov loop derives from the temporal component of 
the gauge field $A_\mu$. The expectation value of the latter $\langle A_0\rangle$ relates to 
another Polyakov-loop order parameter, $\Phi\slr{\langle A_0\rangle}$.

LQCD results are for $\langle\Phi\rangle$ since this is the easily accessible quantity in 
these calculations.
In Polyakov-loop extended effective models for QCD the Polyakov-loop potential is adjusted to LQCD results on $\langle\Phi\rangle$ in pure gauge theory while it is actually the gauge field that appears initially in the fermionic determinant (see \Eq{omega}) which is then rewritten into a dependence on the Polyakov loop $\Phi$ in \Eqs{zplus} and \eq{zmoins}.
Both Polyakov-loop order parameters are related in general by %Jensen's inequality in the following way: 
$\langle\Phi\slr{A_0}\rangle\leq\Phi\slr{\langle A_0\rangle}$ 
\cite{Braun:2007bx}.\footnote{This relation has to be taken insofar with care as the lattice results involve a non-trivial renormalization factor such that for pure gauge theory $\langle\Phi\slr{A_0}\rangle$ exceeds unity in a certain temperature range, whereas $\Phi\slr{\langle A_0\rangle}\leq1$.}
This condition shows that the transition scales of both 
Polyakov-loop order parameters might differ with that of $\langle\Phi\rangle$ being larger than 
that of $\Phi\slr{\langle A_0\rangle}$. The results shown in \Fig{fig:PQMvsLQCD} are consistent 
considering this condition.

\begin{figure}%[b]
	\centering
	\includegraphics[width=0.47\textwidth]{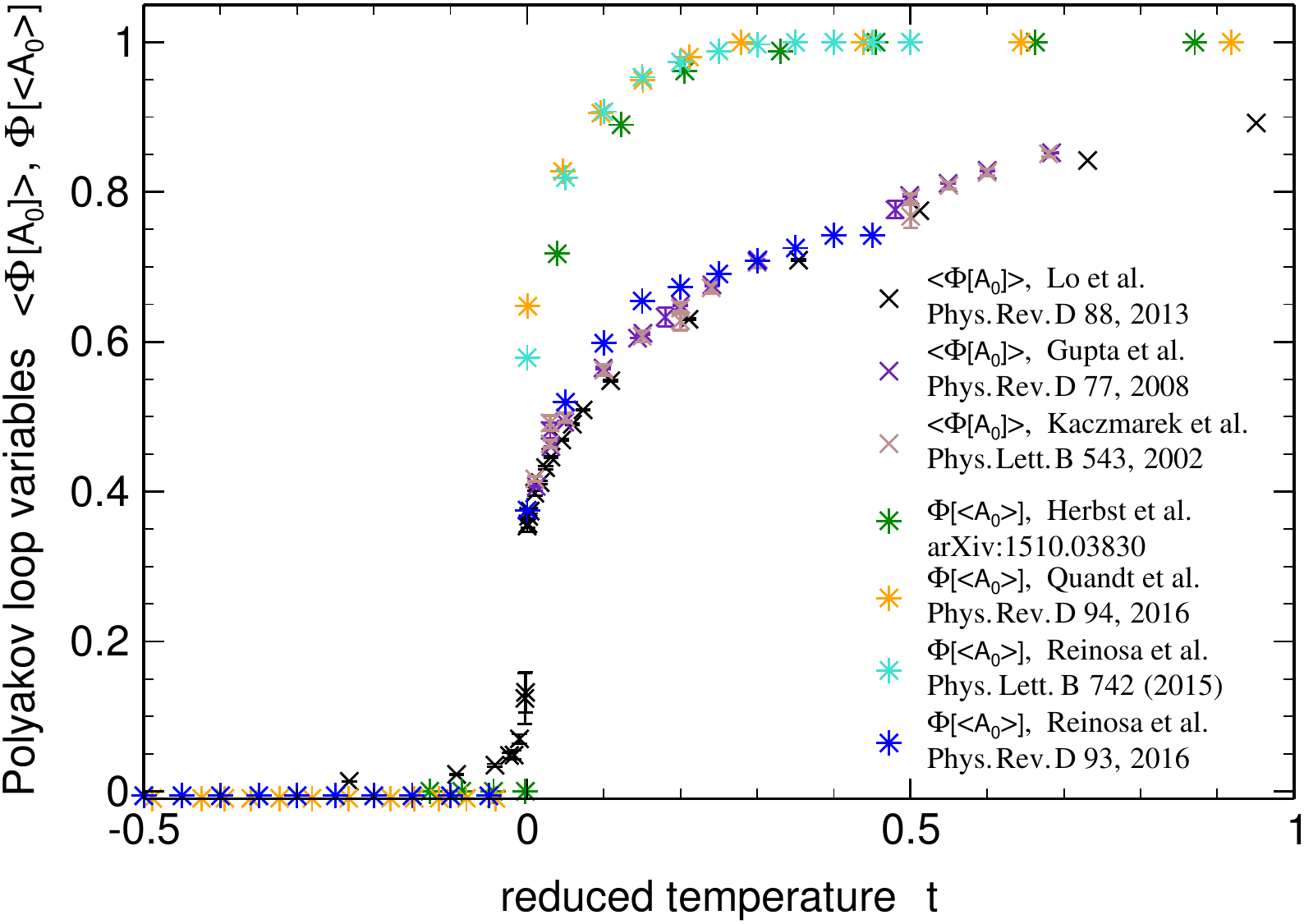}
	\caption{Comparison of the temperature dependence of the Polyakov-loop order parameter $\langle\Phi\slr{A_0}\rangle$ on the lattice \cite{Kaczmarek:2002mc,Gupta:2007ax,Lo:2013hla} and $\Phi\slr{\langle A_0\rangle}$ in different continuum approaches \cite{Herbst:2015ona,Quandt:2016ykm,Reinosa:2014ooa,Reinosa:2015gxn}.}
	\label{fig:PloopYM_latticeVScont}
\end{figure}

The origin of both these Polyakov-loop order parameters, their derivations, and their relation
is discussed in detail in \Ref{Herbst:2015ona}.
The temperature dependence of $\Phi\slr{\langle A_0\rangle}$ is calculated in continuum 
approaches such as the Functional Renormalization Group, with Dyson-Schwinger equations and the 
2PI formalism as well as in the Hamiltonian approach and covariant variational approach, and in 
perturbative approaches.
For pure gauge theory, available results on $\Phi\slr{\langle A_0\rangle}$ in different 
continuum approaches are shown together with those for $\langle\Phi\slr{A_0}\rangle$ of 
different LQCD calculations in \Fig{fig:PloopYM_latticeVScont}. 
%%%%%%%%%%%%%%%%%%%%%%
%%%%%%%%%%%%%%%%%%%%%%
Interestingly enough, for the perturbation theory calculations, the leading-order result for $\Phi\slr{\langle A_0\rangle}$ agrees 
with the results of the other continuum approaches while 
the evaluation of the next-to-leading-order 
result for $\langle A_0\rangle$ with the leading-order relation between the expectation value of the gauge field and the Polyakov loop $\Phi\slr{\langle A_0\rangle}$ (see also \Ref{Reinosa:2015gxn}) 
is very close to the LQCD results for $\langle\Phi\slr{A_0}\rangle$. %This justifies the considerations in the following that one has to consider to adjust the chiral and deconfinement phase scales.

The information given by LQCD calculations is that the scales of chiral-symmetry restoration 
and center-symmetry breaking do not match perfectly. In general, in the PNJL/PQM-model 
families,  this scale  separation is not achieved. Even worse, with common parametrizations as 
the one we use in the PNJL model ($T_0 = 190$ MeV), the scale hierarchy is inverted; see 
\Fig{fig:PNJLPhases}. 

In effective models it is possible to have control over this  hierarchy as we will show in the following. 
The shortcoming, however, is that the model output for the absolute temperature of the transitions will not be  correct. This is a problem but not to the point of making the model meaningless.  Indeed, in  the spirit  of  Landau theory,  two systems with the  same universal behavior cannot  be directly compared but they have to be compared relatively to a given  scale. This is of course not a complete justification for what we will do in the following but it will illustrate which information can be obtained about some mechanism that is probably missing in the model. We recall  that a scale adjustment was already done at the beginning of PNJL/PQM-model studies: for example, in Ref.~\cite{Ratti:2005jh} the scale $T_0$ was modified to get a complete agreement with the chiral transition temperature given by LQCD results.

\begin{figure}
  \centering  
\includegraphics[width=0.47\textwidth]{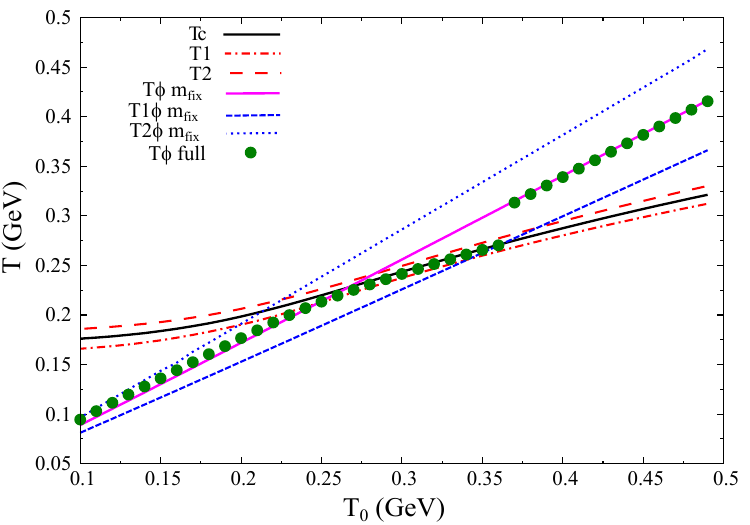}
  \caption{Chiral (black line) and  deconfinement (magenta line and green dots) transition temperatures (together with the size of the transition 
	regions $T_{1,2}$ and for deconfinement, the  full  (green) and  fixed mass  (magenta)
	calculations  are  presented) as  functions  of   the  pure  gauge deconfinement 
	scale  $T_0$ for the PNJL model. 
	We use  the polynomial Polyakov-loop potential (Eq.~(\ref{Ueff})).  We point out that we 
	have the 	
	same feature with the $U_{Log}$ potential (Eq.~(\ref{UeffLog})) but at a much higher scale.
	}
  \label{fig:T0variation}
\end{figure}

So, the point  is that we want to have in the model the correct scale hierarchy  for  the transitions. In both, PNJL and PQM models, we have two sectors  to determine  the respective scales, the  NJL/QM one (fitted  to chiral-symmetry breaking phenomenology in vacuum, a scale related to the strength of the condensate) and the gluonic one (fixed by pure gauge results at finite temperature with  the  scale $T_0$).   The  coupling  is  done via  the  covariant derivative but if we allow $T_0$ to vary then we can control the relative scales of the  transitions. 
This  is done in Fig.~\ref{fig:T0variation}:
we see  that the correct  hierarchy can  indeed be achieved  with $T_0\simeq  400$  MeV  but the  price  to  pay  is  that both, chiral and deconfinement transitions, occur at too high temperatures.

%%%%%%%%%%%%%%%%%%%%%%
%%%%%%%%%%%%%%%%%%%%%%

The message is that some mechanism is missing to get the
correct scale  hierarchy and we should try to understand which mechanism
would allow to have the same effect as our rather artificial
increase of  $T_0$. Since up to now no model can reproduce this fact, we probably 
need to look outside of the existing mechanisms of the PNJL/PQM models.  
Our guess is that introducing a minimal dynamics to the gauge
sector \cite{Alba:2014lda} may be the correct way.

We also notice that in PQM-model calculations, shown in Fig.~\ref{fig:PQMvsLQCD},
the pressure and quark density are probably a little bit overestimated for  $t>0$ 
when compared to LQCD. 
If the  correct  scale hierarchy could be implemented, it is possible that the later 
liberation of quarks via statistical deconfinement could reduce this excess.  
This is a somewhat far fetched conclusion since the excess is low (statistically it seems to be a small but constant $\simeq$1 sigma deviation) and in the PQM model, mesonic
degrees of freedom tend to survive in the QGP phase \cite{Herbst:2013ufa}.

Finally, for what concerns this aspect of quark and  Polyakov-loop
correlations, we also notice in Fig.~\ref{fig:T0variation} that the transition for the full calculation for $\Phi$ from \Eq{eq:domegadfi} and for the one keeping the mass  $m$ at a fixed value $m_{\text{fix}}$ in \Eq{eq:domegadfi}\footnote{This is obtained by solving the mean-field equations, not self-consistently as usual, but simply by fixing the mass $m$ to a given value $m_{\text{fix}}$ in this equation. We will explore further this method in the following.} coincide,  except  when  the
difference  between   the  chiral   temperature  and  the   fixed  $m$
deconfinement temperature  is smaller  than about  25 MeV.  This means
that the  order parameters $\Phi$ and the chiral condensate  are weakly
coupled by the  mean-field equations because, except for a small 25 MeV
$\ll \Lambda_{QCD}$ scale difference, the uncorrelated calculation (fixed $m$) gives the
same result as the correlated one. We will discuss this kind of analysis in more detail in 
the next section.

At low $T_0$ the chiral transition is  below its NJL/QM
value (around 220 MeV) and it happens at a large (but below 1) value
of the Polyakov loop while at high $T_0$ the chiral transition occurs for
a small value (around zero) of the Polyakov loop. Thus, in the PNJL/PQM model
the chiral transition always  behaves differently from what happens
in the  NJL/QM model due to  the back-reaction of the  Polyakov loop, explaining
the success of  the model. The shortcoming of the 
model is essentially in the behavior of the Polyakov loop with respect to quarks but 
nevertheless, it seems that the action of the Polyakov loop on quarks, namely, the statistical confinement,
captures most of the effect needed to correctly describe LQCD thermodynamics.

%%%%%%%%%%%%%%%%%%%%%%
\begin{figure*}[t!]
  \centering
  \includegraphics[width=0.47\textwidth]{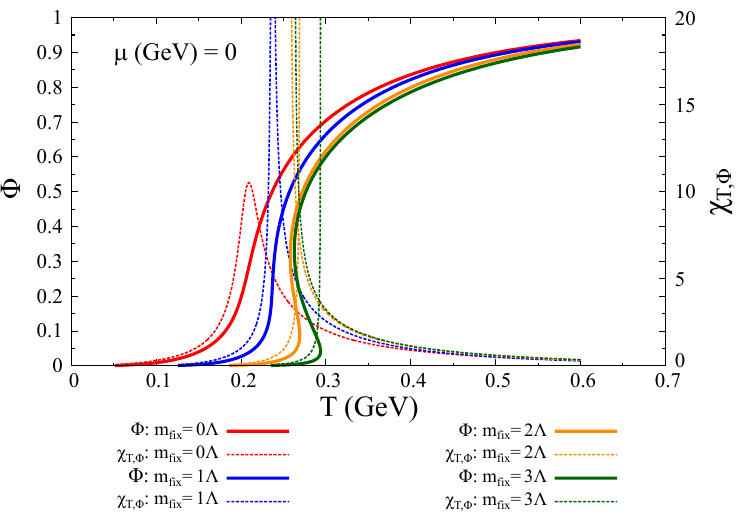}
  \includegraphics[width=0.47\textwidth]{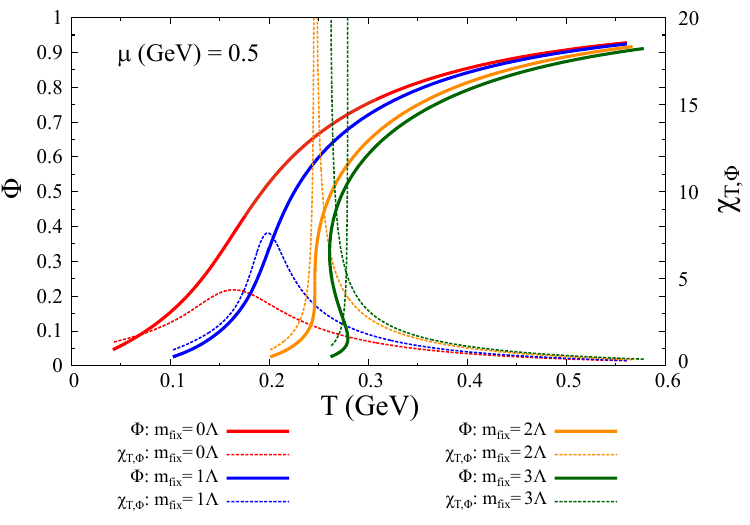}
  \caption{$\Phi$ and $\chi_\Phi=d\Phi/dT$ (Polyakov-loop susceptibility) as 
	functions  of $T$ for different values of the constituent quark mass $m$
	in the PNJL model at zero density (left) and at finite density (right) where the Fermi 
	momentum contributes to the kinetic energy. 
	Varying the mass changes the strength of explicit breaking of  the $\Zed_3$ 
	symmetry: when the mass is lower, the transition becomes a larger crossover.
	The higher the chemical potential, the higher the mass is needed  to restore the  
	symmetry and to have a first-order phase transition.}
 \label{fig:Z3ExplicitBreaking_m}
\end{figure*}
%%%%%%%%%%%%%%%%%%%%%%
%%%%%%%%%%%%%%%%%%%%%%
\begin{figure*}
  \centering
  \includegraphics[width=0.47\textwidth]{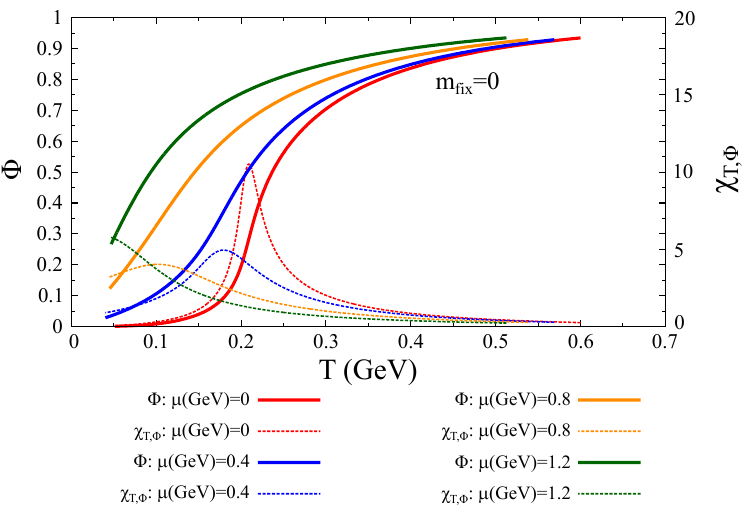}
  \includegraphics[width=0.47\textwidth]{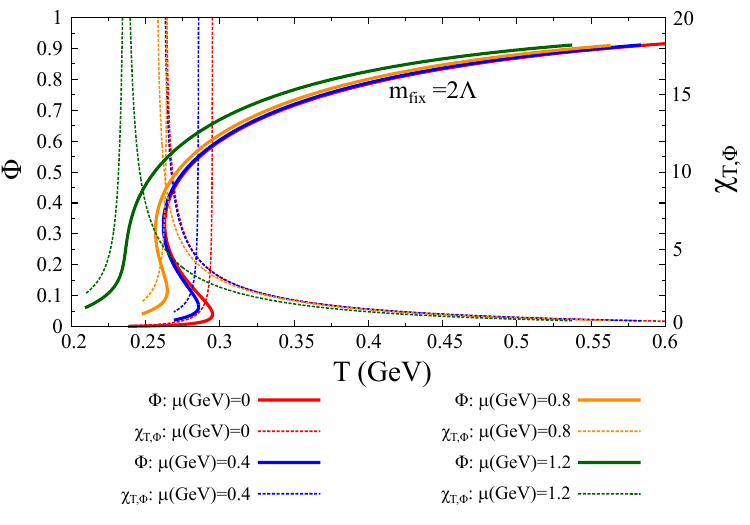}
  \caption{$\Phi$ and $\chi_\Phi=d\Phi/dT$ in the PNJL model as functions  
	of $T$  for  different   values  of   the  chemical potential at zero quark mass (left) and large quark mass $m=2\Lambda$ (right). 
	A higher chemical potential implies a larger kinetic term and explicit 
	$\Zed_3$-symmetry  breaking. A  larger  mass is  required to  get  back a  
	weak explicit breaking and a first-order phase transition.}
\label{fig:Z3ExplicitBreaking_mu} 
\end{figure*}
%%%%%%%%%%%%%%%%%%%%%%

The general conclusion is that quite certainly this class of models
underestimates some effects for the lack of dynamical quark loop
effect on the Polyakov loop. The only effect present is due to the
presence of thermodynamical quarks (via the quark Boltzmann factor in
the mean-field equation) that results from the minimal coupling
$\bar q A_\mu q$. But, since $A_\mu$ is not dynamical this term will
not generate quark loops in a diagrammatic approach ($A_\mu$ acts as
an external constant gauge field only).

%%% %%% %%% %%% %%% %%% %%% %%% %%% %%% %%% %%% %%% %%% %%% %%% %%% %%% %%% %%%
\section{Dynamical thermal quarks effect
} \label{sec:thermaleffect}
%%% %%% %%% %%% %%% %%% %%% %%% %%% %%% %%% %%% %%% %%% %%% %%% %%% %%% %%% %%%

%{\question Henceforth, we will use the PNJL model, except when explicitly stated.}

In this section we will consider the effect of the kinetic
contribution of quarks in Eq.~(\ref{eq:domegadfi}). At zero density
($\mu=0$) and with infinite quark mass, the second term disappears as
it should. Indeed, quarks are then no more dynamical (they are
quenched) and the system is in its pure gauge or Yang-Mills limit.
The $\Zed_3$ symmetry is then exact (for the grand potential) but
spontaneously broken at $T > T_0$.

When quarks are considered dynamical again, by lowering  their masses approaching the respective
physical values, the $\Zed_3$ symmetry gets also an explicit breaking term
(see Secs.~2.2.2 and 2.3.2 in \Ref{Stiele:2014gks}) due to the kinetic term in the
Lagrangian.
This mechanism can be understood in detail by solving the mean-field equations of the PNJL or PQM model, 
not self-consistently  as usual, but simply by fixing $m$ to a given value $m_\text{fix}$ in 
Eq.~(\ref{eq:domegadfi}),  as already mentioned previously. In Figs.~\ref{fig:Z3ExplicitBreaking_m} 
and 
\ref{fig:Z3ExplicitBreaking_mu} we plot the behavior of  the Polyakov loop for 
different  masses, and hence changing the ratio  between the mass energy $E_m$ and  
the kinetic energy $E_k$ of a  quark: increasing the mass reduces  the ratio 
$E_k/E_m$;  when this ratio is  zero (quenched quarks) the symmetry is restored.
To better illustrate if there is a stronger or weaker explicit 
breaking of the $\Zed_3$ symmetry, we use as Polyakov-loop potential a parametrization that has 
a strong first-order phase transition in the pure gauge sector (the $U_{Log}$ potential, 
Eq.~(\ref{UeffLog})). If the  transition becomes  smoother (crossover like transition) 
it indicates that we have a stronger explicit breaking of the $\Zed_3$ symmetry. 

Figures \ref{fig:Z3ExplicitBreaking_m} and \ref{fig:Z3ExplicitBreaking_mu} also 
show the  impact of  varying the chemical  potential.
Indeed, there is a contribution to  the kinetic energy due to the Fermi
momentum  of quarks:  increasing  $\mu$ increases  the  ratio $E_k/E_m$ and  the
symmetry is more strongly broken. We will discuss further this effect  in
Sec.~\ref{sec:CondensPloopcorr}
but for now, we focus on the zero density scenario where there is only one 
contribution to the kinetic energy.

We observe  that a lowering of the mass term introduces  a stronger
explicit breaking of $\Zed_3$ symmetry as  seen by the large crossover obtained
and the respective smooth Polyakov-loop susceptibility,
\be
\left.\chi_\Phi = \frac {d\Phi}{dT}\right|_{\mu=0}.
\label{eq:chi_Phi}
\ee

\begin{figure*}
  \centering
  \includegraphics[width=0.47\textwidth]{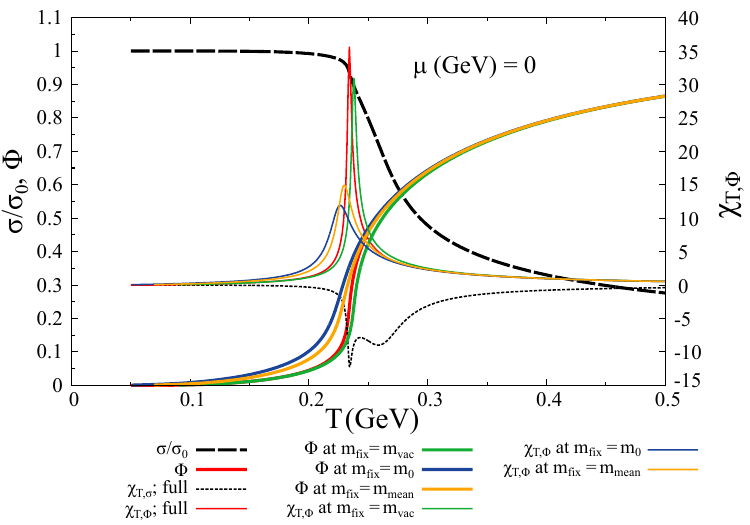}
  \includegraphics[width=0.47\textwidth]{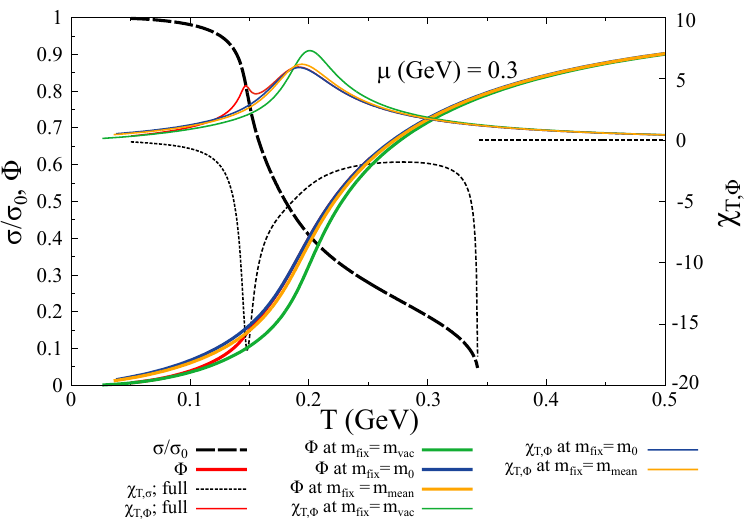}
  \caption{Quark   condensate   normalized   to   its   vacuum   value
($\sigma/\sigma_0$) and the Polyakov loop as a function of $T$ at zero
and non-vanishing chemical potential (left  and right).
The Polyakov loop is computed either with  a fixed mass ($m_0$, the mass in the vacuum $m_{vac}$ or $m_{mean}=\frac{m_0+m_{vac}}{2}$) or by fully solving
the mean-field equations.
}
\label{fig:PhiMFix}
\end{figure*}

On Fig.~\ref{fig:PhiMFix} we notice that, except when very close to the
chiral transition (where $m$ changes quickly) the Polyakov-loop
susceptibility is not much altered if the mass is kept constant. Thus, the transition properties are hardly affected by the exact value of the mass. Again, it is probable that this lack of sensitivity
to the mass of the quark indicates a missing mechanism  to account for
the quark back-reaction on $\Phi$ because of  the lack of dynamical quark loops in
its calculation. For this reason, in the literature it  is
proposed that one adds phenomenologically more sources of back-reactions by using, for
example in the entangled PNJL
(EPNJL) model a scalar coupling $G_S$ that depends on  $\Phi$ \cite{Sakai:2010rp}. Fitting this dependence is quite difficult (due
to the  lack of  enough data)  and some authors  proposed to use the 
Roberge-Weiss transition at imaginary chemical potential \cite{Roberge:1986mm} 
to fit  this dependence.  Other sources of  back-reaction may  be  added  by  
using  a $\mu$-dependent  gauge transition temperature  $T_0(\mu)$ 
\cite{Schaefer:2007pw} or even a  $\mu$-dependent Polyakov-loop potential 
${\cal U}(\Phi ;  T, \mu)$ \cite{Dexheimer:2009hi}.

We will not discuss further the EPNJL model but we would like to
emphasize that it is interesting to understand the multiple aspects of
back reaction already in the PNJL/PQM models before resorting to this
type of model with reinforced back-reaction. Furthermore, models with
many sources of back-reaction ($G(\Phi)$, $T_0(\mu)$) introduce many
new parameters and it is not easy to parametrize them consistently
without overestimating the back-reaction. In this regard, LQCD data at
finite density would be very helpful or high density constraints (e.g., the measurement of the position of the CEP in heavy ion collision (HIC) experiments). It is
interesting to note that LQCD data at finite imaginary chemical
potential exists
\cite{deForcrand:2002hgr,Wu:2006su,Nagata:2011yf,Cea:2012ev,Bonati:2018fvg}
and it is assumed that the continuation $i \mu \rightarrow \mu$ is
analytic (this can be easily seen in the grand canonical potential of
the PNJL/PQM model). Therefore, constraints at imaginary $\mu$ are
relevant for models to be used at real $\mu$.

Let us mention a final comment on Fig.~\ref{fig:PhiMFix}: 
as already discussed in Sec.~\ref{sec:StatisticalConf}, the main effect of the Polyakov loop
on the quark condensate is to obtain a steeper crossover transition and in the left panel of
Fig.~\ref{fig:PhiMFix}  we see that, indeed, when  $\chi_{\Phi}$ (see Eq.~(\ref{eq:chi_Phi})) is at its maximum it also introduces a dominant peak in $\chi_{\sigma}$.

%%%%%%%%%%%%%%%%%%%%%%
%%%%%%%%%%%%%%%%%%%%%%

%%% %%% %%% %%% %%% %%% %%% %%% %%% %%% %%% %%% %%% %%% %%% %%% %%% %%% %%% %%%
\section{Mean-field phase diagram} \label{sec:MFPhaseDiag}
%%% %%% %%% %%% %%% %%% %%% %%% %%% %%% %%% %%% %%% %%% %%% %%% %%% %%% %%% %%%

Using  the techniques  described  in \cite{Costa:2010zw}  we will now
investigate the phase structure of a  common two-flavor PNJL model, 
computed in this case with the logarithmic Polyakov-loop potential\footnote{The results in 
this section do not depend on the precise details of the model. 
They are features shared by all Polyakov-loop type models (see Refs.~\cite{Costa:2008gr,Costa:2009ae} for results with the polynomial 
Polyakov-loop potential).}. To investigate the deconfinement transition we study the Polyakov-loop susceptibility at finite chemical potential,
\be
\left.\chi_\Phi = \frac {d\Phi}{dT}\right|_{\mu},
\ee
and extract from it the characteristic pseudo-critical transition
temperature associated to deconfinement, defined as the maximum of the
susceptibility. In fact, we could also use
$\chi_{\bar\Phi} = d\bar\Phi/dT$ to obtain the transition temperature
being that $\chi_{\Phi} = \chi_{\bar\Phi}$ at $\mu=0$. At finite
density $\Phi \ne \bar\Phi$ and the transitions temperatures obtained
by the two susceptibilities are slightly different but very close to
each other.

%%%%%%%%%%%%%%%%%%%%%%
\begin{figure}%[t]
  \centering
  \includegraphics[width=0.47\textwidth]{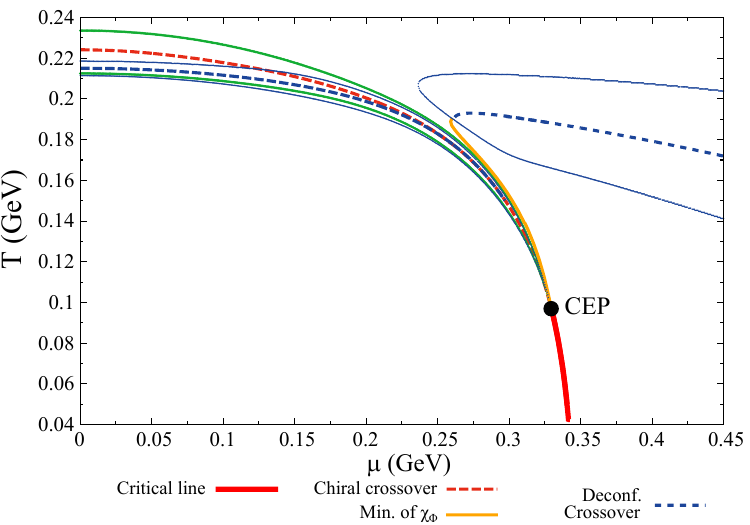}
  \caption{Phase diagram  obtained in  the ($T,  \mu$) plane for the  PNJL model.
  The red dashed and solid lines represent the chiral crossover and critical line. 
	The blue dashed line marks the deconfinement crossover and is within the transition 
	region which is outlined by the blue solid lines (and defined within the main text).
  The yellow  line represents  the minimum  of the Polyakov-loop susceptibility,
  $\chi_\Phi$.
  The transition region of the chiral crossover is within the green solid lines.
  }
\label{fig:PNJLPhases}
\end{figure}
%%%%%%%%%%%%%%%%%%%%%%

Some details of the numerical calculations for a similar model, which
are not of importance for our quantitative discussions, can be found
in \cite{Biguet:2014sga}. To obtain the susceptibilities the ``easy
way'' is to compute them as numerical derivatives. This is both
inefficient (it is quite long since for each step in temperature the
mean-field equations have to be solved several times) and inaccurate.
The inaccuracy will show up as a numerical fluctuation in the
transition line plot which shows the maxima of the susceptibilities. A
more accurate and faster way is to compute analytically (or at least
semi-analytically in the sense that the non analytic thermal integrals
are kept as numerical integrals) all first- and second-order partial
derivatives of the grand potential with respect to the order
parameters and the thermodynamic
parameters $T$ and $\mu$ and then combine them to get the desired
susceptibilities (see \cite{Wagner:2009pm,Biguet:2014sga}).

%%%%%%%%%%%%%%%%%%%%%%
\begin{figure}%[t]
  \centering
  \includegraphics[width=0.47\textwidth]{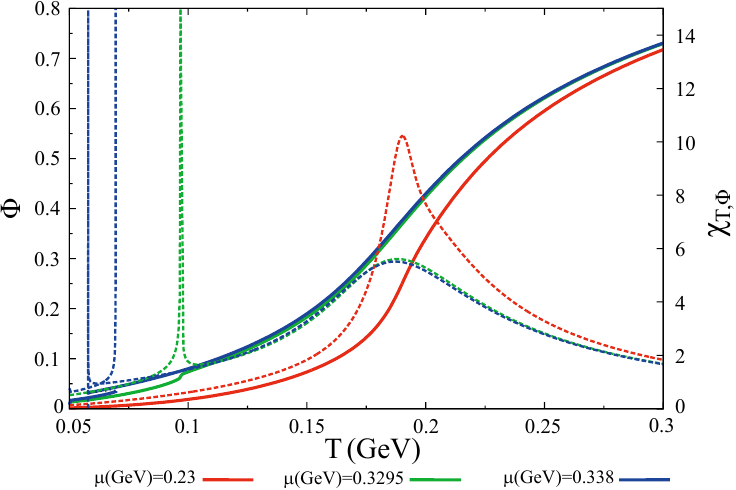}
    \caption{Polyakov loop, $\Phi$, and its susceptibility for chemical potentials
 below the one of the chiral CEP in the PNJL model (red), at
the CEP chemical potential (green) and above the CEP chemical potential (blue).
    }
\label{fig:PhiSusc}
\end{figure}
%%%%%%%%%%%%%%%%%%%%%%

Results   for   the  order   parameters   $m$   and  $\Phi$   are presented in
Fig.~\ref{fig:PNJLPhases}. Concerning the chiral properties, they behave as discussed in the 
introduction and we observe a first-order phase transition line that 
starts at zero temperature and ends with increasing temperature at the CEP, where a second-order phase transition occurs at ($T^\mrm{CEP},\mu^\mrm{CEP}$), continued by a crossover for further increasing temperatures and decreasing chemical potentials.

For what concerns the deconfinement transition we see that
for chemical potentials $\gtrsim250\,\mrm{MeV}$ $\chi_\Phi$ has several 
maxima or divergent peaks (Fig.~\ref{fig:PhiSusc}) but the order  parameter always shows a  crossover
transition.
Indeed if we look at the Polyakov-loop susceptibility in Fig.~\ref{fig:PhiSusc}, we see that  there is a divergence at the  chiral transition temperature but physically this is
not the signal of the deconfinement transition. 
We see that at this chiral first-order transition,
$\Phi$ increases only about 5\%  and stays below 0.2: this  is not a (first-order)
deconfinement transition because quarks are still statistically confined when $\Phi = 0.2$.
There is also a bump in $\chi_\Phi$, and there $\Phi$  changes between 0.4 and 0.7 and the system reaches the statistically deconfined phase. Hence, we will consider the temperature at the maximum as the one of the deconfinement crossover.

For the sake of completeness, we also have plotted in the phase diagram of 
\Fig{fig:PNJLPhases}, besides the maximum of $\chi_{\Phi}$, its inflection points
(which estimate the ``extent'' of the crossover region) and also the minimum of
$\chi_{\Phi}$, when it exists, between the bump of the deconfinement crossover and 
the peak of the first-order phase transition of chiral symmetry. 

It should be noted that both transitions appear to be strongly correlated below $\mu\lesssim250\,\mrm{MeV}$ and then become uncorrelated at higher chemical potentials, indicating the mutual influence
between chiral and deconfinement sectors.   
The topic of the  quark - Polyakov-loop  correlation deserves therefore special
attention as discussed in \secref{sec:backreaction}.

%%% %%% %%% %%% %%% %%% %%% %%% %%% %%% %%% %%% %%% %%% %%% %%% %%% %%% %%% %%%
\section{Quark condensate and Polyakov-loop correlation \label{sec:CondensPloopcorr}}
%%% %%% %%% %%% %%% %%% %%% %%% %%% %%% %%% %%% %%% %%% %%% %%% %%% %%% %%% %%%

Here we continue our study of the finite density (finite chemical potential) region using the PNJL model
and we compare results obtained with and without fixing the mass.
At zero density, the pseudo-critical transition temperatures for the partial chiral-symmetry restoration and deconfinement coming from LQCD results are very close, differing 
by approximately 20 MeV \cite{Borsanyi:2010bp}.
At finite  density the  status  is not  clear (different  model parametrizations 
give very  different results). 
Experimentally, there is  a lack of high  temperature and density  experiments. 
Furthermore, it is also difficult  to extract information about the chiral sector 
only or the deconfinement sector only.

We will study the high density regions and the possible opening of a new phase near the CEP (this  feature  is  model  dependent  and will eventually
disappear by considering a $T_0(\mu)$ parametrization, see Ref.~\cite{Kahara:2010wh}).
From Fig.~\ref{fig:PNJLPhases} and starting at $\mu=0$, it is seen that as the chemical potential increases the chiral and the deconfinement crossover lines approach each other as one approaches the CEP. Then they merge for $\mu<\mu^{CEP}$. This property of the QCD phase diagram has been used to estimate a lower bound for the chemical potential at the CEP (see for example Fig. 1 of Ref.~\cite{Endrodi:2011gv}). Still, near the CEP, chiral and deconfinement transitions decouple.

\begin{figure}[t]
  \centering  
\includegraphics[width=0.49\textwidth]{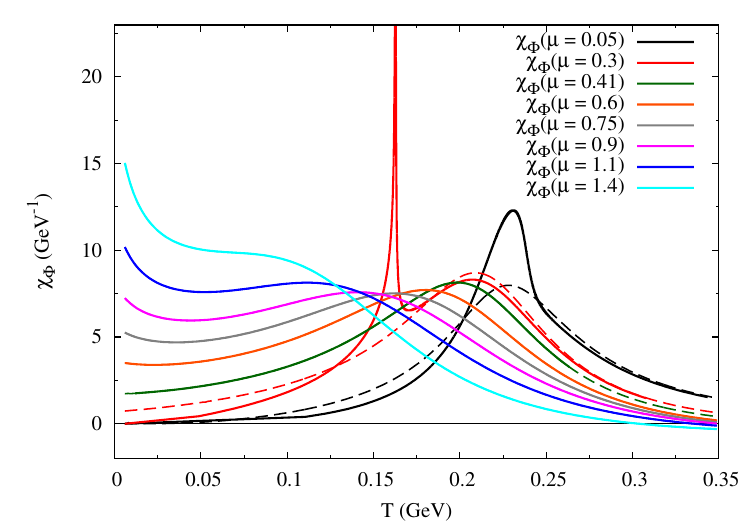}
  \caption{Polyakov-loop  susceptibility as  a  function  of $T$  for  different
	chemical potentials (given in GeV).  The dashed lines are the calculation with fixed 
	$m = m_{mean}$ (see definition in the main text); the full line is the fully 
	consistent calculation. Notice in the region of low temperature that for  high 
	chemical potentials,  $\chi_{\Phi}$ has no more maximum.
	}
\label{fig:dPhiMFix} 
\end{figure}

In Fig.~\ref{fig:dPhiMFix} the  Polyakov-loop susceptibility is computed 
(using the polynomial Polyakov-loop potential)
with the full mean-field equations, and with a mass fixed  to the mean value
of its value in the vacuum and the current quark mass $m_0$, $m_{mean} = \frac{m_0 + m_{vac}}{2}$ (dashed lines).
We can see that for a chemical potential that gives a first-order phase transition  
for  the  chiral transition (red full line),  there is a peak  in the Polyakov-loop susceptibility at the chiral transition.
Having in mind the discussion of the previous section,
at this peak, the Polyakov-loop only varies from 0.2  to 0.3.
There is also a bump in $d\Phi/dT$ at higher temperatures and around this bump, 
$\Phi$  goes  from  0.4 (essentially  confined phase)  to 0.7  (essentially deconfined).   
Our interpretation is the following: 
the peak in $\chi_{\Phi}$ is not a sign of a first-order deconfinement transition
because the value of the Polyakov-loop remains small.
But there  is a  smooth crossover towards the deconfined phase at  higher temperatures (at the bump).
These two transitions are relevant because they are related to physical
effects of the thermodynamic potential we are using, the grand canonical
potential: the peak (the chiral first-order transition) manifests
itself as a singularity of the derivative of the grand potential; the bump
gives a small region where there is a sharp increase then decrease
(but continuous) of the derivative of the grand potential with respect to the 
order parameters. It is a   manifestation of the deconfinement crossover. 
These two transitions show their imprint on the pressure as we will discuss in 
\Sec{sec:CCS} where a plot of the relative pressure difference (compared to the 
Stefan-Boltzmann one, \Fig{fig:DeltaP}) clearly shows that both transitions 
delimit three distinct regions in the phase diagram.
This interpretation opens the possibility to describe an intermediate
phase, that we call confined chirally restored phase (CCS). Indeed, it
can be explicitly shown that in this phase, quark matter is still
(statistically) confined. However, we postpone the detailed discussion
to \secref{sec:CCS}.

%%%%%%%%%%%%%%%%%%%%%%
%%%%%%%%%%%%%%%%%%%%%%

In Fig.~\ref{fig:dPhiMFix} it can also be seen that when using a constant mass, the
chiral  peak  in  the Polyakov-loop susceptibility disappears (dashed lines).    
This is understandable since we argued that there is not a significant variation in
$\Phi$ at these temperatures but only a very strong mass change that creates
the large derivative. $\Phi$ itself does not vary much since we saw that
it is  not very  sensitive to mass variations below $\Lambda_{QCD}$.

When the mass is fixed the maximum is approximately the same
as in the full calculation, not taking into account the possible divergent peak. This shows that
taking $m$ fixed really allows to decorrelate the chiral condensate
effects because the shape of the Polyakov-loop susceptibility does not change significantly 
when the mass varies strongly  (in the  interval $T \in [100 \text{MeV}, 250 \text{MeV}]$ the mass goes approximately from  320 MeV to
150  MeV). All  the  remaining  effects are  only  due  to the  chemical
potential, an aspect that we will discuss in 
the next section about the Fermi momentum effect. 

Another important aspect is that  the use of a fixed  mass, $m_{\text{fix}}$,  is also
very useful  numerically. If we want  to draw a phase diagram, but the 
exact behavior of the Polyakov loop near the CEP is not so relevant, we can 
use this trick and  get a faster algorithm since only two
mean-field equations have to be solved\footnote{We also expect that in
some reasonable limits an analytical  solution can be found.}. 
This  method can  be  helpful  even  to compute  the  full
solution: it can  greatly increase the calculation speed if we  start to compute
the  approximated $m_{\text{fix}}$  value and  then incorporate  it  in the solving
algorithm  for  the full  equations.   This  kind of  algorithm  (root
polishing  when the approximate solution is known)  is particularly helpful when 
the dimension  of the  problem increases (here three dimensions since there are
three mean-field equation equations; including the strange  sector and without isospin
symmetry it would be five dimensions).

\begin{figure}
  \centering
  \includegraphics[width=0.49\textwidth]{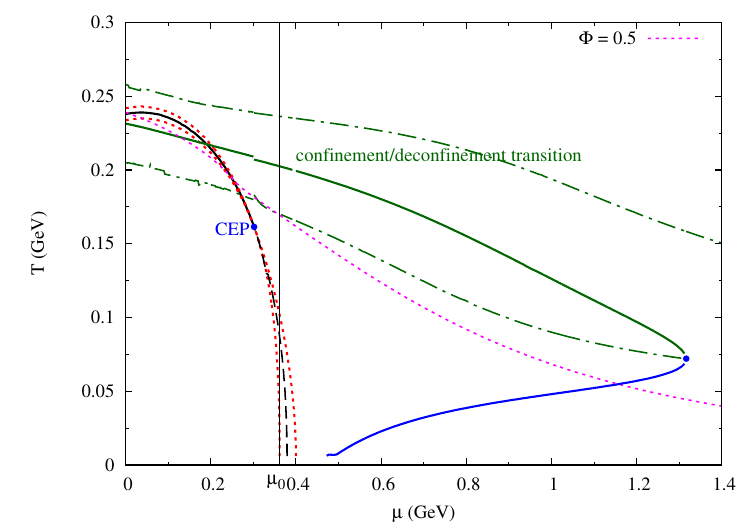}
  \caption{The PNJL-model phase   diagram   using   a  constant   mass   
	to   compute $\Phi$. 
  Other  features are added: the magenta dashed line represents $\Phi =0.5$; 
  the blue line is given by the minimum of the Polyakov-loop susceptibility.
  When this minimum and the maximum (the crossover curve, represented by the
  green line) meet, it is no  more possible to strictly define even  a crossover 
  transition at higher chemical potentials. 
	$\mu_0$ is the  chemical potential where $\chi_\Phi(T=0)$ starts to be non zero, 
	see \Fig{fig:dPhiMFix}. The red dashed lines represent the transition region 
	and metastable region of the chiral transition and the dash-dotted green lines 
	the transition region of the deconfinement crossover. 
  }
\label{fig:mfixPhaseD}
\end{figure}

It is possible to qualitatively compare both methods, fixed mass and full 
calculation, by looking  to  the phase  diagram  in  Fig.~\ref{fig:mfixPhaseD} 
(here the Polyakov-loop potential is the polynomial one but qualitatively our 
results are independent of the choice of this potential).
The conclusion is that except  in the region of the CEP  it does  not differ 
from the previous one found in Fig.~\ref{fig:PNJLPhases}, \secref{sec:MFPhaseDiag}. 
The extra features will be discussed in  the next section since they concern
large density effects.

%%%%%%%%%%%%%%%%%%%%%%
%%%%%%%%%%%%%%%%%%%%%%

%%% %%% %%% %%% %%% %%% %%% %%% %%% %%% %%% %%% %%% %%% %%% %%% %%% %%% %%% %%%
\section{Fermi momentum effect \label{sec:FermiMomEffect}}
%%% %%% %%% %%% %%% %%% %%% %%% %%% %%% %%% %%% %%% %%% %%% %%% %%% %%% %%% %%%

In the previous section, we have seen in detail one important aspect
of the chiral transition/deconfinement transition correlation by
carefully considering the effect of the quark masses on the field
equations. As mentioned, there is another quark effect in the mean-field equations via the quark chemical potential, which originates from
a non zero Fermi momentum of the quarks.

Because of the covariant derivative and the inclusion of $\mu$ in the
Lagrangian, there is a dependence on $\mu$ in Eq.~(\ref{eq:domegadfi})
that can have a strong effect on the system. In fact, the Fermi energy
will act exactly as any kinetic energy (in the Lagrangian
$\mu\gamma^0$ acts as $\partial_0\gamma^0$) and increases the explicit
breaking of $\Zed_3$ symmetry. As seen in Fig.~\ref{fig:dPhiMFix},
where we computed $\chi_{\Phi}$ for several chemical potentials, at
some point, the breaking of the symmetry is so strong that the
transition cannot be defined even as a crossover anymore (at least
according to our definition). From the moment that the
Polyakov-loop susceptibility does not have a maximum anymore, our
interpretation is that the transition goes so fast from $\Phi = 0$ to
$1$ (and without the characteristic of a crossover transition where
correlation lengths quickly increase then decrease) that we can almost
say that the system is deconfined even at zero temperature (in Fig.
\ref{fig:mfixPhaseD}, $\Phi$ is already 0.5 at $T=50$ MeV for large
values of $\mu$). $\Phi$ is forced to be zero at zero temperature but
in the large $\mu$ limit, $\Phi=1$ at any non-vanishing temperature.
Hence, the model basically tells us that, apart from a small missing
ingredient, a
form of effective deconfinement only induced by the density is
possible. This missing ingredient is again probably related to quark
back-reaction on the gauge fields.

Here, we want to point out that it is quite  remarkable that, in their
simple  versions,  the  PNJL and PQM models  already  allow  a  deconfinement
transition caused only by the density (with a small temperature), 
without any sort of {\it ad hoc} back-reaction added. 

In the phase diagram presented in Fig.~\ref{fig:mfixPhaseD}, more relevant features 
can be found.  
It is interesting to have an idea of the value of the order parameter, $\Phi$, with
respect to  the transition  lines, so we add the $\Phi  = 0.5$ 
characteristic line (magenta dashed curve).  More importantly, we show in 
this diagram that there is a  kind of ``crossover end point'' in this 
model calculation. At very high $\mu$
the explicit breaking of $\Zed_3$ symmetry is so strong that $\chi_\Phi$ 
has  no longer  a maximum  (except at $T=0$, see also Fig.~\ref{fig:dPhiMFix}).
Somehow, this explicit symmetry breaking is dominant in the Lagrangian and after
this point it is no longer possible to talk about a transition, even
a crossover one. The symmetry is completely destroyed by this breaking term. 

As a conclusion, we emphasize that only the careful assessment of
the sources of correlations in the mean-field equations allowed us
to see  those effects. We advocate that even in  this simple
model some mechanisms, such as the Fermi momentum action, can be  seen.  The
use of more realistic models (e.g., SU(3)  PNJL/PQM models) would also be
interesting to get more faithful results. However, we think  
that those  models, with many  interplays  between  physical  sectors  
whose  contributions are controlled by parameters 
adjusted to phenomenology, are complementary to the simpler models and both have 
their merit.

%%% %%% %%% %%% %%% %%% %%% %%% %%% %%% %%% %%% %%% %%% %%% %%% %%% %%% %%% %%%
\section{Combined effects of chiral condensate and Fermi momentum\label{sec:CombindedEffects}}
%%% %%% %%% %%% %%% %%% %%% %%% %%% %%% %%% %%% %%% %%% %%% %%% %%% %%% %%% %%%

\begin{figure}%
\begin{center}
\includegraphics[width=0.49\textwidth]{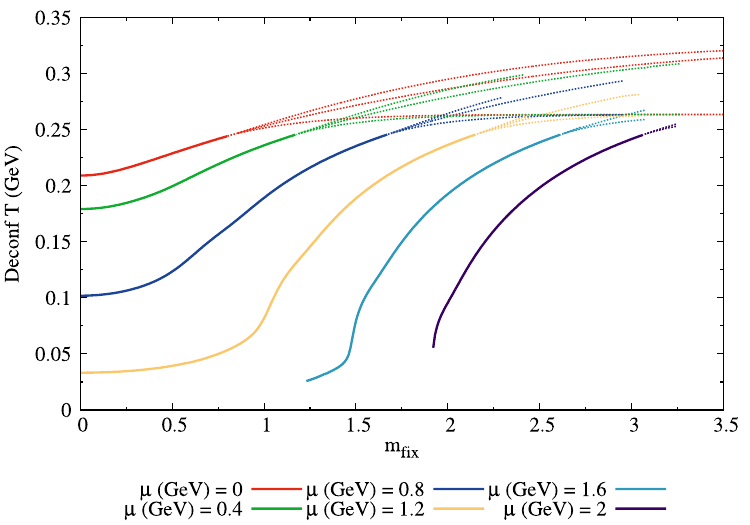}
\end{center}
\caption{\label{fig:Tmumfix} Deconfinement  characteristic temperature
as  a  function   of  the  fixed  mass  and   for  different  chemical
potentials. The full  line is the crossover temperature  and the dotted
lines are the metastable limit of a first-order phase transition.
}
\end{figure}

To complement our discussion on the influence of changing the chiral condensate 
(by varying  $m$) and the Fermi  momentum (by varying $\mu$) on $\Phi$, 
Fig.~\ref{fig:Tmumfix}  presents the results for  the deconfinement
temperature obtained when varying  both $\mu$  and the  fixed  mass $m_{\text{fix}}$.

%%%%%%%%%%%%%%%%%%%%%%
%%%%%%%%%%%%%%%%%%%%%%

Taking $\mu  = 0$,  the deconfinement  transition for $m_{\text{fix}} = 0$  is a
crossover with a transition temperature that is about the  same as the chiral one (for this
model parametrization  it is about  220 MeV). This corresponds  to the
usual situation  of the PNJL/PQM  model when both transitions almost coincide  at zero
density.       As      already     explained      when      discussing
Fig.~\ref{fig:Z3ExplicitBreaking_m},  by  increasing the  mass  (thus
decreasing the  explicit $\Zed_3$-symmetry breaking) the  temperature increases
until  it reaches a  value  where  the   crossover  becomes a first-order phase transition,  at  a
temperature  of about  270  MeV (which  is  the value  of  $T_0$ in  this
parametrization).

Now, looking at $m_{\text{fix}} = 0$ and  increasing $\mu$, we see a lowering of
the  deconfinement  crossover  temperature  until a  value  where  the
crossover cannot be defined anymore.  The symmetry breaking induced by
the Fermi momentum is so strong  that after this point, $\Zed_3$ is no
more a valid symmetry of the model.

Finally, by varying both $\mu$ and $m_{\text{fix}}$ it is always possible to go to a
first-order phase transition if the  mass is sufficiently high compared to
$\mu$ in order to counteract the explicit breaking of $\Zed_3$ symmetry.

%%% %%% %%% %%% %%% %%% %%% %%% %%% %%% %%% %%% %%% %%% %%% %%% %%% %%% %%% %%%
\section{Confined chirally restored (CCS) phase}\label{sec:CCS}
%%% %%% %%% %%% %%% %%% %%% %%% %%% %%% %%% %%% %%% %%% %%% %%% %%% %%% %%% %%%

In this section we will take a closer look to the CCS phase, a phase where
chiral symmetry and deconfinement are decorrelated. 

In
PNJL/PQM models it is always necessary to remember that quarks are not
confined in the strict sense and when we talk about confinement 
we mean statistical confinement as given by the definition in 
\secref{sec:StatisticalConf}. These types of models cannot tell us the true nature  
of quark confinement.

To explore the CCS phase, we want to count the number of thermal  quark degrees of
freedom  (the only one affected by statistical  confinement).  
For this reason we need to subtract the quarks in the vacuum, present due to the lack 
of confinement (we have seen that their contribution is quite reasonable  by looking  
to the quark  number density in  \secref{sec:StatisticalConf}).
Hence, we calculate a relative pressure difference by subtracting
the pressure at zero temperature and normalizing it to the Stefan-Boltzmann pressure:
\begin{equation}
  \label{eq:DeltaP}
  \Delta P \equiv \frac{P(T, \mu) - P(T=0, \mu)} {P_{SB}(T, \mu) - P_{SB}(T=0, \mu)}.
\end{equation}

\begin{figure}
  \centering 
\includegraphics[width=0.49\textwidth]{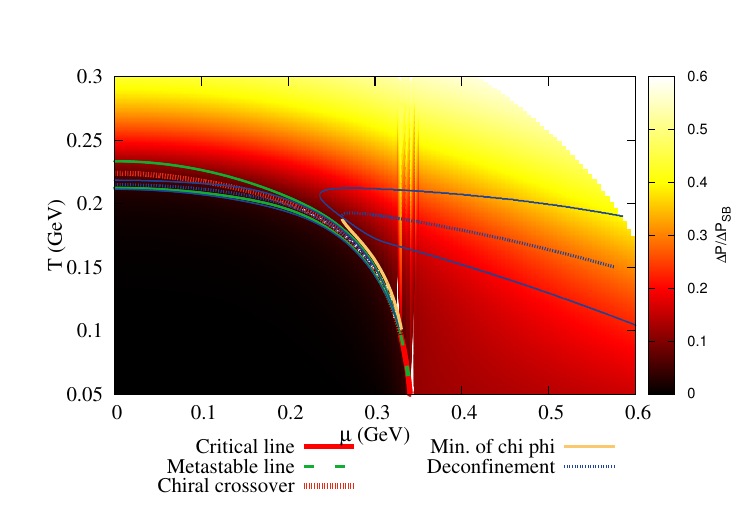}
  \caption{ \label{fig:DeltaP} 
Relative  pressure difference as defined in \Eq{eq:DeltaP} (color map), in  the ($T,  \mu$) plane, superimposed  on the PNJL-model phase diagram of \Fig{fig:PNJLPhases}.
}
\end{figure}

This quantity  is adequately related  to the thermal quark  degrees of
freedom and  indeed it shows  signs of statistical confinement  in the
CCS phase (see Fig.~\ref{fig:DeltaP}).
%%%%%%%%%%%%%%%%%%%%%%
%%%%%%%%%%%%%%%%%%%%%%
In the hadronic phase, $\Delta P$  is zero (the black region), meaning  that the PNJL/PQM
model will  not thermodynamically create  any quarks in  the hadronic phase  (besides those
already present in vacuum). In the QGP phase (the yellow region), thermal quarks are very
active thermodynamically. Finally, in the CCS phase  a quite stable ``plateau''
(the red/orange region at large chemical potential and small temperature)
at around 0.2 is found.  
At a given (small) temperature, there is a discontinuous jump from 0 to 0.2
when increasing $\mu$ across  the first-order phase transition.  So, in the  CCS phase some quarks
have been liberated  by the system due to  the chiral transition but their contribution to thermodynamics remains low.  
It seems that in this phase, where the chiral symmetry is restored, the model tells us 
that quarks still are far from being asymptotically  free. Phenomenologically, this could  have 
rather  important  effects on  experimental  observables  in HICs  (viscosity,  hadronic
abundances, etc).  
If this hypothetical phase exists in QCD, the PNJL/PQM model cannot say if quarks  are
really confined (no asymptotic  states) or if they are deconfined but the model predicts that the interaction
is still strong allowing the existence of mesonic resonance as we will show.
Performing the same calculation in the pure NJL/QM model, i.e.~without statistical  confinement, one observes
at small  temperature  that  $\Delta P$  has the  
same  jump at  the chiral transition  and  then  increases slowly  without a  ``plateau'' behavior.

It is also relevant to notice that, at those densities, an important
missing effect is the presence of diquark condensates. The strange
sector is also important and it deserves further investigation. \\

To confirm this finding and that it may lead to significant
differences in HICs, we study a more direct observable, namely, we check
if mesonic probes act as it is expected in a confined but chirally
symmetric phase.

%%%%%%%%%%%%%%%%%%%%%%
\begin{figure*}
  \centering
 \includegraphics[width=0.32\textwidth]{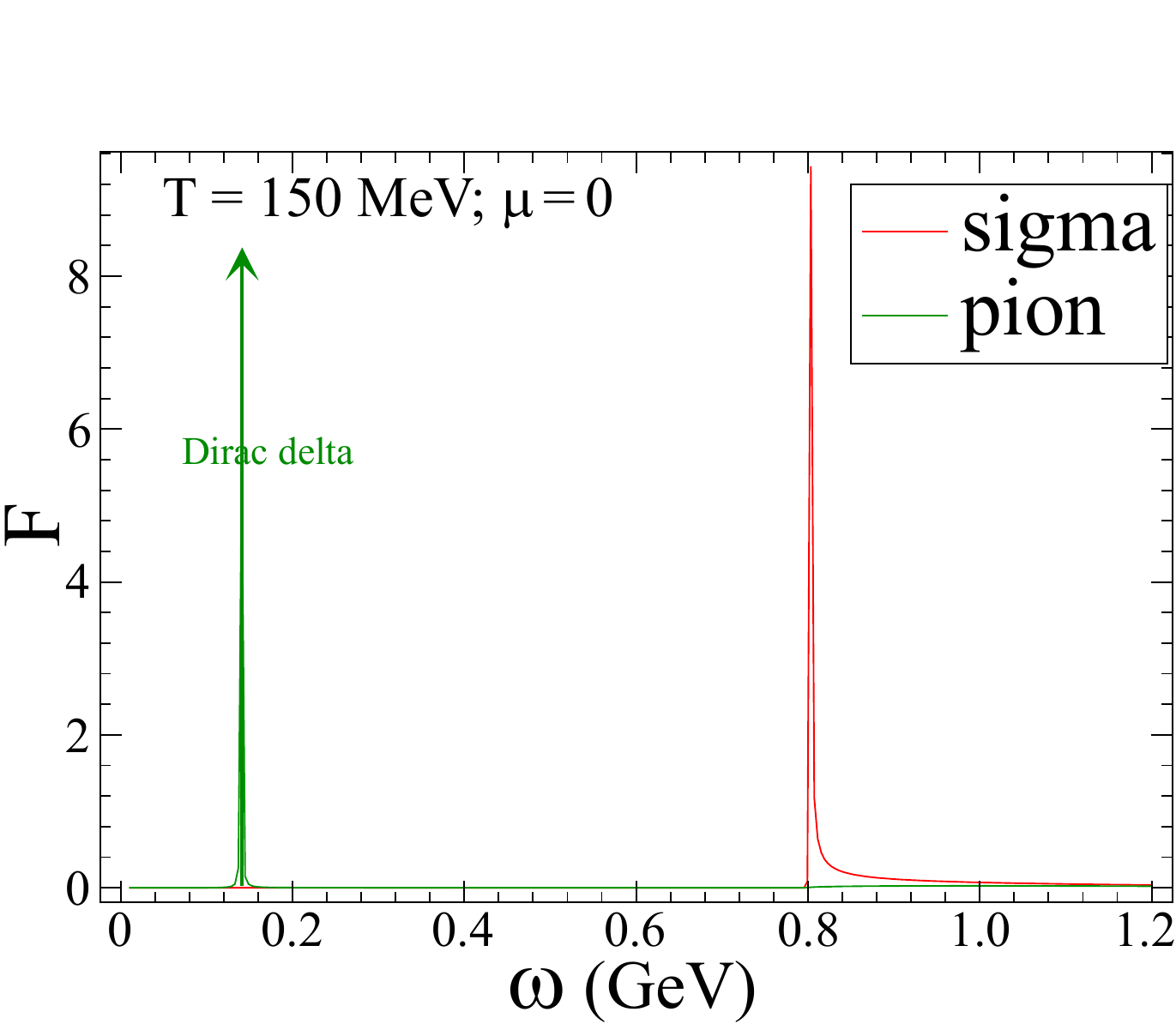}
  \hfill
 \includegraphics[width=0.32\textwidth]{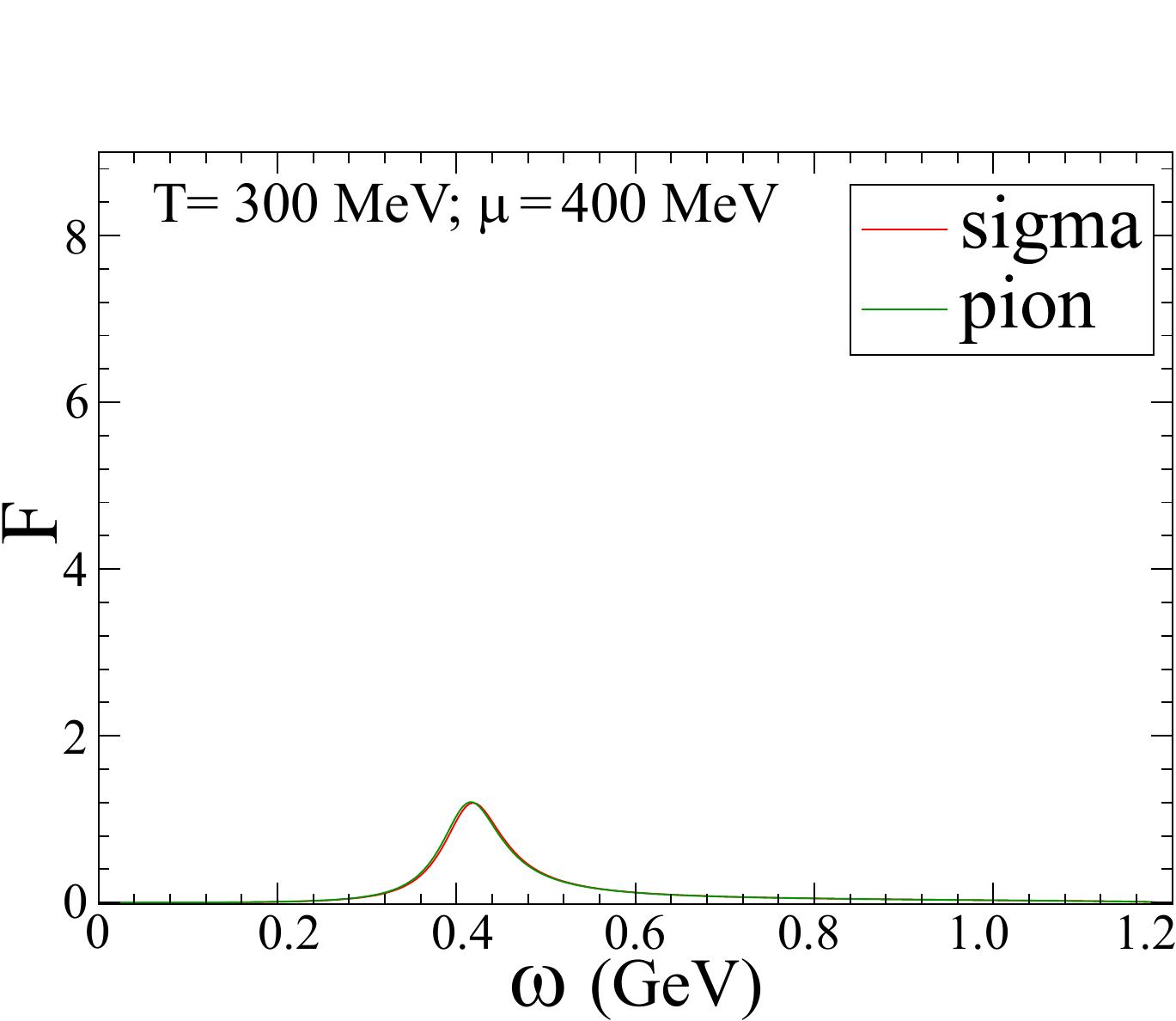}
 \includegraphics[width=0.32\textwidth]{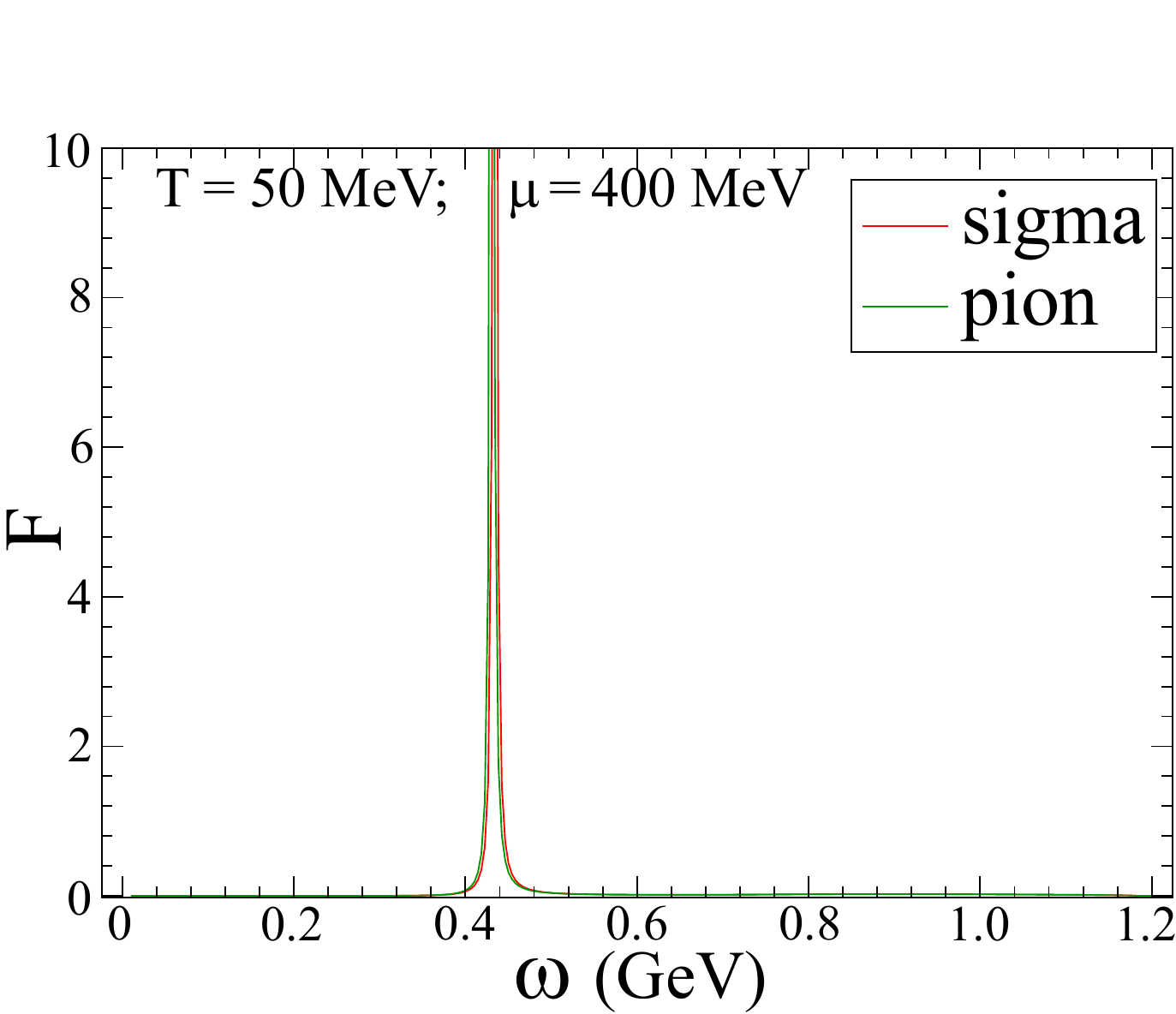}
  \caption{Pion and  sigma spectral function in the PNJL model for the hadronic phase
  (left) and the QGP phase (middle). Confining regime: $\sigma$ is a narrow resonance;
  the chiral symmetry is broken and $\langle\bar qq\rangle  \neq 0$; $\Phi \simeq 0 $.
  Deconfinement: $\sigma$  and $\pi$  spectral functions are  almost the same  as an
  uncorrelated $\bar  q q$  pair;  the chiral  symmetry is restored and
  $ \langle\bar qq\rangle \simeq 0$; $\Phi \neq 0 $.\\
  Right panel: Pion and  sigma spectral function in the CCS phase.
  The system is in the statistically confined phase since $\sigma$
    and $\pi$ are still narrow resonances. But chiral symmetry is
    restored: $ \langle\bar qq\rangle \simeq 0$; $\Phi \simeq 0$.
  }
\label{fig:MesonicCCS1}
\end{figure*}
%%%%%%%%%%%%%%%%%%%%%%

We compute in the PNJL model the spectral functions (see Ref.~\cite{Hansen:2006ee}) in the
meson to quarks channel at zero momentum and show them in
Fig.~\ref{fig:MesonicCCS1}. The finite width of the sigma in vacuum
indicates that quarks are not confined (this channel is open) but there is a
strong interaction and thus it is a sharp resonance.

The expected behavior for mesons in a CCS is observed in the same
Fig.~\ref{fig:MesonicCCS1}: chiral symmetry is effectively restored when the 
sigma and pion spectral functions essentially coincide \cite{Hansen:2006ee} but
the statistical confinement is still strong as can be seen by the fact
that their width is small (right panel) when compared to the fully deconfined phase
(central panel).

Finally, as already mentioned, other parametrizations, for example a $T_0(\mu)$,
will essentially close this CCS phase \cite{Kahara:2010wh}. 
Yet, we think it is still interesting to try to understand better the CCS phase. 
We also think that $T_0(\mu$) is a too strong {\it a priori} generalization
to introduce in the model.  
This dependence is  justified theoretically in Ref.~\cite{Schaefer:2007pw}, where
it is argued that this generalization is needed, because otherwise the confinement-deconfinement
transition has a higher pseudo-critical temperature than the chiral transition, 
which is an unphysical scenario since QCD with dynamic massless 
quarks in the chirally restored phase cannot be confining because the string 
breaking scale would be zero.
However, the problem is that it is poorly constrained. 
Using this parametrization, the hypothesis is basically that
deconfinement and chiral  restoration coincide.
Of course, the CCS phase will then not exist anymore but in our opinion this is not 
a sufficient argument against the existence of a CCS phase in general.
It can eventfully be an argument against the CCS phase only if this {\it a priori}
generalization is correct but this has to be put on a firmer basis  
and, in particular, be better constrained.

In the  meantime, we think that it is still interesting to consider models with a 
CCS phase even if in the future it will be contradicted by experimental data, which will be the ultimate judge. 
Anyway, it is still a nice exercise to see what physical information can be 
gathered by using effective models.

%%% %%% %%% %%% %%% %%% %%% %%% %%% %%% %%% %%% %%% %%% %%% %%% %%% %%% %%% %%%
\section{Conclusions and outlook}\label{sec:CaO}
%%% %%% %%% %%% %%% %%% %%% %%% %%% %%% %%% %%% %%% %%% %%% %%% %%% %%% %%% %%%

At the beginning of this work we present the current status on how accurately 
effective models based on the restoration of chiral symmetry and on the breaking 
of center symmetry (to describe the deconfinement transition of QCD) can reproduce the results coming from LQCD calculations.
Based on the observation that there is a substantial quantitative difference in 
the temperature dependence of the Polyakov loop relative to that of the chiral condensate,
we have analyzed several aspects of the correlation between the
chiral sector and the Polyakov loop.\\
By examining how the pseudo-critical temperatures for chiral and the Polyakov-loop 
transitions depend on the critical scale of the Polyakov-loop potential, $T_0$, 
we see that the correct ordering of chiral and Polyakov-loop transitions (and their
difference in temperature), as given by LQCD calculations, can be achieved with 
a $T_0$ larger than the one given by pure gauge theory (270 MeV). 
The absolute temperature scale of the QCD transition would then be larger than the one 
in LQCD. In the spirit of Landau theory, however, effective models should be considered on relative scales.

For studying the effect of the kinetic contribution of quarks to the correlation between quarks and the Polyakov loop, we compared the behavior of the latter between calculations using a self-consistent medium-dependent quark mass and for several values of constant quark masses, $m_{\text{fix}}$. Except very close to the chiral transition (where $m$ changes quickly), the Polyakov-loop susceptibility is not much altered by keeping the mass constant at the value $m_{mean}$ so that even the transition properties are almost not affected by the precise value of the mass. We interpret this lack of sensitivity to the mass of the quark as a sign of a missing mechanism to take into account the quark back-reaction on the Polyakov loop due to the lack of dynamical quark loops in its calculation.
Comparing the Polyakov-loop susceptibility for a medium-dependent and a constant mass at finite density allows to disentangle a smooth crossover proper to the breaking of center symmetry from a peak in the susceptibility introduced purely by a chiral first-order phase transition but at which the value of the Polyakov loop remains small, so in the confined regime. This picture opens the possibility to discuss an intermediate phase,
which, in our model, is a confined chirally restored phase.
Probing the number of thermal quark degrees of freedom within this phase, indeed, one sees signs of statistical confinement with a quite stable ``plateau'' where some quarks have been liberated due to the chiral transition but where their contribution to thermodynamics remains low. Furthermore, meson spectral functions behave as they are expected in a confined but chirally symmetric phase: the sigma and pion spectral functions essentially coincide but their width is small.\\
Another contribution to the correlation between quarks and the Polyakov loop at non-vanishing density comes from the fact that the Fermi energy acts exactly as any kinetic energy and increases the explicit breaking of center symmetry. From some large chemical potential on this explicit breaking of the symmetry is so strong that the Polyakov-loop susceptibility does not have a maximum anymore except at $T = 0$. Hence, apart from a missing ingredient, deconfinement induced only by the density seems to be realizable within the model. This missing ingredient is again probably related to quark back-reaction on the gauge fields.

Overall, the goal of this work is to shed light on some lesser known aspects of Polyakov-loop--extended chiral models to motivate certain paths for future improvements and detailed quantitative analysis.

%%%%%%%%%%%%%%%%%%%%%%%%%%%%%%%%%%%%%%%%%%%%%%%%%%%%%%%%%%%%%%%%%%%%%%%%%%%%%%%%
%%%%%%%%%%%%%%%%%%%%%%%%%%%%%%%%%%%%%%%%%%%%%%%%%%%%%%%%%%%%%%%%%%%%%%%%%%%%%%%%
\begin{acknowledgments}
  The authors are grateful to the Laboratoire d'Excellence (LABEX) Lyon Institute of Origins
  (ANR-10-LABX-0066) of the Universit\'e de Lyon for its financial
  support within the program ``Investissements d'Avenir''
  (ANR-11-IDEX-0007) of the French government operated by the National
  Research Agency (ANR).
  This work was supported by FCT (Fundação para a Ciência e
  Tecnologia), Portugal, under Grant No. SFRH/BPD/102273/2014 (P.~C.)
  and by an annual INFN Post Doctoral Fellowship (competition INFN
  notice n.~18372/2016, R.~S.).
  R.~S.~acknowledges hospitality at the Laboratoire de Math\'ematiques et
  Physique Th\'eorique (LMPT), today's Institut Denis Poisson at
  Universit\'e de Tours, France and thanks Szabolcs Borsanyi, Gergo
  Endrodi, Tina K.~Herbst, Peter Petreczky, Markus Quandt, and Urko
  Reinosa for providing their respective data and correspondence.
  This work was supported by STSM grants from the COST Action CA15213
  ``Theory of hot matter and relativistic heavy-ion collisions''
  (THOR) and H.H.~and R.S.~thank CFisUC (program UID/FIS/04564/2016) for their support and hospitality.
\end{acknowledgments}

%%%%%%%%%%%%%%%%%%%%%%%%%%%%%%%%%%%%%%%%%%%%%%%%%%%%%%%%%%%%%%%%%%%%%%%
%%%%%%%%%%%%%%%%%%%%%%%%%%%%%%%%%%%%%%%%%%%%%%%%%%%%%%%%%%%%%%%%%%%%%%%

\bibliography{references}

\begin{thebibliography}{74}%
\makeatletter
\providecommand \@ifxundefined [1]{%
 \@ifx{#1\undefined}
}%
\providecommand \@ifnum [1]{%
 \ifnum #1\expandafter \@firstoftwo
 \else \expandafter \@secondoftwo
 \fi
}%
\providecommand \@ifx [1]{%
 \ifx #1\expandafter \@firstoftwo
 \else \expandafter \@secondoftwo
 \fi
}%
\providecommand \natexlab [1]{#1}%
\providecommand \enquote  [1]{``#1''}%
\providecommand \bibnamefont  [1]{#1}%
\providecommand \bibfnamefont [1]{#1}%
\providecommand \citenamefont [1]{#1}%
\providecommand \href@noop [0]{\@secondoftwo}%
\providecommand \href [0]{\begingroup \@sanitize@url \@href}%
\providecommand \@href[1]{\@@startlink{#1}\@@href}%
\providecommand \@@href[1]{\endgroup#1\@@endlink}%
\providecommand \@sanitize@url [0]{\catcode `\\12\catcode `\$12\catcode
  `\&12\catcode `\#12\catcode `\^12\catcode `\_12\catcode `\%12\relax}%
\providecommand \@@startlink[1]{}%
\providecommand \@@endlink[0]{}%
\providecommand \url  [0]{\begingroup\@sanitize@url \@url }%
\providecommand \@url [1]{\endgroup\@href {#1}{\urlprefix }}%
\providecommand \urlprefix  [0]{URL }%
\providecommand \Eprint [0]{\href }%
\providecommand \doibase [0]{http://dx.doi.org/}%
\providecommand \selectlanguage [0]{\@gobble}%
\providecommand \bibinfo  [0]{\@secondoftwo}%
\providecommand \bibfield  [0]{\@secondoftwo}%
\providecommand \translation [1]{[#1]}%
\providecommand \BibitemOpen [0]{}%
\providecommand \bibitemStop [0]{}%
\providecommand \bibitemNoStop [0]{.\EOS\space}%
\providecommand \EOS [0]{\spacefactor3000\relax}%
\providecommand \BibitemShut  [1]{\csname bibitem#1\endcsname}%
\let\auto@bib@innerbib\@empty
%</preamble>
\bibitem [{\citenamefont {Ye}(2019)}]{Ye:2018oqi}%
  \BibitemOpen
  \bibfield  {author} {\bibinfo {author} {\bibfnamefont {Z.}~\bibnamefont {Ye}}
  (\bibinfo {collaboration} {STAR}),\ }\href {\doibase
  10.1016/j.nuclphysa.2018.09.086} {\bibfield  {journal} {\bibinfo  {journal}
  {Nucl.~Phys.}\ }\textbf {\bibinfo {volume} {A 982}},\ \bibinfo {pages} {29}
  (\bibinfo {year} {2019})},\ \Eprint {http://arxiv.org/abs/1809.05713}
  {arXiv:1809.05713 [nucl-ex]} \BibitemShut {NoStop}%
%%CITATION = ARXIV:1809.05713;%%
\bibitem [{\citenamefont {Andronov}(2019)}]{Andronov:2018ccl}%
  \BibitemOpen
  \bibfield  {author} {\bibinfo {author} {\bibfnamefont {E.}~\bibnamefont
  {Andronov}} (\bibinfo {collaboration} {NA61/SHINE}),\ }\href {\doibase
  10.1016/j.nuclphysa.2018.09.019} {\bibfield  {journal} {\bibinfo  {journal}
  {Nucl.~Phys.}\ }\textbf {\bibinfo {volume} {A 982}},\ \bibinfo {pages} {835}
  (\bibinfo {year} {2019})},\ \Eprint {http://arxiv.org/abs/1807.10737}
  {arXiv:1807.10737 [nucl-ex]} \BibitemShut {NoStop}%
%%CITATION = ARXIV:1807.10737;%%
\bibitem [{\citenamefont {Kekelidze}\ \emph {et~al.}(2016)\citenamefont
  {Kekelidze}, \citenamefont {Kovalenko}, \citenamefont {Lednicky},
  \citenamefont {Matveev}, \citenamefont {Meshkov}, \citenamefont {Sorin},\
  and\ \citenamefont {Trubnikov}}]{Kekelidze:2016wkp}%
  \BibitemOpen
  \bibfield  {author} {\bibinfo {author} {\bibfnamefont {V.}~\bibnamefont
  {Kekelidze}}, \bibinfo {author} {\bibfnamefont {A.}~\bibnamefont
  {Kovalenko}}, \bibinfo {author} {\bibfnamefont {R.}~\bibnamefont {Lednicky}},
  \bibinfo {author} {\bibfnamefont {V.}~\bibnamefont {Matveev}}, \bibinfo
  {author} {\bibfnamefont {I.}~\bibnamefont {Meshkov}}, \bibinfo {author}
  {\bibfnamefont {A.}~\bibnamefont {Sorin}}, \ and\ \bibinfo {author}
  {\bibfnamefont {G.}~\bibnamefont {Trubnikov}},\ }\href {\doibase
  10.1016/j.nuclphysa.2016.03.019} {\bibfield  {journal} {\bibinfo  {journal}
  {Nucl.~Phys.}\ }\textbf {\bibinfo {volume} {A 956}},\ \bibinfo {pages} {846}
  (\bibinfo {year} {2016})}\BibitemShut {NoStop}%
%%CITATION = NUPHA,A956,846;%%
\bibitem [{\citenamefont {Senger}(2017)}]{Senger:2017oqn}%
  \BibitemOpen
  \bibfield  {author} {\bibinfo {author} {\bibfnamefont {P.}~\bibnamefont
  {Senger}} (\bibinfo {collaboration} {CBM}),\ }\href {\doibase
  10.1016/j.nuclphysa.2017.06.056} {\bibfield  {journal} {\bibinfo  {journal}
  {Nucl.~Phys.}\ }\textbf {\bibinfo {volume} {A 967}},\ \bibinfo {pages} {892}
  (\bibinfo {year} {2017})}\BibitemShut {NoStop}%
%%CITATION = NUPHA,A967,892;%%
\bibitem [{\citenamefont {Abbott}\ \emph {et~al.}(2017)\citenamefont {Abbott}
  \emph {et~al.}}]{TheLIGOScientific:2017qsa}%
  \BibitemOpen
  \bibfield  {author} {\bibinfo {author} {\bibfnamefont {B.~P.}\ \bibnamefont
  {Abbott}} \emph {et~al.} (\bibinfo {collaboration} {LIGO Scientific,
  Virgo}),\ }\href {\doibase 10.1103/PhysRevLett.119.161101} {\bibfield
  {journal} {\bibinfo  {journal} {Phys.~Rev.~Lett.}\ }\textbf {\bibinfo
  {volume} {119}},\ \bibinfo {pages} {161101} (\bibinfo {year} {2017})},\
  \Eprint {http://arxiv.org/abs/1710.05832} {arXiv:1710.05832 [gr-qc]}
  \BibitemShut {NoStop}%
%%CITATION = ARXIV:1710.05832;%%
\bibitem [{\citenamefont {Soltz}\ \emph {et~al.}(2015)\citenamefont {Soltz},
  \citenamefont {DeTar}, \citenamefont {Karsch}, \citenamefont {Mukherjee},\
  and\ \citenamefont {Vranas}}]{Soltz:2015ula}%
  \BibitemOpen
  \bibfield  {author} {\bibinfo {author} {\bibfnamefont {R.~A.}\ \bibnamefont
  {Soltz}}, \bibinfo {author} {\bibfnamefont {C.}~\bibnamefont {DeTar}},
  \bibinfo {author} {\bibfnamefont {F.}~\bibnamefont {Karsch}}, \bibinfo
  {author} {\bibfnamefont {S.}~\bibnamefont {Mukherjee}}, \ and\ \bibinfo
  {author} {\bibfnamefont {P.}~\bibnamefont {Vranas}},\ }\href {\doibase
  10.1146/annurev-nucl-102014-022157} {\bibfield  {journal} {\bibinfo
  {journal} {Ann.~Rev.~Nucl.~Part.~Sci.}\ }\textbf {\bibinfo {volume} {65}},\
  \bibinfo {pages} {379} (\bibinfo {year} {2015})},\ \Eprint
  {http://arxiv.org/abs/1502.02296} {arXiv:1502.02296 [hep-lat]} \BibitemShut
  {NoStop}%
%%CITATION = ARXIV:1502.02296;%%
\bibitem [{\citenamefont {Aarts}\ \emph {et~al.}(2016)\citenamefont {Aarts},
  \citenamefont {Attanasio}, \citenamefont {J\"ager},\ and\ \citenamefont
  {Sexty}}]{Aarts:2016qrv}%
  \BibitemOpen
  \bibfield  {author} {\bibinfo {author} {\bibfnamefont {G.}~\bibnamefont
  {Aarts}}, \bibinfo {author} {\bibfnamefont {F.}~\bibnamefont {Attanasio}},
  \bibinfo {author} {\bibfnamefont {B.}~\bibnamefont {J\"ager}}, \ and\
  \bibinfo {author} {\bibfnamefont {D.}~\bibnamefont {Sexty}},\ }\href
  {\doibase 10.1007/JHEP09(2016)087} {\bibfield  {journal} {\bibinfo  {journal}
  {JHEP}\ }\textbf {\bibinfo {volume} {09}},\ \bibinfo {pages} {087} (\bibinfo
  {year} {2016})},\ \Eprint {http://arxiv.org/abs/1606.05561} {arXiv:1606.05561
  [hep-lat]} \BibitemShut {NoStop}%
%%CITATION = ARXIV:1606.05561;%%
\bibitem [{\citenamefont {Bloch}\ \emph {et~al.}(2018)\citenamefont {Bloch},
  \citenamefont {Glesaaen}, \citenamefont {Verbaarschot},\ and\ \citenamefont
  {Zafeiropoulos}}]{Bloch:2017sex}%
  \BibitemOpen
  \bibfield  {author} {\bibinfo {author} {\bibfnamefont {J.}~\bibnamefont
  {Bloch}}, \bibinfo {author} {\bibfnamefont {J.}~\bibnamefont {Glesaaen}},
  \bibinfo {author} {\bibfnamefont {J.~J.~M.}\ \bibnamefont {Verbaarschot}}, \
  and\ \bibinfo {author} {\bibfnamefont {S.}~\bibnamefont {Zafeiropoulos}},\
  }\href {\doibase 10.1007/JHEP03(2018)015} {\bibfield  {journal} {\bibinfo
  {journal} {JHEP}\ }\textbf {\bibinfo {volume} {03}},\ \bibinfo {pages} {015}
  (\bibinfo {year} {2018})},\ \Eprint {http://arxiv.org/abs/1712.07514}
  {arXiv:1712.07514 [hep-lat]} \BibitemShut {NoStop}%
%%CITATION = ARXIV:1712.07514;%%
\bibitem [{\citenamefont {Kogut}\ and\ \citenamefont
  {Sinclair}(2019)}]{Kogut:2019qmi}%
  \BibitemOpen
  \bibfield  {author} {\bibinfo {author} {\bibfnamefont {J.~B.}\ \bibnamefont
  {Kogut}}\ and\ \bibinfo {author} {\bibfnamefont {D.~K.}\ \bibnamefont
  {Sinclair}},\ }\href {\doibase 10.1103/PhysRevD.100.054512} {\bibfield
  {journal} {\bibinfo  {journal} {Phys.~Rev.}\ }\textbf {\bibinfo {volume} {D
  100}},\ \bibinfo {pages} {054512} (\bibinfo {year} {2019})},\ \Eprint
  {http://arxiv.org/abs/1903.02622} {arXiv:1903.02622 [hep-lat]} \BibitemShut
  {NoStop}%
%%CITATION = ARXIV:1903.02622;%%
\bibitem [{\citenamefont {Mitter}\ \emph {et~al.}(2015)\citenamefont {Mitter},
  \citenamefont {Pawlowski},\ and\ \citenamefont
  {Strodthoff}}]{Mitter:2014wpa}%
  \BibitemOpen
  \bibfield  {author} {\bibinfo {author} {\bibfnamefont {M.}~\bibnamefont
  {Mitter}}, \bibinfo {author} {\bibfnamefont {J.~M.}\ \bibnamefont
  {Pawlowski}}, \ and\ \bibinfo {author} {\bibfnamefont {N.}~\bibnamefont
  {Strodthoff}},\ }\href {\doibase 10.1103/PhysRevD.91.054035} {\bibfield
  {journal} {\bibinfo  {journal} {Phys.~Rev.}\ }\textbf {\bibinfo {volume} {D
  91}},\ \bibinfo {pages} {054035} (\bibinfo {year} {2015})},\ \Eprint
  {http://arxiv.org/abs/1411.7978} {arXiv:1411.7978 [hep-ph]} \BibitemShut
  {NoStop}%
%%CITATION = ARXIV:1411.7978;%%
\bibitem [{\citenamefont {Maelger}\ \emph {et~al.}(2018)\citenamefont
  {Maelger}, \citenamefont {Reinosa},\ and\ \citenamefont
  {Serreau}}]{Maelger:2018vow}%
  \BibitemOpen
  \bibfield  {author} {\bibinfo {author} {\bibfnamefont {J.}~\bibnamefont
  {Maelger}}, \bibinfo {author} {\bibfnamefont {U.}~\bibnamefont {Reinosa}}, \
  and\ \bibinfo {author} {\bibfnamefont {J.}~\bibnamefont {Serreau}},\ }\href
  {\doibase 10.1103/PhysRevD.98.094020} {\bibfield  {journal} {\bibinfo
  {journal} {Phys.~Rev.}\ }\textbf {\bibinfo {volume} {D 98}},\ \bibinfo
  {pages} {094020} (\bibinfo {year} {2018})},\ \Eprint
  {http://arxiv.org/abs/1805.10015} {arXiv:1805.10015 [hep-th]} \BibitemShut
  {NoStop}%
%%CITATION = ARXIV:1805.10015;%%
\bibitem [{\citenamefont {Maelger}\ \emph {et~al.}(2020)\citenamefont
  {Maelger}, \citenamefont {Reinosa},\ and\ \citenamefont
  {Serreau}}]{Maelger:2019cbk}%
  \BibitemOpen
  \bibfield  {author} {\bibinfo {author} {\bibfnamefont {J.}~\bibnamefont
  {Maelger}}, \bibinfo {author} {\bibfnamefont {U.}~\bibnamefont {Reinosa}}, \
  and\ \bibinfo {author} {\bibfnamefont {J.}~\bibnamefont {Serreau}},\ }\href
  {\doibase 10.1103/PhysRevD.101.014028} {\bibfield  {journal} {\bibinfo
  {journal} {Phys.~Rev.}\ }\textbf {\bibinfo {volume} {D 101}},\ \bibinfo
  {pages} {014028} (\bibinfo {year} {2020})},\ \Eprint
  {http://arxiv.org/abs/1903.04184} {arXiv:1903.04184 [hep-th]} \BibitemShut
  {NoStop}%
%%CITATION = ARXIV:1903.04184;%%
\bibitem [{\citenamefont {Quandt}\ \emph {et~al.}(2018)\citenamefont {Quandt},
  \citenamefont {Ebadati}, \citenamefont {Reinhardt},\ and\ \citenamefont
  {Vastag}}]{Quandt:2018bbu}%
  \BibitemOpen
  \bibfield  {author} {\bibinfo {author} {\bibfnamefont {M.}~\bibnamefont
  {Quandt}}, \bibinfo {author} {\bibfnamefont {E.}~\bibnamefont {Ebadati}},
  \bibinfo {author} {\bibfnamefont {H.}~\bibnamefont {Reinhardt}}, \ and\
  \bibinfo {author} {\bibfnamefont {P.}~\bibnamefont {Vastag}},\ }\href
  {\doibase 10.1103/PhysRevD.98.034012} {\bibfield  {journal} {\bibinfo
  {journal} {Phys.~Rev.}\ }\textbf {\bibinfo {volume} {D 98}},\ \bibinfo
  {pages} {034012} (\bibinfo {year} {2018})},\ \Eprint
  {http://arxiv.org/abs/1806.04493} {arXiv:1806.04493 [hep-lat]} \BibitemShut
  {NoStop}%
%%CITATION = ARXIV:1806.04493;%%
\bibitem [{\citenamefont {Nambu}\ and\ \citenamefont
  {Jona-Lasinio}(1961)}]{Nambu:1961tp}%
  \BibitemOpen
  \bibfield  {author} {\bibinfo {author} {\bibfnamefont {Y.}~\bibnamefont
  {Nambu}}\ and\ \bibinfo {author} {\bibfnamefont {G.}~\bibnamefont
  {Jona-Lasinio}},\ }\href {\doibase 10.1103/PhysRev.122.345} {\bibfield
  {journal} {\bibinfo  {journal} {Phys.~Rev.}\ }\textbf {\bibinfo {volume}
  {122}},\ \bibinfo {pages} {345} (\bibinfo {year} {1961})},\ \bibinfo {note}
  {[,127(1961)]}\BibitemShut {NoStop}%
%%CITATION = PHRVA,122,345;%%
\bibitem [{\citenamefont {Gell-Mann}\ and\ \citenamefont
  {Levy}(1960)}]{GellMann:1960np}%
  \BibitemOpen
  \bibfield  {author} {\bibinfo {author} {\bibfnamefont {M.}~\bibnamefont
  {Gell-Mann}}\ and\ \bibinfo {author} {\bibfnamefont {M.}~\bibnamefont
  {Levy}},\ }\href {\doibase 10.1007/BF02859738} {\bibfield  {journal}
  {\bibinfo  {journal} {Nuovo Cim.}\ }\textbf {\bibinfo {volume} {16}},\
  \bibinfo {pages} {705} (\bibinfo {year} {1960})}\BibitemShut {NoStop}%
%%CITATION = NUCIA,16,705;%%
\bibitem [{\citenamefont {Metzger}\ \emph {et~al.}(1994)\citenamefont
  {Metzger}, \citenamefont {Meyer-Ortmanns},\ and\ \citenamefont
  {Pirner}}]{Metzger:1993cu}%
  \BibitemOpen
  \bibfield  {author} {\bibinfo {author} {\bibfnamefont {D.}~\bibnamefont
  {Metzger}}, \bibinfo {author} {\bibfnamefont {H.}~\bibnamefont
  {Meyer-Ortmanns}}, \ and\ \bibinfo {author} {\bibfnamefont {H.~J.}\
  \bibnamefont {Pirner}},\ }\href {\doibase 10.1016/0370-2693(94)90328-X,
  10.1016/0370-2693(94)91516-4} {\bibfield  {journal} {\bibinfo  {journal}
  {Phys.~Lett.}\ }\textbf {\bibinfo {volume} {B 321}},\ \bibinfo {pages} {66}
  (\bibinfo {year} {1994})},\ \bibinfo {note} {[Erratum: Phys.
  Lett.B328,547(1994)]},\ \Eprint {http://arxiv.org/abs/hep-ph/9312252}
  {arXiv:hep-ph/9312252 [hep-ph]} \BibitemShut {NoStop}%
%%CITATION = HEP-PH/9312252;%%
\bibitem [{\citenamefont {Jungnickel}\ and\ \citenamefont
  {Wetterich}(1996)}]{Jungnickel:1995fp}%
  \BibitemOpen
  \bibfield  {author} {\bibinfo {author} {\bibfnamefont {D.~U.}\ \bibnamefont
  {Jungnickel}}\ and\ \bibinfo {author} {\bibfnamefont {C.}~\bibnamefont
  {Wetterich}},\ }\href {\doibase 10.1103/PhysRevD.53.5142} {\bibfield
  {journal} {\bibinfo  {journal} {Phys.~Rev.}\ }\textbf {\bibinfo {volume} {D
  53}},\ \bibinfo {pages} {5142} (\bibinfo {year} {1996})},\ \Eprint
  {http://arxiv.org/abs/hep-ph/9505267} {arXiv:hep-ph/9505267 [hep-ph]}
  \BibitemShut {NoStop}%
%%CITATION = HEP-PH/9505267;%%
\bibitem [{\citenamefont {Mocsy}\ \emph {et~al.}(2004)\citenamefont {Mocsy},
  \citenamefont {Sannino},\ and\ \citenamefont {Tuominen}}]{Mocsy:2003qw}%
  \BibitemOpen
  \bibfield  {author} {\bibinfo {author} {\bibfnamefont {A.}~\bibnamefont
  {Mocsy}}, \bibinfo {author} {\bibfnamefont {F.}~\bibnamefont {Sannino}}, \
  and\ \bibinfo {author} {\bibfnamefont {K.}~\bibnamefont {Tuominen}},\ }\href
  {\doibase 10.1103/PhysRevLett.92.182302} {\bibfield  {journal} {\bibinfo
  {journal} {Phys.~Rev.~Lett.}\ }\textbf {\bibinfo {volume} {92}},\ \bibinfo
  {pages} {182302} (\bibinfo {year} {2004})},\ \Eprint
  {http://arxiv.org/abs/hep-ph/0308135} {arXiv:hep-ph/0308135 [hep-ph]}
  \BibitemShut {NoStop}%
%%CITATION = HEP-PH/0308135;%%
\bibitem [{\citenamefont {Fukushima}(2004)}]{Fukushima:2003fw}%
  \BibitemOpen
  \bibfield  {author} {\bibinfo {author} {\bibfnamefont {K.}~\bibnamefont
  {Fukushima}},\ }\href {\doibase 10.1016/j.physletb.2004.04.027} {\bibfield
  {journal} {\bibinfo  {journal} {Phys.~Lett.}\ }\textbf {\bibinfo {volume} {B
  591}},\ \bibinfo {pages} {277} (\bibinfo {year} {2004})},\ \Eprint
  {http://arxiv.org/abs/hep-ph/0310121} {arXiv:hep-ph/0310121 [hep-ph]}
  \BibitemShut {NoStop}%
%%CITATION = HEP-PH/0310121;%%
\bibitem [{\citenamefont {Scavenius}\ \emph {et~al.}(2001)\citenamefont
  {Scavenius}, \citenamefont {Mocsy}, \citenamefont {Mishustin},\ and\
  \citenamefont {Rischke}}]{Scavenius:2000qd}%
  \BibitemOpen
  \bibfield  {author} {\bibinfo {author} {\bibfnamefont {O.}~\bibnamefont
  {Scavenius}}, \bibinfo {author} {\bibfnamefont {A.}~\bibnamefont {Mocsy}},
  \bibinfo {author} {\bibfnamefont {I.~N.}\ \bibnamefont {Mishustin}}, \ and\
  \bibinfo {author} {\bibfnamefont {D.~H.}\ \bibnamefont {Rischke}},\ }\href
  {\doibase 10.1103/PhysRevC.64.045202} {\bibfield  {journal} {\bibinfo
  {journal} {Phys.~Rev.}\ }\textbf {\bibinfo {volume} {C 64}},\ \bibinfo
  {pages} {045202} (\bibinfo {year} {2001})},\ \Eprint
  {http://arxiv.org/abs/nucl-th/0007030} {arXiv:nucl-th/0007030 [nucl-th]}
  \BibitemShut {NoStop}%
%%CITATION = NUCL-TH/0007030;%%
\bibitem [{\citenamefont {Buballa}(2005)}]{Buballa:2003qv}%
  \BibitemOpen
  \bibfield  {author} {\bibinfo {author} {\bibfnamefont {M.}~\bibnamefont
  {Buballa}},\ }\href {\doibase 10.1016/j.physrep.2004.11.004} {\bibfield
  {journal} {\bibinfo  {journal} {Phys.~Rept.}\ }\textbf {\bibinfo {volume}
  {407}},\ \bibinfo {pages} {205} (\bibinfo {year} {2005})},\ \Eprint
  {http://arxiv.org/abs/hep-ph/0402234} {arXiv:hep-ph/0402234 [hep-ph]}
  \BibitemShut {NoStop}%
%%CITATION = HEP-PH/0402234;%%
\bibitem [{\citenamefont {Costa}\ \emph {et~al.}(2008)\citenamefont {Costa},
  \citenamefont {Ruivo},\ and\ \citenamefont {de~Sousa}}]{Costa:2008yh}%
  \BibitemOpen
  \bibfield  {author} {\bibinfo {author} {\bibfnamefont {P.}~\bibnamefont
  {Costa}}, \bibinfo {author} {\bibfnamefont {M.~C.}\ \bibnamefont {Ruivo}}, \
  and\ \bibinfo {author} {\bibfnamefont {C.~A.}\ \bibnamefont {de~Sousa}},\
  }\href {\doibase 10.1103/PhysRevD.77.096001} {\bibfield  {journal} {\bibinfo
  {journal} {Phys.~Rev.}\ }\textbf {\bibinfo {volume} {D 77}},\ \bibinfo
  {pages} {096001} (\bibinfo {year} {2008})},\ \Eprint
  {http://arxiv.org/abs/0801.3417} {arXiv:0801.3417 [hep-ph]} \BibitemShut
  {NoStop}%
%%CITATION = ARXIV:0801.3417;%%
\bibitem [{\citenamefont {Polyakov}(1978)}]{Polyakov:1978vu}%
  \BibitemOpen
  \bibfield  {author} {\bibinfo {author} {\bibfnamefont {A.~M.}\ \bibnamefont
  {Polyakov}},\ }\href {\doibase 10.1016/0370-2693(78)90737-2} {\bibfield
  {journal} {\bibinfo  {journal} {Phys.~Lett.}\ }\textbf {\bibinfo {volume} {B
  72}},\ \bibinfo {pages} {477} (\bibinfo {year} {1978})}\BibitemShut {NoStop}%
%%CITATION = PHLTA,72B,477;%%
\bibitem [{\citenamefont {Hansen}\ \emph {et~al.}(2007)\citenamefont {Hansen},
  \citenamefont {Alberico}, \citenamefont {Beraudo}, \citenamefont {Molinari},
  \citenamefont {Nardi},\ and\ \citenamefont {Ratti}}]{Hansen:2006ee}%
  \BibitemOpen
  \bibfield  {author} {\bibinfo {author} {\bibfnamefont {H.}~\bibnamefont
  {Hansen}}, \bibinfo {author} {\bibfnamefont {W.~M.}\ \bibnamefont
  {Alberico}}, \bibinfo {author} {\bibfnamefont {A.}~\bibnamefont {Beraudo}},
  \bibinfo {author} {\bibfnamefont {A.}~\bibnamefont {Molinari}}, \bibinfo
  {author} {\bibfnamefont {M.}~\bibnamefont {Nardi}}, \ and\ \bibinfo {author}
  {\bibfnamefont {C.}~\bibnamefont {Ratti}},\ }\href {\doibase
  10.1103/PhysRevD.75.065004} {\bibfield  {journal} {\bibinfo  {journal} {Phys.
  Rev.}\ }\textbf {\bibinfo {volume} {D75}},\ \bibinfo {pages} {065004}
  (\bibinfo {year} {2007})},\ \Eprint {http://arxiv.org/abs/hep-ph/0609116}
  {arXiv:hep-ph/0609116 [hep-ph]} \BibitemShut {NoStop}%
%%CITATION = HEP-PH/0609116;%%
\bibitem [{\citenamefont {Meisinger}\ and\ \citenamefont
  {Ogilvie}(1996)}]{Meisinger:1995ih}%
  \BibitemOpen
  \bibfield  {author} {\bibinfo {author} {\bibfnamefont {P.~N.}\ \bibnamefont
  {Meisinger}}\ and\ \bibinfo {author} {\bibfnamefont {M.~C.}\ \bibnamefont
  {Ogilvie}},\ }\href {\doibase 10.1016/0370-2693(96)00447-9} {\bibfield
  {journal} {\bibinfo  {journal} {Phys.~Lett.}\ }\textbf {\bibinfo {volume} {B
  379}},\ \bibinfo {pages} {163} (\bibinfo {year} {1996})},\ \Eprint
  {http://arxiv.org/abs/hep-lat/9512011} {arXiv:hep-lat/9512011 [hep-lat]}
  \BibitemShut {NoStop}%
%%CITATION = HEP-LAT/9512011;%%
\bibitem [{\citenamefont {Pisarski}(2000)}]{Pisarski:2000eq}%
  \BibitemOpen
  \bibfield  {author} {\bibinfo {author} {\bibfnamefont {R.~D.}\ \bibnamefont
  {Pisarski}},\ }\href {\doibase 10.1103/PhysRevD.62.111501} {\bibfield
  {journal} {\bibinfo  {journal} {Phys. Rev.}\ }\textbf {\bibinfo {volume}
  {D62}},\ \bibinfo {pages} {111501} (\bibinfo {year} {2000})},\ \Eprint
  {http://arxiv.org/abs/hep-ph/0006205} {arXiv:hep-ph/0006205 [hep-ph]}
  \BibitemShut {NoStop}%
%%CITATION = HEP-PH/0006205;%%
\bibitem [{\citenamefont {Meisinger}\ \emph {et~al.}(2002)\citenamefont
  {Meisinger}, \citenamefont {Miller},\ and\ \citenamefont
  {Ogilvie}}]{Meisinger:2001cq}%
  \BibitemOpen
  \bibfield  {author} {\bibinfo {author} {\bibfnamefont {P.~N.}\ \bibnamefont
  {Meisinger}}, \bibinfo {author} {\bibfnamefont {T.~R.}\ \bibnamefont
  {Miller}}, \ and\ \bibinfo {author} {\bibfnamefont {M.~C.}\ \bibnamefont
  {Ogilvie}},\ }\href {\doibase 10.1103/PhysRevD.65.034009} {\bibfield
  {journal} {\bibinfo  {journal} {Phys.~Rev.}\ }\textbf {\bibinfo {volume} {D
  65}},\ \bibinfo {pages} {034009} (\bibinfo {year} {2002})},\ \Eprint
  {http://arxiv.org/abs/hep-ph/0108009} {arXiv:hep-ph/0108009 [hep-ph]}
  \BibitemShut {NoStop}%
%%CITATION = HEP-PH/0108009;%%
\bibitem [{\citenamefont {Ratti}\ \emph {et~al.}(2006)\citenamefont {Ratti},
  \citenamefont {Thaler},\ and\ \citenamefont {Weise}}]{Ratti:2005jh}%
  \BibitemOpen
  \bibfield  {author} {\bibinfo {author} {\bibfnamefont {C.}~\bibnamefont
  {Ratti}}, \bibinfo {author} {\bibfnamefont {M.~A.}\ \bibnamefont {Thaler}}, \
  and\ \bibinfo {author} {\bibfnamefont {W.}~\bibnamefont {Weise}},\ }\href
  {\doibase 10.1103/PhysRevD.73.014019} {\bibfield  {journal} {\bibinfo
  {journal} {Phys. Rev.}\ }\textbf {\bibinfo {volume} {D73}},\ \bibinfo {pages}
  {014019} (\bibinfo {year} {2006})},\ \Eprint
  {http://arxiv.org/abs/hep-ph/0506234} {arXiv:hep-ph/0506234 [hep-ph]}
  \BibitemShut {NoStop}%
%%CITATION = HEP-PH/0506234;%%
\bibitem [{\citenamefont {Kaczmarek}\ \emph {et~al.}(2002)\citenamefont
  {Kaczmarek}, \citenamefont {Karsch}, \citenamefont {Petreczky},\ and\
  \citenamefont {Zantow}}]{Kaczmarek:2002mc}%
  \BibitemOpen
  \bibfield  {author} {\bibinfo {author} {\bibfnamefont {O.}~\bibnamefont
  {Kaczmarek}}, \bibinfo {author} {\bibfnamefont {F.}~\bibnamefont {Karsch}},
  \bibinfo {author} {\bibfnamefont {P.}~\bibnamefont {Petreczky}}, \ and\
  \bibinfo {author} {\bibfnamefont {F.}~\bibnamefont {Zantow}},\ }\href
  {\doibase 10.1016/S0370-2693(02)02415-2} {\bibfield  {journal} {\bibinfo
  {journal} {Phys. Lett.}\ }\textbf {\bibinfo {volume} {B543}},\ \bibinfo
  {pages} {41} (\bibinfo {year} {2002})},\ \Eprint
  {http://arxiv.org/abs/hep-lat/0207002} {arXiv:hep-lat/0207002 [hep-lat]}
  \BibitemShut {NoStop}%
%%CITATION = HEP-LAT/0207002;%%
\bibitem [{\citenamefont {Boyd}\ \emph {et~al.}(1996)\citenamefont {Boyd},
  \citenamefont {Engels}, \citenamefont {Karsch}, \citenamefont {Laermann},
  \citenamefont {Legeland} \emph {et~al.}}]{Boyd:1996bx}%
  \BibitemOpen
  \bibfield  {author} {\bibinfo {author} {\bibfnamefont {G.}~\bibnamefont
  {Boyd}}, \bibinfo {author} {\bibfnamefont {J.}~\bibnamefont {Engels}},
  \bibinfo {author} {\bibfnamefont {F.}~\bibnamefont {Karsch}}, \bibinfo
  {author} {\bibfnamefont {E.}~\bibnamefont {Laermann}}, \bibinfo {author}
  {\bibfnamefont {C.}~\bibnamefont {Legeland}},  \emph {et~al.},\ }\href
  {\doibase 10.1016/0550-3213(96)00170-8} {\bibfield  {journal} {\bibinfo
  {journal} {Nucl.~Phys.}\ }\textbf {\bibinfo {volume} {B 469}},\ \bibinfo
  {pages} {419} (\bibinfo {year} {1996})},\ \Eprint
  {http://arxiv.org/abs/hep-lat/9602007} {arXiv:hep-lat/9602007 [hep-lat]}
  \BibitemShut {NoStop}%
%%CITATION = HEP-LAT/9602007;%%
\bibitem [{\citenamefont {Roessner}\ \emph {et~al.}(2007)\citenamefont
  {Roessner}, \citenamefont {Ratti},\ and\ \citenamefont
  {Weise}}]{Roessner:2006xn}%
  \BibitemOpen
  \bibfield  {author} {\bibinfo {author} {\bibfnamefont {S.}~\bibnamefont
  {Roessner}}, \bibinfo {author} {\bibfnamefont {C.}~\bibnamefont {Ratti}}, \
  and\ \bibinfo {author} {\bibfnamefont {W.}~\bibnamefont {Weise}},\ }\href
  {\doibase 10.1103/PhysRevD.75.034007} {\bibfield  {journal} {\bibinfo
  {journal} {Phys. Rev.}\ }\textbf {\bibinfo {volume} {D75}},\ \bibinfo {pages}
  {034007} (\bibinfo {year} {2007})},\ \Eprint
  {http://arxiv.org/abs/hep-ph/0609281} {arXiv:hep-ph/0609281 [hep-ph]}
  \BibitemShut {NoStop}%
%%CITATION = HEP-PH/0609281;%%
\bibitem [{\citenamefont {Haas}\ \emph {et~al.}(2013)\citenamefont {Haas},
  \citenamefont {Stiele}, \citenamefont {Braun}, \citenamefont {Pawlowski},\
  and\ \citenamefont {Schaffner-Bielich}}]{Haas:2013qwp}%
  \BibitemOpen
  \bibfield  {author} {\bibinfo {author} {\bibfnamefont {L.~M.}\ \bibnamefont
  {Haas}}, \bibinfo {author} {\bibfnamefont {R.}~\bibnamefont {Stiele}},
  \bibinfo {author} {\bibfnamefont {J.}~\bibnamefont {Braun}}, \bibinfo
  {author} {\bibfnamefont {J.~M.}\ \bibnamefont {Pawlowski}}, \ and\ \bibinfo
  {author} {\bibfnamefont {J.}~\bibnamefont {Schaffner-Bielich}},\ }\href
  {\doibase 10.1103/PhysRevD.87.076004} {\bibfield  {journal} {\bibinfo
  {journal} {Phys.~Rev.}\ }\textbf {\bibinfo {volume} {D 87}},\ \bibinfo
  {pages} {076004} (\bibinfo {year} {2013})},\ \Eprint
  {http://arxiv.org/abs/1302.1993} {arXiv:1302.1993 [hep-ph]} \BibitemShut
  {NoStop}%
\bibitem [{\citenamefont {Costa}\ \emph
  {et~al.}(2010{\natexlab{a}})\citenamefont {Costa}, \citenamefont {Hansen},
  \citenamefont {Ruivo},\ and\ \citenamefont {de~Sousa}}]{Costa:2009ae}%
  \BibitemOpen
  \bibfield  {author} {\bibinfo {author} {\bibfnamefont {P.}~\bibnamefont
  {Costa}}, \bibinfo {author} {\bibfnamefont {H.}~\bibnamefont {Hansen}},
  \bibinfo {author} {\bibfnamefont {M.~C.}\ \bibnamefont {Ruivo}}, \ and\
  \bibinfo {author} {\bibfnamefont {C.~A.}\ \bibnamefont {de~Sousa}},\ }\href
  {\doibase 10.1103/PhysRevD.81.016007} {\bibfield  {journal} {\bibinfo
  {journal} {Phys. Rev.}\ }\textbf {\bibinfo {volume} {D81}},\ \bibinfo {pages}
  {016007} (\bibinfo {year} {2010}{\natexlab{a}})},\ \Eprint
  {http://arxiv.org/abs/0909.5124} {arXiv:0909.5124 [hep-ph]} \BibitemShut
  {NoStop}%
%%CITATION = ARXIV:0909.5124;%%
\bibitem [{\citenamefont {Klevansky}(1992)}]{Klevansky:1992qe}%
  \BibitemOpen
  \bibfield  {author} {\bibinfo {author} {\bibfnamefont {S.~P.}\ \bibnamefont
  {Klevansky}},\ }\href {\doibase 10.1103/RevModPhys.64.649} {\bibfield
  {journal} {\bibinfo  {journal} {Rev. Mod. Phys.}\ }\textbf {\bibinfo {volume}
  {64}},\ \bibinfo {pages} {649} (\bibinfo {year} {1992})}\BibitemShut
  {NoStop}%
%%CITATION = RMPHA,64,649;%%
\bibitem [{\citenamefont {Blin}\ \emph {et~al.}(1988)\citenamefont {Blin},
  \citenamefont {Hiller},\ and\ \citenamefont {Schaden}}]{Blin:1987hw}%
  \BibitemOpen
  \bibfield  {author} {\bibinfo {author} {\bibfnamefont {A.~H.}\ \bibnamefont
  {Blin}}, \bibinfo {author} {\bibfnamefont {B.}~\bibnamefont {Hiller}}, \ and\
  \bibinfo {author} {\bibfnamefont {M.}~\bibnamefont {Schaden}},\ }\href@noop
  {} {\bibfield  {journal} {\bibinfo  {journal} {Z. Phys.}\ }\textbf {\bibinfo
  {volume} {A331}},\ \bibinfo {pages} {75} (\bibinfo {year}
  {1988})}\BibitemShut {NoStop}%
%%CITATION = ZEPYA,A331,75;%%
\bibitem [{\citenamefont {Fukushima}\ and\ \citenamefont
  {Hidaka}(2007)}]{Fukushima:2006uv}%
  \BibitemOpen
  \bibfield  {author} {\bibinfo {author} {\bibfnamefont {K.}~\bibnamefont
  {Fukushima}}\ and\ \bibinfo {author} {\bibfnamefont {Y.}~\bibnamefont
  {Hidaka}},\ }\href {\doibase 10.1103/PhysRevD.75.036002} {\bibfield
  {journal} {\bibinfo  {journal} {Phys.~Rev.}\ }\textbf {\bibinfo {volume} {D
  75}},\ \bibinfo {pages} {036002} (\bibinfo {year} {2007})},\ \Eprint
  {http://arxiv.org/abs/hep-ph/0610323} {arXiv:hep-ph/0610323 [hep-ph]}
  \BibitemShut {NoStop}%
%%CITATION = HEP-PH/0610323;%%
\bibitem [{\citenamefont {Roessner}\ \emph {et~al.}(2008)\citenamefont
  {Roessner}, \citenamefont {Hell}, \citenamefont {Ratti},\ and\ \citenamefont
  {Weise}}]{Rossner:2007ik}%
  \BibitemOpen
  \bibfield  {author} {\bibinfo {author} {\bibfnamefont {S.}~\bibnamefont
  {Roessner}}, \bibinfo {author} {\bibfnamefont {T.}~\bibnamefont {Hell}},
  \bibinfo {author} {\bibfnamefont {C.}~\bibnamefont {Ratti}}, \ and\ \bibinfo
  {author} {\bibfnamefont {W.}~\bibnamefont {Weise}},\ }\href {\doibase
  10.1016/j.nuclphysa.2008.10.006} {\bibfield  {journal} {\bibinfo  {journal}
  {Nucl.~Phys.}\ }\textbf {\bibinfo {volume} {A 814}},\ \bibinfo {pages} {118}
  (\bibinfo {year} {2008})},\ \Eprint {http://arxiv.org/abs/0712.3152}
  {arXiv:0712.3152 [hep-ph]} \BibitemShut {NoStop}%
%%CITATION = ARXIV:0712.3152;%%
\bibitem [{\citenamefont {Mintz}\ \emph {et~al.}(2013)\citenamefont {Mintz},
  \citenamefont {Stiele}, \citenamefont {Ramos},\ and\ \citenamefont
  {Schaffner-Bielich}}]{Mintz:2012mz}%
  \BibitemOpen
  \bibfield  {author} {\bibinfo {author} {\bibfnamefont {B.~W.}\ \bibnamefont
  {Mintz}}, \bibinfo {author} {\bibfnamefont {R.}~\bibnamefont {Stiele}},
  \bibinfo {author} {\bibfnamefont {R.~O.}\ \bibnamefont {Ramos}}, \ and\
  \bibinfo {author} {\bibfnamefont {J.}~\bibnamefont {Schaffner-Bielich}},\
  }\href {\doibase 10.1103/PhysRevD.87.036004} {\bibfield  {journal} {\bibinfo
  {journal} {Phys.~Rev.}\ }\textbf {\bibinfo {volume} {D 87}},\ \bibinfo
  {pages} {036004} (\bibinfo {year} {2013})},\ \Eprint
  {http://arxiv.org/abs/1212.1184} {arXiv:1212.1184 [hep-ph]} \BibitemShut
  {NoStop}%
%%CITATION = ARXIV:1212.1184;%%
\bibitem [{\citenamefont {Costa}\ \emph
  {et~al.}(2010{\natexlab{b}})\citenamefont {Costa}, \citenamefont {Ruivo},
  \citenamefont {de~Sousa},\ and\ \citenamefont {Hansen}}]{Costa:2010zw}%
  \BibitemOpen
  \bibfield  {author} {\bibinfo {author} {\bibfnamefont {P.}~\bibnamefont
  {Costa}}, \bibinfo {author} {\bibfnamefont {M.~C.}\ \bibnamefont {Ruivo}},
  \bibinfo {author} {\bibfnamefont {C.~A.}\ \bibnamefont {de~Sousa}}, \ and\
  \bibinfo {author} {\bibfnamefont {H.}~\bibnamefont {Hansen}},\ }\href
  {\doibase 10.3390/sym2031338} {\bibfield  {journal} {\bibinfo  {journal}
  {Symmetry}\ }\textbf {\bibinfo {volume} {2}},\ \bibinfo {pages} {1338}
  (\bibinfo {year} {2010}{\natexlab{b}})},\ \Eprint
  {http://arxiv.org/abs/1007.1380} {arXiv:1007.1380 [hep-ph]} \BibitemShut
  {NoStop}%
%%CITATION = ARXIV:1007.1380;%%
\bibitem [{\citenamefont {Song}\ \emph {et~al.}(2019)\citenamefont {Song},
  \citenamefont {Baym}, \citenamefont {Hatsuda},\ and\ \citenamefont
  {Kojo}}]{Song:2019qoh}%
  \BibitemOpen
  \bibfield  {author} {\bibinfo {author} {\bibfnamefont {Y.}~\bibnamefont
  {Song}}, \bibinfo {author} {\bibfnamefont {G.}~\bibnamefont {Baym}}, \bibinfo
  {author} {\bibfnamefont {T.}~\bibnamefont {Hatsuda}}, \ and\ \bibinfo
  {author} {\bibfnamefont {T.}~\bibnamefont {Kojo}},\ }\href {\doibase
  10.1103/PhysRevD.100.034018} {\bibfield  {journal} {\bibinfo  {journal}
  {Phys.~Rev.}\ }\textbf {\bibinfo {volume} {D 100}},\ \bibinfo {pages}
  {034018} (\bibinfo {year} {2019})},\ \Eprint
  {http://arxiv.org/abs/1905.01005} {arXiv:1905.01005 [astro-ph.HE]}
  \BibitemShut {NoStop}%
%%CITATION = ARXIV:1905.01005;%%
\bibitem [{\citenamefont {Hufner}\ \emph {et~al.}(1994)\citenamefont {Hufner},
  \citenamefont {Klevansky}, \citenamefont {Zhuang},\ and\ \citenamefont
  {Voss}}]{Hufner:1994ma}%
  \BibitemOpen
  \bibfield  {author} {\bibinfo {author} {\bibfnamefont {J.}~\bibnamefont
  {Hufner}}, \bibinfo {author} {\bibfnamefont {S.~P.}\ \bibnamefont
  {Klevansky}}, \bibinfo {author} {\bibfnamefont {P.}~\bibnamefont {Zhuang}}, \
  and\ \bibinfo {author} {\bibfnamefont {H.}~\bibnamefont {Voss}},\ }\href
  {\doibase 10.1006/aphy.1994.1080} {\bibfield  {journal} {\bibinfo  {journal}
  {Annals Phys.}\ }\textbf {\bibinfo {volume} {234}},\ \bibinfo {pages} {225}
  (\bibinfo {year} {1994})}\BibitemShut {NoStop}%
%%CITATION = APNYA,234,225;%%
\bibitem [{\citenamefont {Blaschke}\ \emph {et~al.}(2015)\citenamefont
  {Blaschke}, \citenamefont {Dubinin},\ and\ \citenamefont
  {Buballa}}]{Blaschke:2014zsa}%
  \BibitemOpen
  \bibfield  {author} {\bibinfo {author} {\bibfnamefont {D.}~\bibnamefont
  {Blaschke}}, \bibinfo {author} {\bibfnamefont {A.}~\bibnamefont {Dubinin}}, \
  and\ \bibinfo {author} {\bibfnamefont {M.}~\bibnamefont {Buballa}},\ }\href
  {\doibase 10.1103/PhysRevD.91.125040} {\bibfield  {journal} {\bibinfo
  {journal} {Phys. Rev.}\ }\textbf {\bibinfo {volume} {D91}},\ \bibinfo {pages}
  {125040} (\bibinfo {year} {2015})},\ \Eprint {http://arxiv.org/abs/1412.1040}
  {arXiv:1412.1040 [hep-ph]} \BibitemShut {NoStop}%
%%CITATION = ARXIV:1412.1040;%%
\bibitem [{\citenamefont {Stiele}(2014)}]{Stiele:2014gks}%
  \BibitemOpen
  \bibfield  {author} {\bibinfo {author} {\bibfnamefont {R.}~\bibnamefont
  {Stiele}},\ }\href {\doibase 10.11588/heidok.00016887} {Ph.D. thesis},\
  \bibinfo  {school} {Ruprecht-Karls-Universit\"at Heidelberg} (\bibinfo {year}
  {2014})\BibitemShut {NoStop}%
%%CITATION = INSPIRE-1764818;%%
\bibitem [{\citenamefont {Herbst}\ \emph {et~al.}(2014)\citenamefont {Herbst},
  \citenamefont {Mitter}, \citenamefont {Pawlowski}, \citenamefont {Schaefer},\
  and\ \citenamefont {Stiele}}]{Herbst:2013ufa}%
  \BibitemOpen
  \bibfield  {author} {\bibinfo {author} {\bibfnamefont {T.~K.}\ \bibnamefont
  {Herbst}}, \bibinfo {author} {\bibfnamefont {M.}~\bibnamefont {Mitter}},
  \bibinfo {author} {\bibfnamefont {J.~M.}\ \bibnamefont {Pawlowski}}, \bibinfo
  {author} {\bibfnamefont {B.-J.}\ \bibnamefont {Schaefer}}, \ and\ \bibinfo
  {author} {\bibfnamefont {R.}~\bibnamefont {Stiele}},\ }\href {\doibase
  10.1016/j.physletb.2014.02.045} {\bibfield  {journal} {\bibinfo  {journal}
  {Phys. Lett.}\ }\textbf {\bibinfo {volume} {B731}},\ \bibinfo {pages} {248}
  (\bibinfo {year} {2014})},\ \Eprint {http://arxiv.org/abs/1308.3621}
  {arXiv:1308.3621 [hep-ph]} \BibitemShut {NoStop}%
%%CITATION = ARXIV:1308.3621;%%
\bibitem [{\citenamefont {Torres-Rincon}\ and\ \citenamefont
  {Aichelin}(2017)}]{Torres-Rincon:2017zbr}%
  \BibitemOpen
  \bibfield  {author} {\bibinfo {author} {\bibfnamefont {J.~M.}\ \bibnamefont
  {Torres-Rincon}}\ and\ \bibinfo {author} {\bibfnamefont {J.}~\bibnamefont
  {Aichelin}},\ }\href {\doibase 10.1103/PhysRevC.96.045205} {\bibfield
  {journal} {\bibinfo  {journal} {Phys.~Rev.}\ }\textbf {\bibinfo {volume} {C
  96}},\ \bibinfo {pages} {045205} (\bibinfo {year} {2017})},\ \Eprint
  {http://arxiv.org/abs/1704.07858} {arXiv:1704.07858 [nucl-th]} \BibitemShut
  {NoStop}%
%%CITATION = ARXIV:1704.07858;%%
\bibitem [{\citenamefont {Fukushima}\ and\ \citenamefont
  {Sasaki}(2013)}]{Fukushima:2013rx}%
  \BibitemOpen
  \bibfield  {author} {\bibinfo {author} {\bibfnamefont {K.}~\bibnamefont
  {Fukushima}}\ and\ \bibinfo {author} {\bibfnamefont {C.}~\bibnamefont
  {Sasaki}},\ }\href {\doibase 10.1016/j.ppnp.2013.05.003} {\bibfield
  {journal} {\bibinfo  {journal} {Prog. Part. Nucl. Phys.}\ }\textbf {\bibinfo
  {volume} {72}},\ \bibinfo {pages} {99} (\bibinfo {year} {2013})},\ \Eprint
  {http://arxiv.org/abs/1301.6377} {arXiv:1301.6377 [hep-ph]} \BibitemShut
  {NoStop}%
%%CITATION = ARXIV:1301.6377;%%
\bibitem [{\citenamefont {Fukushima}\ and\ \citenamefont
  {Skokov}(2017)}]{Fukushima:2017csk}%
  \BibitemOpen
  \bibfield  {author} {\bibinfo {author} {\bibfnamefont {K.}~\bibnamefont
  {Fukushima}}\ and\ \bibinfo {author} {\bibfnamefont {V.}~\bibnamefont
  {Skokov}},\ }\href {\doibase 10.1016/j.ppnp.2017.05.002} {\bibfield
  {journal} {\bibinfo  {journal} {Prog. Part. Nucl. Phys.}\ }\textbf {\bibinfo
  {volume} {96}},\ \bibinfo {pages} {154} (\bibinfo {year} {2017})},\ \Eprint
  {http://arxiv.org/abs/1705.00718} {arXiv:1705.00718 [hep-ph]} \BibitemShut
  {NoStop}%
%%CITATION = ARXIV:1705.00718;%%
\bibitem [{\citenamefont {Baym}\ \emph {et~al.}(2018)\citenamefont {Baym},
  \citenamefont {Hatsuda}, \citenamefont {Kojo}, \citenamefont {Powell},
  \citenamefont {Song},\ and\ \citenamefont {Takatsuka}}]{Baym:2017whm}%
  \BibitemOpen
  \bibfield  {author} {\bibinfo {author} {\bibfnamefont {G.}~\bibnamefont
  {Baym}}, \bibinfo {author} {\bibfnamefont {T.}~\bibnamefont {Hatsuda}},
  \bibinfo {author} {\bibfnamefont {T.}~\bibnamefont {Kojo}}, \bibinfo {author}
  {\bibfnamefont {P.~D.}\ \bibnamefont {Powell}}, \bibinfo {author}
  {\bibfnamefont {Y.}~\bibnamefont {Song}}, \ and\ \bibinfo {author}
  {\bibfnamefont {T.}~\bibnamefont {Takatsuka}},\ }\href {\doibase
  10.1088/1361-6633/aaae14} {\bibfield  {journal} {\bibinfo  {journal}
  {Rept.~Prog.~Phys.}\ }\textbf {\bibinfo {volume} {81}},\ \bibinfo {pages}
  {056902} (\bibinfo {year} {2018})},\ \Eprint
  {http://arxiv.org/abs/1707.04966} {arXiv:1707.04966 [astro-ph.HE]}
  \BibitemShut {NoStop}%
%%CITATION = ARXIV:1707.04966;%%
\bibitem [{\citenamefont {Borsanyi}\ \emph
  {et~al.}(2010{\natexlab{a}})\citenamefont {Borsanyi}, \citenamefont {Fodor},
  \citenamefont {Hoelbling}, \citenamefont {Katz}, \citenamefont {Krieg},
  \citenamefont {Ratti},\ and\ \citenamefont {Szabo}}]{Borsanyi:2010bp}%
  \BibitemOpen
  \bibfield  {author} {\bibinfo {author} {\bibfnamefont {S.}~\bibnamefont
  {Borsanyi}}, \bibinfo {author} {\bibfnamefont {Z.}~\bibnamefont {Fodor}},
  \bibinfo {author} {\bibfnamefont {C.}~\bibnamefont {Hoelbling}}, \bibinfo
  {author} {\bibfnamefont {S.~D.}\ \bibnamefont {Katz}}, \bibinfo {author}
  {\bibfnamefont {S.}~\bibnamefont {Krieg}}, \bibinfo {author} {\bibfnamefont
  {C.}~\bibnamefont {Ratti}}, \ and\ \bibinfo {author} {\bibfnamefont {K.~K.}\
  \bibnamefont {Szabo}} (\bibinfo {collaboration} {Wuppertal-Budapest}),\
  }\href {\doibase 10.1007/JHEP09(2010)073} {\bibfield  {journal} {\bibinfo
  {journal} {JHEP}\ }\textbf {\bibinfo {volume} {09}},\ \bibinfo {pages} {073}
  (\bibinfo {year} {2010}{\natexlab{a}})},\ \Eprint
  {http://arxiv.org/abs/1005.3508} {arXiv:1005.3508 [hep-lat]} \BibitemShut
  {NoStop}%
%%CITATION = ARXIV:1005.3508;%%
\bibitem [{\citenamefont {Borsanyi}\ \emph
  {et~al.}(2010{\natexlab{b}})\citenamefont {Borsanyi}, \citenamefont
  {Endrodi}, \citenamefont {Fodor}, \citenamefont {Jakovac}, \citenamefont
  {Katz} \emph {et~al.}}]{Borsanyi:2010cj}%
  \BibitemOpen
  \bibfield  {author} {\bibinfo {author} {\bibfnamefont {S.}~\bibnamefont
  {Borsanyi}}, \bibinfo {author} {\bibfnamefont {G.}~\bibnamefont {Endrodi}},
  \bibinfo {author} {\bibfnamefont {Z.}~\bibnamefont {Fodor}}, \bibinfo
  {author} {\bibfnamefont {A.}~\bibnamefont {Jakovac}}, \bibinfo {author}
  {\bibfnamefont {S.~D.}\ \bibnamefont {Katz}},  \emph {et~al.},\ }\href
  {\doibase 10.1007/JHEP11(2010)077} {\bibfield  {journal} {\bibinfo  {journal}
  {JHEP}\ }\textbf {\bibinfo {volume} {1011}},\ \bibinfo {pages} {077}
  (\bibinfo {year} {2010}{\natexlab{b}})},\ \Eprint
  {http://arxiv.org/abs/1007.2580} {arXiv:1007.2580 [hep-lat]} \BibitemShut
  {NoStop}%
%%CITATION = ARXIV:1007.2580;%%
\bibitem [{\citenamefont {Bazavov}(2013)}]{Bazavov:2012bp}%
  \BibitemOpen
  \bibfield  {author} {\bibinfo {author} {\bibfnamefont {A.}~\bibnamefont
  {Bazavov}} (\bibinfo {collaboration} {HotQCD}),\ }\href {\doibase
  10.1016/j.nuclphysa.2013.02.155} {\bibfield  {journal} {\bibinfo  {journal}
  {Nucl. Phys.}\ }\textbf {\bibinfo {volume} {A904-905}},\ \bibinfo {pages}
  {877c} (\bibinfo {year} {2013})},\ \Eprint {http://arxiv.org/abs/1210.6312}
  {arXiv:1210.6312 [hep-lat]} \BibitemShut {NoStop}%
%%CITATION = ARXIV:1210.6312;%%
\bibitem [{\citenamefont {Bazavov}\ and\ \citenamefont
  {Petreczky}(2013)}]{Bazavov:2013yv}%
  \BibitemOpen
  \bibfield  {author} {\bibinfo {author} {\bibfnamefont {A.}~\bibnamefont
  {Bazavov}}\ and\ \bibinfo {author} {\bibfnamefont {P.}~\bibnamefont
  {Petreczky}},\ }\href {\doibase 10.1103/PhysRevD.87.094505} {\bibfield
  {journal} {\bibinfo  {journal} {Phys. Rev.}\ }\textbf {\bibinfo {volume}
  {D87}},\ \bibinfo {pages} {094505} (\bibinfo {year} {2013})},\ \Eprint
  {http://arxiv.org/abs/1301.3943} {arXiv:1301.3943 [hep-lat]} \BibitemShut
  {NoStop}%
%%CITATION = ARXIV:1301.3943;%%
\bibitem [{\citenamefont {Braun}\ \emph {et~al.}(2010)\citenamefont {Braun},
  \citenamefont {Gies},\ and\ \citenamefont {Pawlowski}}]{Braun:2007bx}%
  \BibitemOpen
  \bibfield  {author} {\bibinfo {author} {\bibfnamefont {J.}~\bibnamefont
  {Braun}}, \bibinfo {author} {\bibfnamefont {H.}~\bibnamefont {Gies}}, \ and\
  \bibinfo {author} {\bibfnamefont {J.~M.}\ \bibnamefont {Pawlowski}},\ }\href
  {\doibase 10.1016/j.physletb.2010.01.009} {\bibfield  {journal} {\bibinfo
  {journal} {Phys. Lett.}\ }\textbf {\bibinfo {volume} {B684}},\ \bibinfo
  {pages} {262} (\bibinfo {year} {2010})},\ \Eprint
  {http://arxiv.org/abs/0708.2413} {arXiv:0708.2413 [hep-th]} \BibitemShut
  {NoStop}%
%%CITATION = ARXIV:0708.2413;%%
\bibitem [{\citenamefont {Herbst}\ \emph {et~al.}(2015)\citenamefont {Herbst},
  \citenamefont {Luecker},\ and\ \citenamefont {Pawlowski}}]{Herbst:2015ona}%
  \BibitemOpen
  \bibfield  {author} {\bibinfo {author} {\bibfnamefont {T.~K.}\ \bibnamefont
  {Herbst}}, \bibinfo {author} {\bibfnamefont {J.}~\bibnamefont {Luecker}}, \
  and\ \bibinfo {author} {\bibfnamefont {J.~M.}\ \bibnamefont {Pawlowski}},\
  }\href@noop {} {\  (\bibinfo {year} {2015})},\ \Eprint
  {http://arxiv.org/abs/1510.03830} {arXiv:1510.03830 [hep-ph]} \BibitemShut
  {NoStop}%
%%CITATION = ARXIV:1510.03830;%%
\bibitem [{\citenamefont {Gupta}\ \emph {et~al.}(2008)\citenamefont {Gupta},
  \citenamefont {Huebner},\ and\ \citenamefont {Kaczmarek}}]{Gupta:2007ax}%
  \BibitemOpen
  \bibfield  {author} {\bibinfo {author} {\bibfnamefont {S.}~\bibnamefont
  {Gupta}}, \bibinfo {author} {\bibfnamefont {K.}~\bibnamefont {Huebner}}, \
  and\ \bibinfo {author} {\bibfnamefont {O.}~\bibnamefont {Kaczmarek}},\ }\href
  {\doibase 10.1103/PhysRevD.77.034503} {\bibfield  {journal} {\bibinfo
  {journal} {Phys. Rev.}\ }\textbf {\bibinfo {volume} {D77}},\ \bibinfo {pages}
  {034503} (\bibinfo {year} {2008})},\ \Eprint {http://arxiv.org/abs/0711.2251}
  {arXiv:0711.2251 [hep-lat]} \BibitemShut {NoStop}%
%%CITATION = ARXIV:0711.2251;%%
\bibitem [{\citenamefont {Lo}\ \emph {et~al.}(2013)\citenamefont {Lo},
  \citenamefont {Friman}, \citenamefont {Kaczmarek}, \citenamefont {Redlich},\
  and\ \citenamefont {Sasaki}}]{Lo:2013hla}%
  \BibitemOpen
  \bibfield  {author} {\bibinfo {author} {\bibfnamefont {P.~M.}\ \bibnamefont
  {Lo}}, \bibinfo {author} {\bibfnamefont {B.}~\bibnamefont {Friman}}, \bibinfo
  {author} {\bibfnamefont {O.}~\bibnamefont {Kaczmarek}}, \bibinfo {author}
  {\bibfnamefont {K.}~\bibnamefont {Redlich}}, \ and\ \bibinfo {author}
  {\bibfnamefont {C.}~\bibnamefont {Sasaki}},\ }\href {\doibase
  10.1103/PhysRevD.88.074502} {\bibfield  {journal} {\bibinfo  {journal}
  {Phys.~Rev.}\ }\textbf {\bibinfo {volume} {D 88}},\ \bibinfo {pages} {074502}
  (\bibinfo {year} {2013})},\ \Eprint {http://arxiv.org/abs/1307.5958}
  {arXiv:1307.5958 [hep-lat]} \BibitemShut {NoStop}%
%%CITATION = ARXIV:1307.5958;%%
\bibitem [{\citenamefont {Quandt}\ and\ \citenamefont
  {Reinhardt}(2016)}]{Quandt:2016ykm}%
  \BibitemOpen
  \bibfield  {author} {\bibinfo {author} {\bibfnamefont {M.}~\bibnamefont
  {Quandt}}\ and\ \bibinfo {author} {\bibfnamefont {H.}~\bibnamefont
  {Reinhardt}},\ }\href {\doibase 10.1103/PhysRevD.94.065015} {\bibfield
  {journal} {\bibinfo  {journal} {Phys. Rev.}\ }\textbf {\bibinfo {volume}
  {D94}},\ \bibinfo {pages} {065015} (\bibinfo {year} {2016})},\ \Eprint
  {http://arxiv.org/abs/1603.08058} {arXiv:1603.08058 [hep-th]} \BibitemShut
  {NoStop}%
%%CITATION = ARXIV:1603.08058;%%
\bibitem [{\citenamefont {Reinosa}\ \emph {et~al.}(2015)\citenamefont
  {Reinosa}, \citenamefont {Serreau}, \citenamefont {Tissier},\ and\
  \citenamefont {Wschebor}}]{Reinosa:2014ooa}%
  \BibitemOpen
  \bibfield  {author} {\bibinfo {author} {\bibfnamefont {U.}~\bibnamefont
  {Reinosa}}, \bibinfo {author} {\bibfnamefont {J.}~\bibnamefont {Serreau}},
  \bibinfo {author} {\bibfnamefont {M.}~\bibnamefont {Tissier}}, \ and\
  \bibinfo {author} {\bibfnamefont {N.}~\bibnamefont {Wschebor}},\ }\href
  {\doibase 10.1016/j.physletb.2015.01.006} {\bibfield  {journal} {\bibinfo
  {journal} {Phys. Lett.}\ }\textbf {\bibinfo {volume} {B742}},\ \bibinfo
  {pages} {61} (\bibinfo {year} {2015})},\ \Eprint
  {http://arxiv.org/abs/1407.6469} {arXiv:1407.6469 [hep-ph]} \BibitemShut
  {NoStop}%
%%CITATION = ARXIV:1407.6469;%%
\bibitem [{\citenamefont {Reinosa}\ \emph {et~al.}(2016)\citenamefont
  {Reinosa}, \citenamefont {Serreau}, \citenamefont {Tissier},\ and\
  \citenamefont {Wschebor}}]{Reinosa:2015gxn}%
  \BibitemOpen
  \bibfield  {author} {\bibinfo {author} {\bibfnamefont {U.}~\bibnamefont
  {Reinosa}}, \bibinfo {author} {\bibfnamefont {J.}~\bibnamefont {Serreau}},
  \bibinfo {author} {\bibfnamefont {M.}~\bibnamefont {Tissier}}, \ and\
  \bibinfo {author} {\bibfnamefont {N.}~\bibnamefont {Wschebor}},\ }\href
  {\doibase 10.1103/PhysRevD.93.105002} {\bibfield  {journal} {\bibinfo
  {journal} {Phys. Rev.}\ }\textbf {\bibinfo {volume} {D93}},\ \bibinfo {pages}
  {105002} (\bibinfo {year} {2016})},\ \Eprint
  {http://arxiv.org/abs/1511.07690} {arXiv:1511.07690 [hep-th]} \BibitemShut
  {NoStop}%
%%CITATION = ARXIV:1511.07690;%%
\bibitem [{\citenamefont {Alba}\ \emph {et~al.}(2014)\citenamefont {Alba},
  \citenamefont {Alberico}, \citenamefont {Bluhm}, \citenamefont {Greco},
  \citenamefont {Ratti},\ and\ \citenamefont {Ruggieri}}]{Alba:2014lda}%
  \BibitemOpen
  \bibfield  {author} {\bibinfo {author} {\bibfnamefont {P.}~\bibnamefont
  {Alba}}, \bibinfo {author} {\bibfnamefont {W.}~\bibnamefont {Alberico}},
  \bibinfo {author} {\bibfnamefont {M.}~\bibnamefont {Bluhm}}, \bibinfo
  {author} {\bibfnamefont {V.}~\bibnamefont {Greco}}, \bibinfo {author}
  {\bibfnamefont {C.}~\bibnamefont {Ratti}}, \ and\ \bibinfo {author}
  {\bibfnamefont {M.}~\bibnamefont {Ruggieri}},\ }\href {\doibase
  10.1016/j.nuclphysa.2014.11.011} {\bibfield  {journal} {\bibinfo  {journal}
  {Nucl. Phys.}\ }\textbf {\bibinfo {volume} {A934}},\ \bibinfo {pages} {41}
  (\bibinfo {year} {2014})},\ \Eprint {http://arxiv.org/abs/1402.6213}
  {arXiv:1402.6213 [hep-ph]} \BibitemShut {NoStop}%
%%CITATION = ARXIV:1402.6213;%%
\bibitem [{\citenamefont {Sakai}\ \emph {et~al.}(2010)\citenamefont {Sakai},
  \citenamefont {Sasaki}, \citenamefont {Kouno},\ and\ \citenamefont
  {Yahiro}}]{Sakai:2010rp}%
  \BibitemOpen
  \bibfield  {author} {\bibinfo {author} {\bibfnamefont {Y.}~\bibnamefont
  {Sakai}}, \bibinfo {author} {\bibfnamefont {T.}~\bibnamefont {Sasaki}},
  \bibinfo {author} {\bibfnamefont {H.}~\bibnamefont {Kouno}}, \ and\ \bibinfo
  {author} {\bibfnamefont {M.}~\bibnamefont {Yahiro}},\ }\href {\doibase
  10.1103/PhysRevD.82.076003} {\bibfield  {journal} {\bibinfo  {journal} {Phys.
  Rev.}\ }\textbf {\bibinfo {volume} {D82}},\ \bibinfo {pages} {076003}
  (\bibinfo {year} {2010})},\ \Eprint {http://arxiv.org/abs/1006.3648}
  {arXiv:1006.3648 [hep-ph]} \BibitemShut {NoStop}%
%%CITATION = ARXIV:1006.3648;%%
\bibitem [{\citenamefont {Roberge}\ and\ \citenamefont
  {Weiss}(1986)}]{Roberge:1986mm}%
  \BibitemOpen
  \bibfield  {author} {\bibinfo {author} {\bibfnamefont {A.}~\bibnamefont
  {Roberge}}\ and\ \bibinfo {author} {\bibfnamefont {N.}~\bibnamefont
  {Weiss}},\ }\href {\doibase 10.1016/0550-3213(86)90582-1} {\bibfield
  {journal} {\bibinfo  {journal} {Nucl. Phys.}\ }\textbf {\bibinfo {volume}
  {B275}},\ \bibinfo {pages} {734} (\bibinfo {year} {1986})}\BibitemShut
  {NoStop}%
%%CITATION = NUPHA,B275,734;%%
\bibitem [{\citenamefont {Schaefer}\ \emph {et~al.}(2007)\citenamefont
  {Schaefer}, \citenamefont {Pawlowski},\ and\ \citenamefont
  {Wambach}}]{Schaefer:2007pw}%
  \BibitemOpen
  \bibfield  {author} {\bibinfo {author} {\bibfnamefont {B.-J.}\ \bibnamefont
  {Schaefer}}, \bibinfo {author} {\bibfnamefont {J.~M.}\ \bibnamefont
  {Pawlowski}}, \ and\ \bibinfo {author} {\bibfnamefont {J.}~\bibnamefont
  {Wambach}},\ }\href {\doibase 10.1103/PhysRevD.76.074023} {\bibfield
  {journal} {\bibinfo  {journal} {Phys.~Rev.}\ }\textbf {\bibinfo {volume} {D
  76}},\ \bibinfo {pages} {074023} (\bibinfo {year} {2007})},\ \Eprint
  {http://arxiv.org/abs/0704.3234} {arXiv:0704.3234 [hep-ph]} \BibitemShut
  {NoStop}%
%%CITATION = ARXIV:0704.3234;%%
\bibitem [{\citenamefont {Dexheimer}\ and\ \citenamefont
  {Schramm}(2010)}]{Dexheimer:2009hi}%
  \BibitemOpen
  \bibfield  {author} {\bibinfo {author} {\bibfnamefont {V.~A.}\ \bibnamefont
  {Dexheimer}}\ and\ \bibinfo {author} {\bibfnamefont {S.}~\bibnamefont
  {Schramm}},\ }\href {\doibase 10.1103/PhysRevC.81.045201} {\bibfield
  {journal} {\bibinfo  {journal} {Phys. Rev.}\ }\textbf {\bibinfo {volume}
  {C81}},\ \bibinfo {pages} {045201} (\bibinfo {year} {2010})},\ \Eprint
  {http://arxiv.org/abs/0901.1748} {arXiv:0901.1748 [astro-ph.SR]} \BibitemShut
  {NoStop}%
%%CITATION = ARXIV:0901.1748;%%
\bibitem [{\citenamefont {de~Forcrand}\ and\ \citenamefont
  {Philipsen}(2002)}]{deForcrand:2002hgr}%
  \BibitemOpen
  \bibfield  {author} {\bibinfo {author} {\bibfnamefont {P.}~\bibnamefont
  {de~Forcrand}}\ and\ \bibinfo {author} {\bibfnamefont {O.}~\bibnamefont
  {Philipsen}},\ }\href {\doibase 10.1016/S0550-3213(02)00626-0} {\bibfield
  {journal} {\bibinfo  {journal} {Nucl. Phys.}\ }\textbf {\bibinfo {volume}
  {B642}},\ \bibinfo {pages} {290} (\bibinfo {year} {2002})},\ \Eprint
  {http://arxiv.org/abs/hep-lat/0205016} {arXiv:hep-lat/0205016 [hep-lat]}
  \BibitemShut {NoStop}%
%%CITATION = HEP-LAT/0205016;%%
\bibitem [{\citenamefont {Wu}\ \emph {et~al.}(2007)\citenamefont {Wu},
  \citenamefont {Luo},\ and\ \citenamefont {Chen}}]{Wu:2006su}%
  \BibitemOpen
  \bibfield  {author} {\bibinfo {author} {\bibfnamefont {L.-K.}\ \bibnamefont
  {Wu}}, \bibinfo {author} {\bibfnamefont {X.-Q.}\ \bibnamefont {Luo}}, \ and\
  \bibinfo {author} {\bibfnamefont {H.-S.}\ \bibnamefont {Chen}},\ }\href
  {\doibase 10.1103/PhysRevD.76.034505} {\bibfield  {journal} {\bibinfo
  {journal} {Phys.~Rev.}\ }\textbf {\bibinfo {volume} {D 76}},\ \bibinfo
  {pages} {034505} (\bibinfo {year} {2007})},\ \Eprint
  {http://arxiv.org/abs/hep-lat/0611035} {arXiv:hep-lat/0611035 [hep-lat]}
  \BibitemShut {NoStop}%
%%CITATION = HEP-LAT/0611035;%%
\bibitem [{\citenamefont {Nagata}\ and\ \citenamefont
  {Nakamura}(2011)}]{Nagata:2011yf}%
  \BibitemOpen
  \bibfield  {author} {\bibinfo {author} {\bibfnamefont {K.}~\bibnamefont
  {Nagata}}\ and\ \bibinfo {author} {\bibfnamefont {A.}~\bibnamefont
  {Nakamura}},\ }\href {\doibase 10.1103/PhysRevD.83.114507} {\bibfield
  {journal} {\bibinfo  {journal} {Phys.~Rev.}\ }\textbf {\bibinfo {volume} {D
  83}},\ \bibinfo {pages} {114507} (\bibinfo {year} {2011})},\ \Eprint
  {http://arxiv.org/abs/1104.2142} {arXiv:1104.2142 [hep-lat]} \BibitemShut
  {NoStop}%
%%CITATION = ARXIV:1104.2142;%%
\bibitem [{\citenamefont {Cea}\ \emph {et~al.}(2012)\citenamefont {Cea},
  \citenamefont {Cosmai}, \citenamefont {D'Elia}, \citenamefont {Papa},\ and\
  \citenamefont {Sanfilippo}}]{Cea:2012ev}%
  \BibitemOpen
  \bibfield  {author} {\bibinfo {author} {\bibfnamefont {P.}~\bibnamefont
  {Cea}}, \bibinfo {author} {\bibfnamefont {L.}~\bibnamefont {Cosmai}},
  \bibinfo {author} {\bibfnamefont {M.}~\bibnamefont {D'Elia}}, \bibinfo
  {author} {\bibfnamefont {A.}~\bibnamefont {Papa}}, \ and\ \bibinfo {author}
  {\bibfnamefont {F.}~\bibnamefont {Sanfilippo}},\ }\href {\doibase
  10.1103/PhysRevD.85.094512} {\bibfield  {journal} {\bibinfo  {journal}
  {Phys.~Rev.}\ }\textbf {\bibinfo {volume} {D 85}},\ \bibinfo {pages} {094512}
  (\bibinfo {year} {2012})},\ \Eprint {http://arxiv.org/abs/1202.5700}
  {arXiv:1202.5700 [hep-lat]} \BibitemShut {NoStop}%
%%CITATION = ARXIV:1202.5700;%%
\bibitem [{\citenamefont {Bonati}\ \emph {et~al.}(2019)\citenamefont {Bonati},
  \citenamefont {Calore}, \citenamefont {D'Elia}, \citenamefont {Mesiti},
  \citenamefont {Negro}, \citenamefont {Sanfilippo}, \citenamefont {Schifano},
  \citenamefont {Silvi},\ and\ \citenamefont {Tripiccione}}]{Bonati:2018fvg}%
  \BibitemOpen
  \bibfield  {author} {\bibinfo {author} {\bibfnamefont {C.}~\bibnamefont
  {Bonati}}, \bibinfo {author} {\bibfnamefont {E.}~\bibnamefont {Calore}},
  \bibinfo {author} {\bibfnamefont {M.}~\bibnamefont {D'Elia}}, \bibinfo
  {author} {\bibfnamefont {M.}~\bibnamefont {Mesiti}}, \bibinfo {author}
  {\bibfnamefont {F.}~\bibnamefont {Negro}}, \bibinfo {author} {\bibfnamefont
  {F.}~\bibnamefont {Sanfilippo}}, \bibinfo {author} {\bibfnamefont {S.~F.}\
  \bibnamefont {Schifano}}, \bibinfo {author} {\bibfnamefont {G.}~\bibnamefont
  {Silvi}}, \ and\ \bibinfo {author} {\bibfnamefont {R.}~\bibnamefont
  {Tripiccione}},\ }\href {\doibase 10.1103/PhysRevD.99.014502} {\bibfield
  {journal} {\bibinfo  {journal} {Phys.~Rev.}\ }\textbf {\bibinfo {volume} {D
  99}},\ \bibinfo {pages} {014502} (\bibinfo {year} {2019})},\ \Eprint
  {http://arxiv.org/abs/1807.02106} {arXiv:1807.02106 [hep-lat]} \BibitemShut
  {NoStop}%
%%CITATION = ARXIV:1807.02106;%%
\bibitem [{\citenamefont {Costa}\ \emph {et~al.}(2009)\citenamefont {Costa},
  \citenamefont {de~Sousa}, \citenamefont {Ruivo},\ and\ \citenamefont
  {Hansen}}]{Costa:2008gr}%
  \BibitemOpen
  \bibfield  {author} {\bibinfo {author} {\bibfnamefont {P.}~\bibnamefont
  {Costa}}, \bibinfo {author} {\bibfnamefont {C.~A.}\ \bibnamefont {de~Sousa}},
  \bibinfo {author} {\bibfnamefont {M.~C.}\ \bibnamefont {Ruivo}}, \ and\
  \bibinfo {author} {\bibfnamefont {H.}~\bibnamefont {Hansen}},\ }\href
  {\doibase 10.1209/0295-5075/86/31001} {\bibfield  {journal} {\bibinfo
  {journal} {Europhys. Lett.}\ }\textbf {\bibinfo {volume} {86}},\ \bibinfo
  {pages} {31001} (\bibinfo {year} {2009})},\ \Eprint
  {http://arxiv.org/abs/0801.3616} {arXiv:0801.3616 [hep-ph]} \BibitemShut
  {NoStop}%
%%CITATION = ARXIV:0801.3616;%%
\bibitem [{\citenamefont {Biguet}\ \emph {et~al.}(2015)\citenamefont {Biguet},
  \citenamefont {Hansen}, \citenamefont {Brugi{\`e}re}, \citenamefont {Costa},\
  and\ \citenamefont {Borgnat}}]{Biguet:2014sga}%
  \BibitemOpen
  \bibfield  {author} {\bibinfo {author} {\bibfnamefont {A.}~\bibnamefont
  {Biguet}}, \bibinfo {author} {\bibfnamefont {H.}~\bibnamefont {Hansen}},
  \bibinfo {author} {\bibfnamefont {T.}~\bibnamefont {Brugi{\`e}re}}, \bibinfo
  {author} {\bibfnamefont {P.}~\bibnamefont {Costa}}, \ and\ \bibinfo {author}
  {\bibfnamefont {P.}~\bibnamefont {Borgnat}},\ }\href {\doibase
  10.1140/epja/i2015-15121-1} {\bibfield  {journal} {\bibinfo  {journal} {Eur.
  Phys. J.}\ }\textbf {\bibinfo {volume} {A51}},\ \bibinfo {pages} {121}
  (\bibinfo {year} {2015})},\ \Eprint {http://arxiv.org/abs/1409.0990}
  {arXiv:1409.0990 [hep-ph]} \BibitemShut {NoStop}%
%%CITATION = ARXIV:1409.0990;%%
\bibitem [{\citenamefont {Wagner}\ \emph {et~al.}(2010)\citenamefont {Wagner},
  \citenamefont {Walther},\ and\ \citenamefont {Schaefer}}]{Wagner:2009pm}%
  \BibitemOpen
  \bibfield  {author} {\bibinfo {author} {\bibfnamefont {M.}~\bibnamefont
  {Wagner}}, \bibinfo {author} {\bibfnamefont {A.}~\bibnamefont {Walther}}, \
  and\ \bibinfo {author} {\bibfnamefont {B.-J.}\ \bibnamefont {Schaefer}},\
  }\href {\doibase 10.1016/j.cpc.2009.12.008} {\bibfield  {journal} {\bibinfo
  {journal} {Comput. Phys. Commun.}\ }\textbf {\bibinfo {volume} {181}},\
  \bibinfo {pages} {756} (\bibinfo {year} {2010})},\ \Eprint
  {http://arxiv.org/abs/0912.2208} {arXiv:0912.2208 [hep-ph]} \BibitemShut
  {NoStop}%
%%CITATION = ARXIV:0912.2208;%%
\bibitem [{\citenamefont {Kahara}\ and\ \citenamefont
  {Tuominen}(2010)}]{Kahara:2010wh}%
  \BibitemOpen
  \bibfield  {author} {\bibinfo {author} {\bibfnamefont {T.}~\bibnamefont
  {Kahara}}\ and\ \bibinfo {author} {\bibfnamefont {K.}~\bibnamefont
  {Tuominen}},\ }\href {\doibase 10.1103/PhysRevD.82.114026} {\bibfield
  {journal} {\bibinfo  {journal} {Phys. Rev.}\ }\textbf {\bibinfo {volume}
  {D82}},\ \bibinfo {pages} {114026} (\bibinfo {year} {2010})},\ \Eprint
  {http://arxiv.org/abs/1006.3931} {arXiv:1006.3931 [hep-ph]} \BibitemShut
  {NoStop}%
%%CITATION = ARXIV:1006.3931;%%
\bibitem [{\citenamefont {Endrodi}\ \emph {et~al.}(2011)\citenamefont
  {Endrodi}, \citenamefont {Fodor}, \citenamefont {Katz},\ and\ \citenamefont
  {Szabo}}]{Endrodi:2011gv}%
  \BibitemOpen
  \bibfield  {author} {\bibinfo {author} {\bibfnamefont {G.}~\bibnamefont
  {Endrodi}}, \bibinfo {author} {\bibfnamefont {Z.}~\bibnamefont {Fodor}},
  \bibinfo {author} {\bibfnamefont {S.~D.}\ \bibnamefont {Katz}}, \ and\
  \bibinfo {author} {\bibfnamefont {K.~K.}\ \bibnamefont {Szabo}},\ }\href
  {\doibase 10.1007/JHEP04(2011)001} {\bibfield  {journal} {\bibinfo  {journal}
  {JHEP}\ }\textbf {\bibinfo {volume} {04}},\ \bibinfo {pages} {001} (\bibinfo
  {year} {2011})},\ \Eprint {http://arxiv.org/abs/1102.1356} {arXiv:1102.1356
  [hep-lat]} \BibitemShut {NoStop}%
%%CITATION = ARXIV:1102.1356;%%
\end{thebibliography}%

\end{document}